\documentclass[12pt,oneside]{article}
\usepackage{graphicx}
\usepackage{breqn}

\begin{document}

\begin{center}
  {\large\bf Improvement of a provisional solution of the quantum-corrected
  field equations of $d = 11$ supergravity on\\
  flat $\mathbf{R}^4$ times a compact hyperbolic 7-manifold, and modes that
  decay along the beam line outside the interaction\\
  region at the LHC\\}
\vspace{0.14cm}
\vspace*{.05in}
{Chris Austin\footnote{Email: chris@chrisaustin.info}\\
\small 33 Collins Terrace, Maryport, Cumbria CA15 8DL, England\\
}
\end{center}

\begin{center}
{\bf Abstract}
\end{center}

\noindent A recent provisional solution of the quantum-corrected field equations of $d =
11$ \mbox{super}-gravity on flat $\mathbf{{R}}^4$ times a compact
hyperbolic 7-manifold $\bar{H}^7$, in the presence of magnetic 4-form fluxes
wrapping 4-cycles of $\bar{H}^7$, is improved by showing that the curvature
radius $B$ of $\bar{H}^7$ and the r.m.s.\hspace{-2.5pt} 4-form flux strength
$h$ each have a
single stationary point as the field redefinition parameter $c$ is varied. \
Application of the principle of minimal sensitivity then fixes $c$
in a moderate range such
that $B$ and $h^{1 / 3}$ vary by only 4\% and 5\% respectively over this
range. \ The new best value of $B$ is $\simeq 0.28 \kappa_{11}^{2 / 9}
\simeq 1.2 M_{11}^{- 1}$. \ The low-lying bosonic Kaluza-Klein modes of the
bulk are studied. \ The classically massless
harmonic 3-form modes of the
3-form gauge field, whose number is estimated as roughly $10^{32}$
if the intrinsic volume of $\bar{H}^7$ is $\sim 10^{35}$, acquire
a mass $\simeq 0.2 \frac{A}{B}$ from quantum corrections, where the warp
factor $A$ is fixed by the boundary conditions at the Ho\v{r}ava-Witten (HW)
boundary to lie between about 0.7 and 0.9. \ They have axion-like couplings
to the SM gauge bosons. \ Their lifetimes range from about
\mbox{$10^{-27}$ seconds} to several hours depending on the
distance of their centre from the HW boundary, and they can decay along the beam
line outside the interaction region at the LHC, with a distribution that shows
a power law rather than exponential decrease with distance from the IP. \
Approximate exclusion limits are obtained from recent LHC data, and discovery
prospects at ATLAS and CMS are studied.
\vspace{2.0cm}
\tableofcontents

\section{Introduction}
\label{Introduction}

The weakness of the gravitational interaction relative to the strong and
electroweak interactions would have a natural explanation if there existed $n
\geq 2$ compact extra spatial dimensions of volume $\sim 10^{31}$ TeV$^{- n}$
in which the gravitational field is diluted, while the Standard Model (SM)
fields are confined to a very small part of this region. \ The mass defining
the strength of gravity in $4 + n$ dimensions would then be around a TeV
{\cite{ADD1, AADD, ADD2}}.

The confinement of the SM fields to a small part of the $3 + n$ spatial
dimensions is natural in $d = 11$ supergravity {\cite{CJS}}, because the $d =
11$ supergravity multiplet does not couple to any matter fields in $10 + 1$
smooth dimensions, but can couple to matter fields on various types of
localized impurity, the simplest of which is a Ho\v{r}ava-Witten (HW) boundary
{\cite{Horava Witten 1, Horava Witten 2, Witten Strong Coupling, Moss 1, Moss
2, Moss 3, Moss 4}}.

If the 7 compact dimensions have the topology of a compact orientable
hyperbolic 7-manifold $\bar{H}^7$ that admits a spin structure {\cite{KMST}},
then when the metric on them is locally maximally symmetric, and their
curvature is fixed, they are completely rigid {\cite{Mostow 1,Mostow
2,Prasad,Thurston,Gromov Hyperbolic}}. \ Their shape and size is determined by
their topology, and they can be arbitrarily large. \ The number of distinct
topologies for which their volume is $< V$ grows as $V^{\sigma V}$ at large $V$,
where $\sigma > 0$ is a constant {\cite{Gelander, Burger Gelander Lubotzky Mozes}}.
\ The SM fields can be accommodated on a HW boundary
$\mathbf{{R}}^4 \times \bar{H}^6$ of
$\mathbf{{R}}^4 \times \bar{H}^7$, with a closed hyperbolic
Cartesian factor $\bar{H}^6$.

I shall use the notation and results of {\cite{Almost Flat}}, with some
improvements as follows.

In section 3 of {\cite{Almost Flat}}, a provisional solution of the
quantum-corrected Einstein equations of $d = 11$ supergravity on flat
$\mathbf{{R}}^4$ times $\bar{H}^7$, in the presence of magnetic
4-form fluxes $H_{I J K L} = \partial_I C_{J K L} - \partial_J C_{K L I} +
\partial_K C_{L I J} - \partial_L C_{I J K}$ of the 3-form gauge field $C_{I J
K}$ of $d = 11$ supergravity wrapping 4-cycles of $\bar{H}^7$, was obtained in
the approximation of working to leading order in the Lukas-Ovrut-Waldram (LOW)
harmonic expansion of the energy-momentum tensor on $\bar{H}^7$ {\cite{Lukas
Ovrut Waldram}}. \ The 4-form fluxes were assumed to be proportional to
harmonic 4-forms on $\bar{H}^7$, and thus to solve the classical CJS field
equations for $H_{I J K L}$, and the quantum corrections to those field
equations were neglected. \ The fluxes were assumed to be approximately
uniformly distributed across $\bar{H}^7$, so that the LOW expansion only
needed to be applied over relatively small local regions of $\bar{H}^7$, and
to leading order in the LOW expansion, the flux bilinears were assumed to have
the form:
\begin{equation}
\label{flux bilinears}
   H_{ABEF} H_{CDGH} G^{EG} G^{FH} = \frac{h^2}{B^8} 
  \left( G_{AC} G_{BD} - G_{BC} G_{AD} \right),
\end{equation}
where $h \geq 0$ is a constant of dimension length$^3$, $B$ is the curvature
radius of $\bar{H}^7$, and $G_{I J}$ is the $d = 11$ metric, which on
$\bar{H}^7$ has the form $G_{A B} = B^2 \bar{g}_{A B}$, where $\bar{g}_{A B}$
is a metric of constant sectional curvature $- 1$ on $\bar{H}^7$. \ Coordinate
indices $I, J, K, \ldots$ run over all 11 dimensions, coordinate indices $\mu,
\nu, \sigma, \ldots$ are tangential to the four extended space-time
dimensions, and coordinate indices $A, B, C, \ldots$ are tangential to
$\bar{H}^7$. \ The coordinates are $x^I = \left( \check{x}^{\mu}, \bar{x}^A
\right)$. \ The metric is mostly $+$, and units such that $\hbar = c = 1$ are
used. \ The dependence on $B$ is fixed by the fact that $H_{A B C D}$ is
independent of $B$, because the integral of $H_{A B C D} d \bar{x}^A d
\bar{x}^B d \bar{x}^C d \bar{x}^D$ over a 4-cycle of $\bar{H}^7$ is quantized,
and independent of $B$ {\cite{Rohm Witten, Witten flux quantization}}.

The bosonic part of the quantum-corrected action on the 11-dimensional bulk
was assumed to have the form:
\begin{equation}
  \label{Gamma SG} \Gamma^{\left( \mathrm{bos} \right)}_{\mathrm{SG}} =
  S^{\left( \mathrm{bos} \right)}_{\mathrm{CJS}} + \Gamma^{\left( 8,
  \mathrm{bos} \right)}_{\mathrm{SG}},
\end{equation}
where $S^{\left( \mathrm{bos} \right)}_{\mathrm{CJS}}$ is the bosonic part of
the classical action of $d = 11$ supergravity {\cite{CJS}}:
\begin{equation}
  \label{CJS action} S^{\left( \mathrm{bos} \right)}_{\mathrm{CJS}} =
  \frac{1}{2 \kappa_{11}^2}  \int_{\mathcal{B}} d^{11} xe \left( R -
  \frac{1}{48} H_{IJKL} H^{IJKL} - \frac{1}{144^2} \epsilon^{I_1 \ldots
  I_{11}}_{\left( 11 \right)} C_{I_1 I_2 I_3} H_{I_4 \ldots I_7} H_{I_8 \ldots
  I_{11}} \right),
\end{equation}
and for covariantly constant fluxes, the leading quantum correction
$\Gamma^{\left( 8, \mathrm{bos} \right)}_{\mathrm{SG}}$ is a dimension 8 local
term of the form {\cite{Tseytlin}}:
\begin{equation}
  \label{Gamma 8 SG} \Gamma^{\left( 8, \mathrm{bos} \right)}_{\mathrm{SG}} =
  \frac{1}{147456 \pi^2 \kappa_{11}^2} \left( \frac{\kappa_{11}}{4 \pi}
  \right)^{4 / 3}  \int_{\mathcal{B}} d^{11} xe \left( t_8 t_8  \breve{R}^4 +
  Z - \frac{1}{6} \epsilon_{11} t_8 CR^4 + c \left( \tilde{Z} - Z \right)
  \right) .
\end{equation}
Here $\mathcal{B}$ means the bulk, $\kappa_{11}$ is the gravitational coupling
constant in 11 dimensions, \ $e = \sqrt{- G}$ is the determinant of the
vielbein $e_{I \hat{J}}$, where $G$ is the determinant of the metric $G_{IJ}$,
and the antisymmetric tensor $\epsilon^{I_1 \ldots I_{11}}_{\left( 11
\right)}$ is related to the $\mathrm{SO} \left( 10, 1 \right)$ invariant
tensor $\epsilon^{\hat{I}_1 \ldots \hat{I}_{11}}_{\left( 11 \right)}$, with
components $0, \pm 1$, by $\epsilon^{I_1 \ldots I_{11}}_{\left( 11 \right)} =
e^{I_1}  \hspace{0.25em} \hspace{-0.25em}_{\hat{J}_1} \ldots e^{I_{11}} 
\hspace{0.25em} \hspace{-0.25em}_{\hat{J}_{11}} \epsilon^{\hat{J}_1 \ldots
\hat{J}_{11}}_{\left( 11 \right)}$. \ Hatted indices are local Lorentz
indices. \ The Riemann tensor for the metric $G_{IJ}$ is defined by:
\begin{equation}
  \label{Riemann tensor} R_{IJ} \hspace{0.25em} \hspace{-0.25em}^K
  \hspace{0.25em} \hspace{-0.25em}_L = \partial_I \Gamma_J \hspace{0.25em}
  \hspace{-0.25em}^K \hspace{0.25em} \hspace{-0.25em}_L - \partial_J \Gamma_I
  \hspace{0.25em} \hspace{-0.25em}^K \hspace{0.25em} \hspace{-0.25em}_L +
  \Gamma_I  \hspace{0.25em} \hspace{-0.25em}^K \hspace{0.25em}
  \hspace{-0.25em}_M \Gamma_J \hspace{0.25em} \hspace{-0.25em}^M
  \hspace{0.25em} \hspace{-0.25em}_L - \Gamma_J  \hspace{0.25em}
  \hspace{-0.25em}^K \hspace{0.25em} \hspace{-0.25em}_M \Gamma_I
  \hspace{0.25em} \hspace{-0.25em}^M \hspace{0.25em} \hspace{-0.25em}_L,
\end{equation}
where the standard Christoffel connection is $\Gamma_I \hspace{0.25em}
\hspace{-0.25em}^J \hspace{0.25em} \hspace{-0.25em}_K = \frac{1}{2} G^{JL} 
\left( \partial_I G_{LK} + \partial_K G_{LI} - \partial_L G_{IK} \right)$, and
the Ricci tensor and scalar are defined by $R_{IJ} = R_{KI} \hspace{0.25em}
\hspace{-0.25em}^K \hspace{0.25em} \hspace{-0.25em}_J$, and $R = G^{IJ}
R_{IJ}$. \ The modified Riemann tensor $\breve{R}_{I J K L}$ is defined by:
\begin{equation}
  \label{R cup I J K L} \breve{R}_{IJKL} = R_{IJKL} - \frac{1}{8} H_{IKMN}
  H_{JL} \hspace{0.25em} \hspace{-0.25em}^{MN} + \frac{1}{8} H_{JKMN} H_{IL}
  \hspace{0.25em} \hspace{-0.25em}^{MN} .
\end{equation}
The notation $t_8 t_8  \breve{R}^4$ is a shorthand for
\begin{equation}
  \label{t8 t8 R4} t^{I_1 I_2 J_1 J_2 K_1 K_2 L_1 L_2}_8 t^{M_1 M_2 N_1 N_2
  O_1 O_1 P_1 P_2}_8  \breve{R}_{I_1 I_2 M_1 M_2}  \breve{R}_{J_1 J_2 N_1 N_2}
  \breve{R}_{K_1 K_2 O_1 O_2}  \breve{R}_{L_1 L_2 P_1 P_2},
\end{equation}
where $t^{IJKLMNOP}_8$ is a tensor built from $G^{IJ}$ and antisymmetric in
each successive pair of indices, such that for antisymmetric tensors $A_{IJ}$,
$B_{IJ}$, $C_{IJ}$, and $D_{IJ}$:
\[ t^{IJKLMNOP}_8 A_{IJ} B_{KL} C_{MN} D_{OP} = \hspace{6cm} \]
\[ = 8 \left( \mathrm{tr} \left( ABCD \right) + \mathrm{tr} \left( ACBD
   \right) + \mathrm{tr} \left( ACDB \right) \right) \]
\[ - 2 \left( \mathrm{tr} \left( AB \right) \mathrm{tr} \left( CD \right) +
   \mathrm{tr} \left( AC \right) \mathrm{tr} \left( BD \right) + \mathrm{tr}
   \left( AD \right) \mathrm{tr} \left( BC \right) \right) = \]
\[ = 8 \left( A_{IJ} B_{JK} C_{KL} D_{LI} + A_{IJ} C_{JK} B_{KL} D_{LI} +
   A_{IJ} C_{JK} D_{KL} B_{LI} \right) \]
\begin{equation}
  \label{t8 A B C D} - 2 \left( A_{IJ} B_{JI} C_{KL} D_{LK} + A_{IJ} C_{JI}
  B_{KL} D_{LK} + A_{IJ} D_{JI} B_{KL} C_{LK} \right)
\end{equation}
Repeated lower coordinate indices are understood to be contracted with an
inverse metric, for example $A_{IJ} B_{JK} \equiv A_I  \hspace{0.25em}
\hspace{-0.25em}^J B_{JK} = G^{JL} A_{IL} B_{JK}$. \ Thus:
\begin{dmath}[compact, spread=3pt]
  \label{t8 t8 X4} t_8 t_8  \breve{R}^4 = 12 \hspace{0.25em} \breve{R}_{IJMN} 
  \breve{R}_{I JM N}  \breve{R}_{KLOP}  \breve{R}_{K LO P} + 24 \breve{R}_{I
  JMN}  \hspace{0.25em} \breve{R}_{IJOP}  \breve{R}_{KLO P}  \breve{R}_{K LM
  N} - 96 \breve{R}_{I JM N} \hspace{0.25em} \breve{R}_{IJM P}  \breve{R}_{KLO
  N}  \breve{R}_{K LO P} - 48 \breve{R}_{I JMN} \hspace{0.25em}
  \breve{R}_{IJOP}  \breve{R}_{KLM P}  \breve{R}_{K LO N} - 96 \hspace{0.25em}
  \breve{R}_{IJMN}  \breve{R}_{K JM N}  \breve{R}_{I LO P}  \breve{R}_{KLOP} -
  48 \hspace{0.25em} \breve{R}_{IJMN}  \breve{R}_{KLM N}  \breve{R}_{I LOP} 
  \breve{R}_{K JO P} + 192 \hspace{0.25em} \breve{R}_{IJMN}  \breve{R}_{K JO
  N}  \breve{R}_{KLOP}  \breve{R}_{I LM P} + 384 \hspace{0.25em} \breve{R}_{I
  JM N}  \breve{R}_{ILMP}  \breve{R}_{KJO P}  \breve{R}_{K LON} .
\end{dmath}
Similarly:
\begin{dmath}[compact, spread=3pt]
  \label{eps11 t8 C R4} \frac{1}{6} \epsilon_{11} t_8 CR^4 \equiv 4
  \epsilon_{\left( 11 \right)}^{I_1 \ldots I_{11}} C_{I_1 I_2 I_3} R_{I_4 I_5
  JK} R_{I_6 I_7 KL} R_{I_8 I_9 LM} R_{I_{10} I_{11} MJ} - \epsilon_{\left( 11
  \right)}^{I_1 \ldots I_{11}} C_{I_1 I_2 I_3} R_{I_4 I_5 JK} R_{I_6 I_7 KJ}
  R_{I_8 I_9 LM} R_{I_{10} I_{11} ML} .
\end{dmath}
And:
\begin{dmath}[compact, spread=3pt]
  \label{definition of Z} Z \equiv - \frac{1}{4!} \epsilon_{11} \epsilon_{11} 
  \breve{R}^4 \equiv - \frac{1}{4!} \epsilon_{\left( 11 \right)}^{IJKL_1 L_2
  M_1 M_2 N_1 N_2 O_1 O_2} \epsilon_{\left( 11 \right) IJKP_1 P_2 Q_1 Q_2 R_1
  R_2 S_1 S_2} \times \breve{R}_{L_1 L_2}  \hspace{0.25em}
  \hspace{-0.25em}^{P_1 P_2} \breve{R}_{M_1 M_2}  \hspace{0.25em}
  \hspace{-0.25em}^{Q_1 Q_2} \breve{R}_{N_1 N_2}  \hspace{0.25em}
  \hspace{-0.25em}^{R_1 R_2} \breve{R}_{O_1 O_2} \hspace{0.25em}
  \hspace{-0.25em}^{S_1 S_2} = \frac{8!}{4}  \breve{R}_{L_1 L_2} 
  \hspace{0.25em} \hspace{-0.25em}^{\left[ L_1 L_2 \right.} \breve{R}_{M_1
  M_2}  \hspace{0.25em} \hspace{-0.25em}^{M_1 M_2} \breve{R}_{N_1 N_2} 
  \hspace{0.25em} \hspace{-0.25em}^{N_1 N_2} \breve{R}_{O_1 O_2}
  \hspace{0.25em} \hspace{-0.25em}^{\left. O_1 O_2 \right]},
\end{dmath}
$\tilde{Z}$ is the result of using the classical Einstein equations following
from (\ref{CJS action}):
\begin{equation}
  \label{Einstein equations} R_{IJ} - \frac{1}{2} RG_{IJ} - \frac{1}{12} H_I 
  \hspace{0.25em} \hspace{-0.25em}^{KLM} H_{JKLM} + \frac{1}{96} H^{KLMN}
  H_{KLMN} G_{IJ} = 0,
\end{equation}
to replace all Ricci tensors and Ricci scalars resulting from writing out the
antisymmetrization of the upper indices, in the final form of (\ref{definition
of Z}), by bilinears in $H_{I J K L}$, and $c$ is a coefficient.

The coefficient of the $\epsilon_{11} t_8 CR^4$ term in (\ref{Gamma 8 SG}),
which is known as the Green-Schwarz term because of its role in anomaly
cancellation {\cite{Green Schwarz}}, is fixed absolutely by anomaly
cancellation on five-branes {\cite{Duff Liu Minasian,Witten 5 branes,Freed
Harvey Minasian Moore,Bilal Metzger,Harvey}}, and confirmed by anomaly
cancellation in Ho\v{r}ava-Witten theory {\cite{de Alwis,Conrad,Faux Lust
Ovrut,Lu,Bilal Derendinger Sauser,Harmark,Bilal Metzger,Meissner
Olechowski,Moss 3}}.

The relative coefficients of all terms in (\ref{Gamma 8 SG}) are fixed by
supersymmetry, up to the fact that arbitrary multiples of linear combinations
of terms that vanish when the classical Einstein equations (\ref{Einstein
equations}), and the classical equations:
\begin{equation}
  \label{3 form field equations} D_L H^{LIJK} - \frac{1}{3456}
  \epsilon^{IJKLMNOPQRS}_{\left( 11 \right)} H_{LMNO} H_{PQRS} = 0
\end{equation}
for $C_{I J K}$ that follow from (\ref{CJS action}), are satisfied can be
added, because the overall coefficients of such linear combinations of terms
can be adjusted arbitrarily by making redefinitions of the fields of the form
$G_{IJ}, C_{IJK} \rightarrow G_{IJ} + \kappa_{11}^{4 / 3} X_{IJ}, C_{IJK} +
\kappa_{11}^{4 / 3} Y_{IJK}$, where $X_{IJ}$ and $Y_{IJK}$ are dimension 6
polynomials in the fields and their derivatives {\cite{Hyakutake Ogushi
2,Hyakutake,Metsaev}}. \ The $c \left( \tilde{Z} - Z \right)$ term has been
included to allow for this ambiguity. \ When $c = 0$, compactification of
(\ref{Gamma 8 SG}) on a small $S^1$ gives the form that arises naturally from
type IIA superstring scattering amplitudes {\cite{Richards 1, Richards 2}},
while $c = 1$ gives the form of (\ref{Gamma 8 SG}) that arises naturally when
supersymmetry is systematically implemented by the Noether method
{\cite{Hyakutake Ogushi 1,Hyakutake Ogushi 2,Hyakutake,Hyakutake 2}}.

Field redefinitions of this type are like a change of coordinates in ``field
space'', so they do not change the physical content of the theory. \ In
particular, they do not alter the $S$-matrix {\cite{Diagrammar,Lee Les
Houches,Tseytlin Field Redefinitions}}. \ The collection of all such field
redefinitions, with $X_{I J}$ and $Y_{I J K}$ generalized to expansions of the
form $\sum_{n \geq 0} \kappa^{2 n / 3}_{11} X^{\left( n \right)}_{I J}$ and
$\sum_{n \geq 0} \kappa^{2 n / 3}_{11} Y^{\left( n \right)}_{I J K}$, where
$X^{\left( n \right)}_{I J}$ and $Y^{\left( n \right)}_{I J K}$ are dimension
$6 + 3 n$ polynomials in the fields and their derivatives, forms a ``field
redefinition group'', that generalizes the renormalization group of
renormalizable quantum field theories.

At low orders of perturbation theory field redefinitions do affect physical
quantities, and can even affect whether a particular type of solution of the
quantum-corrected field equations exists or not. \ Sensitivity to field
redefinitions should decrease as higher-order corrections are included, so at
low orders of perturbation theory, we should use the {\itshape{principle of
minimal sensitivity}} (PMS) {\cite{Stevenson}} to choose the best field
redefinition, as in perturbative QCD. \ Doing that should minimize the size of
the higher-order corrections. \ In QCD the PMS resolves the renormalization
scheme ambiguity, and here it means using ``coordinates in field space'' best
suited to the geometry being studied.

With the exception of the $\epsilon_{11} t_8 CR^4$ term, and the 4-field parts
of terms that depend on $H_{I J K L}$ through $D_I H_{J K L M}$, and thus
vanish for covariantly constant fluxes {\cite{Deser Seminara 1,Deser Seminara
2,Peeters Plefka Stern,Metsaev}}, the $C_{I J K}$-dependent terms in
$\Gamma^{\left( 8, \mathrm{bos} \right)}_{\mathrm{SG}}$ are not yet known. \
Their inclusion through the modified Riemann tensor $\breve{R}_{I J K L}$
defined in (\ref{R cup I J K L}) is a guess such that if (\ref{Gamma 8 SG}) is
compactified on a small $S^1$, such that the fields are covariantly constant
and only the fields of the type I supergravity multiplet in 10 dimensions are
nonzero, it agrees with the Kehagias-Partouche (KP) conjecture for the
completion of the dimension 8 local term in 10 dimensions {\cite{Kehagias
Partouche, Gross Sloan}}, up to correction terms that contain factors that
occur in the classical Einstein equation in 10 dimensions. \ The KP conjecture
is supported by recent calculations by Richards {\cite{Richards 2}}, but was
shown in section 2 of {\cite{Almost Flat}} to require a correction, because it
cannot be oxidized as it stands to a generally covariant formula in 11
dimensions.

The metric in the bulk, away from the immediate vicinity of the HW boundary,
is assumed to have the form:
\begin{equation}
  \label{metric ansatz for H7} ds_{11}^2 = G_{IJ} dx^I dx^J = A^2 \eta_{\mu
  \nu} d \check{x}^{\mu} d \check{x}^{\nu} + B^2  \bar{g}_{A B} d \bar{x}^A d
  \bar{x}^B,
\end{equation}
where $\eta_{\mu \nu} = \mathrm{{{diag}}} \left( - 1,
1, 1, 1 \right)$ is the metric on $\left( 3 + 1 \right)$-dimensional Minkowski
space, and $A$ is a constant.

Because the leading quantum correction (\ref{Gamma 8 SG}) is a local term,
independent of the topology of $\bar{H}^7$, the assumption (\ref{flux
bilinears}) means that the quantum-corrected Einstein equations at this order
are consistent with the locally maximally symmetric metric ansatz (\ref{metric
ansatz for H7}), so by Palais's Principle of Symmetric Criticality
{\cite{Palais,Deser Franklin,Torre}}, the Einstein equations can be derived by
substituting (\ref{metric ansatz for H7}) into the action (\ref{Gamma SG}),
and varying with respect to the constants $A$ and $B$.

More directly, when we expand the quantum-corrected action (\ref{Gamma SG})
for a general perturbation $\bar{\bar{G}}_{I J} \equiv G_{I J} + 2 h_{I
J}$ of the metric ansatz (\ref{metric ansatz for H7}) in powers of the
perturbation tensor $h_{I J}$, the result is a sum of local terms built from
$G_{I J}$, its Riemann tensor $R_{I J K L}$, the covariant derivatives $D_I$
built from $G_{I J}$ that satisfy $D_I G_{J K} \equiv 0$, and the tensor $h_{I
J}$, and for the terms linear in $h_{I J}$ we can remove all covariant
derivatives from $h_{I J}$ by integrations by parts, and those covariant
derivatives then give 0 by the local symmetry of (\ref{metric ansatz for H7}).
\ The quantum-corrected Einstein equations for (\ref{metric ansatz for H7})
are thus proportional to the metric blocks $G_{\mu \nu}$ and $G_{A B}$
corresponding to the irreducible locally symmetric space Cartesian factors of
the 11-dimensional product space, and the proportionality factors can be
obtained by using, for example, $\frac{\delta \Gamma^{\left( \mathrm{bos}
\right)}_{\mathrm{SG}}}{\delta B} = \frac{\delta G_{A B}}{\delta B} 
\frac{\delta \Gamma^{\left( \mathrm{bos} \right)}_{\mathrm{SG}}}{\delta G_{A
B}} = \frac{2}{B} G_{A B} \frac{\delta \Gamma^{\left( \mathrm{bos}
\right)}_{\mathrm{SG}}}{\delta G_{A B}}$.

In terms of rescaled parameters:
\begin{equation}
  \label{rescaled parameters} \tilde{B} \equiv \frac{2^{\frac{29}{18}}
  \pi^{\frac{5}{9}}}{21^{\frac{1}{3}}}  \hspace{0.25em}
  \frac{B}{\kappa_{11}^{2 / 9}}, \hspace{1.5cm} \hspace{1cm} \tilde{h} \equiv
  \frac{2^{\frac{29}{6}} \pi^{\frac{5}{3}}}{21}  \hspace{0.25em}
  \frac{h}{\kappa_{11}^{2 / 3}},
\end{equation}
the action density, after substituting in the metric ansatz (\ref{metric
ansatz for H7}), is the square root of the determinant of $\bar{g}_{A B}$,
times:
\begin{dmath}[compact, spread=3pt]
  \label{action with field redefinition term}
  \frac{21^{\frac{5}{3}}}{2^{\frac{163}{18}} \pi^{\frac{25}{9}} \kappa_{11}^{8
  / 9}} A^4  \tilde{B}^7  \left( - \frac{42}{\tilde{B}^2} - \frac{7
  \tilde{h}^2}{8 \tilde{B}^8} + \left( \frac{1}{\tilde{B}^2} +
  \frac{\tilde{h}^2}{8 \tilde{B}^8} \right)^4 + c \left( - \frac{19}{21
  \tilde{B}^8} + \frac{89 \tilde{h}^2}{189 \tilde{B}^{14}} + \frac{23
  \tilde{h}^4}{10368 \tilde{B}^{20}} + \frac{20395 \tilde{h}^6}{870912
  \tilde{B}^{26}} + \frac{2225255 \tilde{h}^8}{3009871872 \tilde{B}^{32}}
  \right) \right) .
\end{dmath}
This is obtained in the approximation of using (\ref{flux bilinears}) for the
$HH$ terms in (\ref{R cup I J K L}) to obtain $\breve{R}_{A B C D} = \left(
\frac{1}{B^2} + \frac{h^2}{8 B^8} \right) \left( G_{AD} G_{BC} - G_{AC} G_{BD}
\right)$, and then substituting this into (\ref{Gamma 8 SG}), together with
the corresponding treatment of $\tilde{Z}$, instead of first expanding
(\ref{Gamma 8 SG}) in powers of $H$ and summing over all pairings of factors
of $H$ before using (\ref{flux bilinears}) and the corresponding equation with
no index contractions, as would be required for $H$ to be a Gaussian random
variable with mean zero and mean square fixed by (\ref{flux bilinears}).

Defining:
\begin{equation}
  \label{definition of eta} \eta \equiv \frac{\tilde{h}}{\tilde{B}^3} =
  \frac{h}{B^3},
\end{equation}
the field equations reduce to:
\begin{dmath}[compact, spread=3pt]
  \label{A field equation} \left( 2633637888 \eta^2 + 126414618624 \right) 
  \tilde{B}^6 - 734832 \eta^8 - 23514624 \eta^6 - 282175488 \eta^4 -
  1504935936 \eta^2 - 3009871872 - \left( 2225255 \eta^8 + 70485120 \eta^6 +
  6676992 \eta^4 + 1417347072 \eta^2 - 2723217408 \right) c = 0,
\end{dmath}
and
\begin{dmath}[compact, spread=3pt]
  \label{B field equation} \left( 2633637888 \eta^2 - 632073093120 \right) 
  \tilde{B}^6 - 18370800 \eta^8 - 446777856 \eta^6 - 3668281344 \eta^4 -
  10534551552 \eta^2 - 3009871872 - \left( 55631375 \eta^8 + 1339217280 \eta^6
  + 86800896 \eta^4 + 9921429504 \eta^2 - 2723217408 \right) c = 0.
\end{dmath}
Solving these as two simultaneous linear equations for $c$ and $\tilde{B}^6$,
we find:
\begin{equation}
  \label{csol} c = - \frac{734832}{\mathcal{P} \left( \eta \right)}  \left(
  \eta^2 + 8 \right)^3  \left( \eta^4 + 60 \eta^2 + 96 \right),
\end{equation}
\begin{equation}
  \label{B tilde to the 6 sol} \tilde{B}^6 = \frac{\eta^2  \left( \eta^2 + 8
  \right)^3  \left( 22595 \eta^6 + 52446528 \eta^4 - 129537792 \eta^2 +
  694738944 \right)}{448\mathcal{P} \left( \eta \right)},
\end{equation}
where
\begin{dmath}[compact, spread=3pt]
  \label{definition of P of x} \mathcal{P} \left( \eta \right) \equiv 2225255
  \eta^{10} + 186379140 \eta^8 + 3386624256 \eta^6 + 594708480 \eta^4 +
  34016329728 \eta^2 - 32678608896.
\end{dmath}
The polynomial $\mathcal{P} \left( \eta \right)$ is positive for $\eta^2 >
\eta^2_{\mathrm{\min}}$ and negative for $\eta^2 < \eta^2_{\mathrm{\min}}$,
where $\eta_{\mathrm{\min}} \simeq 0.9364$, and has no real zeros other than
$\eta = \pm \eta_{\mathrm{\min}}$. \ $c$ is a monotonically increasing
function of $\eta$ for $\eta > \eta_{\mathrm{\min}}$, and tends to the limit
$c_{\mathrm{\max}} \equiv - \frac{734832}{2225255} \simeq - 0.3302$ as $\eta
\rightarrow + \infty$. \ The product $\tilde{B}^6 \mathcal{P} \left( \eta
\right)$ is $\geq 0$ for all $\eta$, so a solution with real $\tilde{B}$ only
exists for $\eta^2 > \eta^2_{\mathrm{\min}}$, and thus only for $c <
c_{\mathrm{\max}}$. \ $\tilde{B}$ and $\tilde{h}$ are positive, so we only
need to consider the region $\eta > \eta_{\mathrm{\min}}$, and in this region,
(\ref{csol}) determines $\eta$ implicitly as a function of the field
redefinition parameter $c$, and (\ref{B tilde to the 6 sol}) then determines
$\tilde{B}$ as a function of $c$.

\subsection{Application of the Principle of Minimal Sensitivity}
\label{subsection PMS}

If the system is physically sensible, then by the PMS, there should be values
of the field redefinition parameter $c$, not too far apart, where the solution
exists, and $\frac{d \tilde{B}}{dc}$ and $\frac{d \tilde{h}}{dc}$ are
respectively 0. \ $c$ should then be chosen somewhere between these values, in
order to minimize the size of the higher order corrections. \ The PMS was not
applied properly in {\cite{Almost Flat}}, because only the dependence of
$\eta$, there called $x$, on $c$ was considered, and it was then necessary to
make an ad hoc choice of $c$.

However $\frac{d \tilde{B}}{d \eta}$ and $\frac{d \tilde{h}}{d \eta}$, and
consequently also $\frac{d \tilde{B}}{dc}$ and $\frac{d \tilde{h}}{dc}$, each
have exactly one zero for $\eta > \eta_{\mathrm{\min}}$. \ For $\frac{d
\tilde{B}}{d \eta}$ and $\frac{d \tilde{B}}{dc}$, the zero is at $\eta \simeq
1.700$, which corresponds to $c \simeq - 1.590$, $\tilde{B} \simeq 0.580$,
$\tilde{h} \simeq 0.332$, $B \simeq 0.277 \kappa^{2 / 9}_{11}$, and $h \simeq
0.0363 \kappa^{2 / 3}_{11}$, and for $\frac{d \tilde{h}}{d \eta}$ and $\frac{d
\tilde{h}}{dc}$, the zero is at $\eta \simeq 1.291$, which corresponds to $c
\simeq - 3.083$, $\tilde{B} \simeq 0.605$, $\tilde{h} \simeq 0.286$, $B \simeq
0.289 \kappa^{2 / 9}_{11}$, and $h \simeq 0.0312 \kappa^{2 / 3}_{11}$. \ These
values give the range of values of the field redefinition parameter $c$, and
the physical quantities $B$ and $h$, selected by the PMS at this order of
perturbation theory. \ The unphysical parameter $c$ varies by almost a factor
of 2 over this range, but $B$ and $h^{1 / 3}$ vary by only 4\% and 5\%
respectively over this range. \ The ad hoc value of $c$ chosen in
{\cite{Almost Flat}} does not lie in this range, so it is necessary to
reconsider some of the conclusions of {\cite{Almost Flat}}.

The mean of the values of $c$ at the ends of the selected interval is $c
\simeq - 2.337$, which corresponds to $\eta \simeq 1.425$, $B \simeq 0.282
\kappa^{2 / 9}_{11}$, and $h \simeq 0.0320 \kappa^{2 / 3}_{11}$. \ I shall use
$B \simeq 0.28 \kappa^{2 / 9}_{11}$ as the best value of $B$.

The conclusion on page 42 of {\cite{Almost Flat}} that $b_1$, the curvature
radius of the closed hyperbolic Cartesian factor $\bar{H}^6$ of the HW
boundary, lies in the range $0.97 B$ to $1.00 B$ is unaltered. \ Thus from
equation (107) on that page, the Giudice-Rattazzi-Wells perturbativity
criterion {\cite{Giudice Rattazzi Wells}} is still satisfied by a large
margin, both in the bulk and on the HW boundary.

The second derivative of (\ref{action with field redefinition term}) with
respect to $\tilde{B}$, when (\ref{csol}) and (\ref{B tilde to the 6 sol}) are
satisfied, is:
\begin{dmath}[compact, spread=3pt]
  \label{second derivative of action wrt B} - \frac{3^{\frac{14}{3}}
  7^{\frac{5}{3}} \eta^2  \left( \eta^2 + 8 \right)^2 A^4}{2^{\frac{253}{18}}
  \pi^{\frac{25}{9}} \kappa_{11}^{8 / 9}  \tilde{B}^3 \mathcal{P} \left( \eta
  \right)}  \left( 22595 \eta^{10} + 36771952 \eta^8 + 3107227616 \eta^6 +
  1531643904 \eta^4 + 1459150848 \eta^2 + 44463292416 \right)
\end{dmath}
Thus the solution is a minimum of the potential energy for all $\eta >
\eta_{\mathrm{\min}}$. \ I shall show in subsection \ref{KK modes of metric},
starting on page \pageref{KK modes of metric}, that (\ref{second derivative of
action wrt B}) gives a first approximation to the mass
$m_{\mathrm{{{dil}}}}$ of the dilaton/radion, as seen on the HW boundary, of
$m_{\mathrm{{{dil}}}} \simeq 9\frac{A}{B} \simeq 30 A
\kappa^{- 2 / 9}_{11} \simeq 7 A M_{11}$, where the warp factor $A$ will be
found below to be fixed by the boundary conditions at the HW boundary to lie
between about 0.7 and 0.9.

The diameter $L$ of a compact manifold is by definition the maximum over all
pairs of points of the manifold of the shortest geodesic distance between
them. \ The {\itshape{intrinsic volume}} and {\itshape{intrinsic diameter}} of
a compact hyperbolic manifold are its volume and diameter when the metric on
it is locally maximally symmetric, with sectional curvature equal to $- 1$.

From equation (111) on page 44 of {\cite{Almost Flat}}, with $b_1 \simeq B$,
the intrinsic volume $\bar{V}_6$ of the closed hyperbolic Cartesian factor
$\bar{H}^6$ of the HW boundary is now estimated to lie in the range from about
270000 to about 580000, corresponding to an Euler number $\chi \left(
\bar{H}^6 \right)$ in the range from about $- 16000$ to about $- 35000$, where
the uncertainty arises mainly from the uncertainty of the value $\alpha_U$ of
the QCD fine structure constant $\alpha_s = \frac{g^2_s}{4 \pi}$ at
unification. \ Thus if $\bar{H}^6$ is reasonably isotropic, in the sense that
it has an approximately spherical fundamental domain in 6-dimensional
hyperbolic space $H^6$, then from equation (9) on page 9 of {\cite{Almost
Flat}}, with $S_5 = \pi^3$, its intrinsic diameter $\bar{L}_6$ lies between
about 5.7 and 6.0.

Moss's improved form of Ho\v{r}ava-Witten theory is used {\cite{Moss 1,Moss
2,Moss 3,Moss 4}}. \ In the region of the HW boundary, the coordinates $x^I$
have the form $\left( \tilde{x}^U, y \right)$, where indices $U, V, W, \ldots$
are tangential to a family of hypersurfaces foliating the $\left( 10 + 1
\right)$-dimensional manifold-with-boundary, one of these hypersurfaces
coinciding with the boundary, and $y$ takes a constant value on each of these
hypersurfaces, with the value of $y$ distinguishing the hypersurfaces. \ $y$
takes the value $y_1$ on the boundary, and $y > y_1$ in the bulk. \ The symbol
$y$ is also used as the coordinate index for the $y$ coordinate.

The boundary is equivalent to a double-sided mirror at $y = y_1$, such that
all the fields on one side of the mirror are exactly copied, up to sign, on
the other side of the mirror, with the fields at $\left( \tilde{x}^U, y
\right)$ mapped to the fields at $\left( \tilde{x}^U, 2 y_1 - y \right)$. \
The Yang-Mills multiplet is adjacent to the mirror, but infinitesimally
displaced from it, so that it has its own reflection infinitesimally on the
other side of the mirror {\cite{Lu}}, and I shall represent this by writing
the $y$ coordinate of the Yang-Mills multiplet as $y = y_{1 +}$.

The bosonic part of the Yang-Mills term in the semi-classical action on the HW
boundary is:
\begin{equation}
  \label{Yang Mills term} S^{\left( \mathrm{bos} \right)}_{\mathrm{YM}} = -
  \frac{1}{16 \pi \kappa^2_{11}} \left( \frac{\kappa_{11}}{4 \pi} \right)^{2 /
  3} \!\! \int_{y = y_{1 +}} \!\!\!\! d^{10} 
  \tilde{x} \, \tilde{e}  \left( \frac{1}{30} \mathrm{tr} F_{UV} F^{UV} -
  \frac{1}{2}  \bar{R}_{UV \hat{W}  \hat{X}}  \bar{R}^{UV \hat{W}  \hat{X}}
  \right) .
\end{equation}
Here $F_{UV} = \partial_U A_V - \partial_V A_U + i \left[ A_U, A_V \right]$ is
the field strength of the $E_8$ Yang-Mills gauge field $A_U = T^{\mathcal{A}}
A_U^{\mathcal{A}}$ localized at $y = y_{1 +}$, indices $\mathcal{A},
\mathcal{B}, \ldots$ run over the 248 generators of $E_8$, and the hermitian
generators $T^{\mathcal{A}}$ in the fundamental/adjoint of $E_8$ satisfy
$\mathrm{tr} T^{\mathcal{A}} T^{\mathcal{B}} = 30
\delta^{\mathcal{A}\mathcal{B}}$. \ In the $\mathrm{SO} \left( 16 \right)$
basis for $E_8$, the $T^{\mathcal{A}}$ are $- \frac{1}{2} i$ times the
generators in Appendix 6.A of {\cite{Green Schwarz Witten}} or subsection 2.1
of {\cite{CCHT}}, and in the $\mathrm{SU} \left( 9 \right)$ basis for $E_8$,
the $T^{\mathcal{A}}$ are the generators in subsection 5.2 of {\cite{CCHT}}. \
$\tilde{e} = \sqrt{- \tilde{G}}$ is the determinant of the vielbein
$\tilde{e}_{U \hat{V}}$, that satisfies $\tilde{e}_{U \hat{W}} \tilde{e}_V \,
\!^{\hat{W}} = \tilde{G}_{U V}$, where $\tilde{G}_{U V}$ is the induced metric
on the boundary, which is obtained from $G_{I J}$ by dropping the row and
column with an index $y$. \ The coefficient of the first term in (\ref{Yang
Mills term}) is fixed by anomaly cancellation {\cite{Horava Witten 2,de
Alwis,Conrad,Faux Lust Ovrut,Lu,Bilal Derendinger Sauser,Harmark,Bilal
Metzger,Meissner Olechowski,Moss 3}} and has the value found by Conrad
{\cite{Conrad}}, which is slightly different from the original value found by
HW. \ The $\bar{R}_{UV \hat{W}  \hat{X}}  \bar{R}^{UV \hat{W}  \hat{X}}$ term
was derived by Moss {\cite{Moss 4}}, with:
\begin{equation}
  \label{R bar} \bar{R}_{UV} \hspace{0.25em} \hspace{-0.25em}^{\hat{W}}
  \hspace{0.25em} \hspace{-0.25em}_{\hat{X}} = \partial_U  \bar{\omega}_V
  \hspace{0.25em} \hspace{-0.25em}^{\hat{W}} \hspace{0.25em}
  \hspace{-0.25em}_{\hat{X}} - \partial_V  \bar{\omega}_U \hspace{0.25em}
  \hspace{-0.25em}^{\hat{W}} \hspace{0.25em} \hspace{-0.25em}_{\hat{X}} +
  \bar{\omega}_U  \hspace{0.25em} \hspace{-0.25em}^{\hat{W}} \hspace{0.25em}
  \hspace{-0.25em}_{\hat{Y}} \bar{\omega}_V \hspace{0.25em}
  \hspace{-0.25em}^{\hat{Y}} \hspace{0.25em} \hspace{-0.25em}_{\hat{X}} -
  \bar{\omega}_V  \hspace{0.25em} \hspace{-0.25em}^{\hat{W}} \hspace{0.25em}
  \hspace{-0.25em}_{\hat{Y}} \bar{\omega}_U \hspace{0.25em}
  \hspace{-0.25em}^{\hat{Y}} \hspace{0.25em} \hspace{-0.25em}_{\hat{X}},
\end{equation}
where
\begin{equation}
  \label{omega bar} \bar{\omega}_{U \hat{V}  \hat{W}} = \tilde{\omega}_{U
  \hat{V}  \hat{W}} \pm \frac{1}{2} H_{\hat{y} U \hat{V}  \hat{W}},
\end{equation}
and $\tilde{\omega}_{U \hat{V}  \hat{W}} = e^X  \hspace{0.25em}
\hspace{-0.25em}_{\hat{W}} \left( \tilde{\Gamma}_U  \hspace{0.25em}
\hspace{-0.25em}^Y \hspace{0.25em} \hspace{-0.25em}_X e_{Y \hat{V}} -
\partial_U e_{X \hat{V}} \right)$ is the Levi-Civita connection for the
vielbein $\tilde{e}_{U \hat{V}}$. \ The sign choice in (\ref{omega bar}) is
correlated with the chirality conditions on the gravitino, gaugino, and
supersymmetry variation parameter on the boundary.

The compact hyperbolic 7-manifold $\bar{H}^7$ of intrinsic volume around
$10^{34}$, with a closed hyperbolic boundary $\bar{H}^6$ of intrinsic volume
in the range from about $3 \times 10^5$ to about $6 \times 10^5$ that
accomodates the SM fields, and possibly also other closed hyperbolic
boundaries that accomodate dark matter fields, is assumed to be obtained from
a closed hyperbolic 7-manifold by cutting it along suitable 6-cycles, and
keeping one connected component of the result.

The SM boundary is near a minimal-area 6-cycle of the compact hyperbolic
7-manifold that was cut to form the boundary, and the metric in the region of
the boundary is a small perturbation of what it would have been if the
boundary was not there. \ The metric in this region has the form:
\begin{equation}
  \label{metric near boundary} ds_{11}^2 = G_{IJ} dx^I dx^J = a \left( y
  \right)^2 \eta_{\mu \nu} d \check{x}^{\mu} d \check{x}^{\nu} + b \left( y
  \right)^2  \hat{g}_{ab} d \hat{x}^a d \hat{x}^b + dy^2
\end{equation}
Here $a \left( y \right) \rightarrow A$ away from the boundary, and $a \left(
y \right) = 1$ on the boundary. \ Indices $a, b, c, \ldots$ are tangential to
$\bar{H}^6$, so that $\bar{x}^A$ in (\ref{metric ansatz for H7}) is now
$\left( \hat{x}^a, y \right)$, and $\tilde{x}^U$ is $\left( \check{x}^{\mu},
\hat{x}^a \right)$. \ $\hat{g}_{ab}$ is a metric of sectional curvature $- 1$
on $\bar{H}^6$.

If the boundary was not there, and $b \left( y \right)$ had its minimum value
at $y = 0$, $b \left( y \right)$ would be $b = B \mathrm{cosh} \left(
\frac{y}{B} \right)$. \ Then (\ref{metric near boundary}), with $a \left( y
\right) = A$, would be in agreement with (\ref{metric ansatz for H7}), for a
particular choice of coordinates on this region of $\bar{H}^7$. \ The
effective energy-momentum tensor $T_{I J}$ in this region is calculated by
requiring that this metric satisfies the classical Einstein equations with
that $T_{I J}$. \ We define $a \left( y \right) \equiv \left( 1 + p \left( y
\right) \right) A$, $b \left( y \right) \equiv \left( 1 + q \left( y \right)
\right) B \mathrm{cosh} \frac{y}{B}$, where $\left| p \left( y \right)
\right|$ and $\left| q \left( y \right) \right|$ are assumed $\ll 1$, and
substitute the perturbed metric into the Einstein equations with the effective
$T_{I J}$. \ Expanding to first order in $p$ and $q$, $p \left( y \right)$ and
$q \left( y \right)$ are found to satisfy:
\begin{equation}
  \label{p dot eqn} \dot{p} = - \frac{5}{4}  \dot{q} + \frac{5 q}{4 B
  \mathrm{sinh} \frac{y}{B} \mathrm{cosh} \frac{y}{B}}
\end{equation}
\begin{equation}
  \label{q double dot eqn} \ddot{q} + 7 \dot{q} \frac{\mathrm{sinh}
  \frac{y}{B}}{B \mathrm{cosh} \frac{y}{B}} - \frac{5 q}{B^2 \mathrm{cosh}^2
  \frac{y}{B}} = 0,
\end{equation}
where a dot denotes differentiation with respect to $y$. \ To find the
solution of (\ref{p dot eqn}) and (\ref{q double dot eqn}) such that $p$ and
$q$ tend to 0 as $y \rightarrow \infty$, we define $\xi \equiv \mathrm{tanh}
\frac{y}{B}$, so that $\xi \rightarrow 1$ as $y \rightarrow \infty$. \ The
equations then become:
\begin{equation}
  \label{p eqn with x} \frac{dp}{d \xi} = - \frac{5}{4}  \frac{dq}{d \xi} +
  \frac{5 q}{4 \xi}
\end{equation}
\begin{equation}
  \label{q eqn with x} \left( 1 - \xi^2 \right)  \frac{d^2 q}{d \xi^2} + 5 \xi
  \frac{dq}{d \xi} - 5 q = 0.
\end{equation}
The solution is:
\begin{equation}
  \label{expansion of q} q \left( \xi \right) = k \left( \left( 1 - \xi
  \right)^{\frac{7}{2}} - \frac{5 \left( 1 - \xi \right)^{\frac{9}{2}}}{12} +
  \frac{35 \left( 1 - \xi \right)^{\frac{11}{2}}}{1056} + \frac{35 \left( 1 -
  \xi \right)^{\frac{13}{2}}}{18304} + \ldots \right),
\end{equation}
\begin{equation}
  \label{solution for p} p \left( \xi \right) = - \frac{5}{4} q \left( \xi
  \right) - \frac{5}{4}  \int^1_{\xi}  \frac{q \left( \xi'  \right)}{\xi'} d
  \xi',
\end{equation}
where $k$ is an arbitrary constant. \ $q \left( \xi \right)$ looks
qualitatively like the base of a parabola centred at $\xi = 1$, and is $\simeq
0.6184 k$ for $\xi = 0$, while $p \left( \xi \right)$ has a logarithmic
singularity as $\xi \rightarrow 0_+$.

For a first estimate of the boundary conditions for the metric
{\cite{Israel,Chamblin Reall,Dyer Hinterbichler,Moss 2}}, I neglected the flux
terms in $\bar{R}_{UVWX}$, defined in (\ref{R bar}), by assuming, if
necessary, that $H_{\hat{y} UVW}$ is smaller than its average value, near the
boundary. \ Then to leading order in the LOW harmonic expansion of the
energy-momentum tensor
\begin{equation}
  \label{T tilde i U V} \tilde{T}^{\left( \mathrm{bos} \right) UV} =
  \frac{2}{\tilde{e}}  \frac{\delta S^{\left( \mathrm{bos}
  \right)}_{\mathrm{YM}}}{\delta \tilde{G}_{UV}}
\end{equation}
on the boundary, and assuming that $\mathrm{tr} F_{ac} F_b \hspace{0.25em}
\hspace{-0.25em}^c$ is a multiple of $\tilde{G}_{ab}$, the boundary conditions
are:
\begin{equation}
  \label{boundary conditions in terms of rho} \left. \frac{\dot{a}}{a}
  \right|_{y = y_{1 +}} = - \rho \frac{\kappa^{2 / 3}_{11}}{b^4_1},
  \hspace{3.2cm} \left. \frac{\dot{b}}{b} \right|_{y = y_{1 +}} = \rho
  \frac{\kappa^{2 / 3}_{11}}{b^4_1},
\end{equation}
where the number $\rho$ is:
\begin{equation}
  \label{definition of rho} \rho \, \equiv \, \frac{1}{\bar{V}_6}
  \int_{\bar{H}^6, y = y_{1 +}} \!\!\!\!\! d^6  \hat{x}
  \sqrt{\hat{g}}  \frac{1}{96 \pi \left( 4 \pi \right)^{2 / 3}}  \hat{g}^{ac} 
  \hat{g}^{bd}  \left( - \frac{1}{30} \mathrm{tr} F_{ab} F_{cd} + \frac{1}{2} 
  \bar{R}_{ab}  \hspace{0.25em} \hspace{-0.25em}^e \hspace{0.25em}
  \hspace{-0.25em}_f \bar{R}_{cde} \hspace{0.25em} \hspace{-0.25em}^f \right),
\end{equation}
where $\bar{V}_6 \equiv \int_{\bar{H}^6} \!\! d^6  \hat{x} \sqrt{\hat{g}}$ is
the intrinsic volume of the $\bar{H}^6$ Cartesian factor of the boundary. \ If
there were no Yang-Mills fluxes on the boundary then $\rho$ would be
$\frac{60}{192 \pi \left( 4 \pi \right)^{2 / 3}} \simeq 0.01840$.

Let $\xi_1 \equiv \mathrm{\tanh} \frac{y_1}{B}$ denote the value of $\xi$ at
the boundary. \ The sum of the boundary conditions (\ref{boundary conditions
in terms of rho}) gives:
\begin{equation}
  \label{equation determining k}  \left( 1 - \xi_1^2 \right)  \left( -
  \frac{1}{4} \left. \frac{dq}{d \xi} \right|_{\xi = \xi_1} + \frac{5 q \left(
  \xi_1 \right)}{4 \xi_1} \right) + \xi_1 = 0,
\end{equation}
where (\ref{p eqn with x}) has been used. \ The logarithmic singularity of $p$
as $\xi \rightarrow 0_+$ means that we require $\xi_1 > 0$ for the assumption
that $\left| p \right| \ll 1$ to be valid, so since $\frac{1}{k}  \frac{dq}{d
\xi} \leq 0$ and $\frac{1}{k} q \geq 0$ for $0 \leq \xi \leq 1$,
(\ref{equation determining k}) implies that $k < 0$. \ Using (\ref{equation
determining k}) to express $k$ in terms of $\xi_1$, we find that $q_1 \equiv q
\left( \xi_1 \right)$, as a function of $\xi_1$, looks qualitatively like an
upside-down parabola, with a peak value of 0 at $\xi_1 = 0$, and $\simeq -
0.25$ at $\xi_1 \simeq 0.63$. \ And $\frac{b_1}{B} = \frac{\left( 1 + q_1
\right)}{\sqrt{1 - \xi_1^2}}$, as a function of $\xi_1$, decreases smoothly
from a peak value of 1 at $\xi_1 = 0$, to a minimum value $\simeq 0.967$ at
$\xi_1 \simeq 0.56$, and then starts increasing at an increasing rate.

Using $B \simeq 0.28 \kappa^{2 / 9}_{11}$ as the best value of $B$ determined
by the PMS, the second equation of (\ref{boundary conditions in terms of rho})
becomes:
\begin{equation}
  \label{second bc in terms of x} \left( 1 - \xi_1^2 \right) \left.
  \frac{dq}{d \xi} \right|_{\xi = \xi_1} + \xi_1 \simeq \frac{729 \rho \left(
  1 - 4 q \left( \xi_1 \right) \right)}{\left( \sqrt{\frac{1 + \xi_1}{1 -
  \xi_1}} + \sqrt{\frac{1 - \xi_1}{1 + \xi_1}} \right)^4},
\end{equation}
so we require $\rho > 0$. \ Thus the number of vacuum Yang-Mills fluxes should
be small enough for the $R^2$ term in (\ref{definition of rho}) to outweigh
the $F^2$ term.

Substituting for $k$ from (\ref{equation determining k}), we find from
(\ref{second bc in terms of x}) that for $\rho = 0.01840$, $\xi_1 \simeq
0.391$, hence $k \simeq - 0.768$, so $p_1 \equiv p \left( \xi_1 \right) \simeq
0.165$, hence $A \simeq 0.858$, and $q_1 \simeq - 0.103$, hence $b_1 \simeq
0.975 B \simeq 0.27 \kappa^{2 / 9}_{11}$. \ Thus working to first order in $p$
and $q$ has been justified.

Moss's derivation of the $R_{UVWX} R^{UVWX}$ term in (\ref{Yang Mills term})
used an expansion scheme in which Ricci tensor and scalar terms, if present,
would only show up at higher orders {\cite{Moss 4}}. \ If the $R_{UVWX}
R^{UVWX}$ term was in fact the first term in a Lovelock-Gauss-Bonnet term of
the form $R_{UVWX} R^{UVWX} - 4 R_{UV} R^{UV} + R^2$, the size of the $RR$
terms in (\ref{definition of rho}) would be increased by a factor of 6, with
the main contribution coming from the square of the Ricci scalar. \ If there
were no Yang-Mills fluxes on the boundary $\rho$ would then be $\simeq
0.1104$, for which (\ref{equation determining k}) and (\ref{second bc in terms
of x}) give $\xi_1 \simeq 0.725$, $k \simeq - 32.75$, $p_1 \simeq 0.430$, $A
\simeq 0.699$, $q_1 \simeq - 0.319$, and $b_1 \simeq 0.989 B \simeq 0.28
\kappa^{2 / 9}_{11}$. \ This is not really within the region where working to
first order in $p$ and $q$ is justified.

The Einstein action on the 4 extended dimensions has the form:
\begin{equation}
  \label{Einstein action} S_{\mathrm{Ein}} = \frac{1}{16 \pi G_N}  \int d^4 
  \check{x} \sqrt{- \check{g}}  \check{g}^{\mu \nu} R_{\mu \nu} \left(
  \check{g} \right),
\end{equation}
where $\check{g}_{\mu \nu}$ differs from $\eta_{\mu \nu}$ by a small
perturbation, that depends on the coordinates $\check{x}^{\sigma}$ on the 4
extended dimensions, but not on the coordinates $\bar{x}^A$ on $\bar{H}^7$. \
$G_N$ is Newton's constant, with the value {\cite{PDG}}:
\begin{equation}
  \label{Newtons constant} G_N = 6.7087 \times 10^{- 33} 
  \mathrm{{{TeV}}}^{- 2},
\end{equation}
so that $\sqrt{G_N} = 8.1907 \times 10^{- 17}$ TeV$^{- 1} \hspace{3pt} =
\hspace{3pt} 1.6160 \times 10^{- 35}$ metres. \ Comparing with (\ref{CJS
action}) and (\ref{metric ansatz for H7}), and noting that $R_{IJ}
\hspace{0.25em} \hspace{-0.25em}^K \hspace{0.25em} \hspace{-0.25em}_L$ and
hence $R_{IJ}$ are unaltered by rescaling the metric by a constant factor, so
that $\sqrt{- G} G^{\mu \nu} R_{\mu \nu} \left( G \right) = A^4 B^7 \sqrt{-
\check{g}}  \sqrt{\bar{g}}  \frac{1}{A^2}  \check{g}^{\mu \nu} R_{\mu \nu}
\left( \check{g} \right)$ everywhere on $\bar{H}^7$ except in the immediate
vicinity of the HW boundary, where $G_{IJ}$ here represents the metric
obtained from (\ref{metric ansatz for H7}) by replacing $\eta_{\mu \nu}$ by
$\check{g}_{\mu \nu}$, we find, in the approximation of neglecting the volume
of the region where $a \left( y \right)$ in (\ref{metric near boundary})
differs appreciably from $A$, that the intrinsic volume $\bar{V}_7 \equiv
\int_{\bar{H}^7} \!\! d^7  \bar{x} \sqrt{\bar{g}}$ \ of $\bar{H}^7$ is given
by {\cite{ADD1,AADD,ADD2,RS1,KMST}}:
\begin{equation}
  \label{V7 in terms of GN} \frac{A^2 B^7  \bar{V}_7}{2 \kappa^2_{11}} =
  \frac{1}{16 \pi G_N}
\end{equation}
The experimental limits on the gravitational coupling constant in $D$
dimensions are expressed in terms of a mass $M_D$, such that for $D = 11$,
$M_{11} = \left( 2 \pi \right)^{7 / 9} \kappa^{- 2 / 9}_{11} = 4.1764
\kappa^{- 2 / 9}_{11}$ {\cite{Giudice Rattazzi Wells, PDG}}. \ The latest
limits from searches for virtual graviton exchange and graviton emission at
the LHC {\cite{Franceschini et al, CMS-EXO-10-026, ATLAS-CONF-2011-096, CMS
PAS EXO-11-038, CMS PAS EXO-11-039, CMS PAS EXO-11-058, 11120688 CMS
diphoton, 11122194 ATLAS}}, and searches for
microscopic black holes at the LHC {\cite{CMS PAS EXO-11-071,
ATLAS-CONF-2011-065, ATLAS-CONF-2011-068}}, suggest that the experimental
lower bound on $M_{11}$, for 7 flat extra dimensions, is now roughly $M_{11}
\geq 2.3 \pm 0.7$ TeV, corresponding to $\kappa^{- 2 / 9}_{11} \geq 0.55 \pm
0.2$ TeV.

From above, $A$ is expected to lie in the range from about 0.7 to about 0.9,
and the best value of $B$, determined by the PMS, is $B \simeq 0.28 \kappa^{2
/ 9}_{11}$. \ If $\kappa^{- 2 / 9}_{11}$ was about 0.55 TeV, so that $B$ was
around $0.51$ TeV$^{- 1}$, $A = 0.7$ would give $\bar{V}_7 \simeq 3.0 \times
10^{35}$, and $A = 0.9$ would give $\bar{V}_7 \simeq 1.8 \times 10^{35}$. \
Thus if $\bar{H}^7$ is reasonably isotropic, in the sense that it has an
approximately spherical fundamental domain in 7-dimensional hyperbolic space
$H^7$, then from page 9 of {\cite{Almost Flat}}, the current upper bound on
the intrinsic diameter $\bar{L}_7$ of $\bar{H}^7$ is about 28, hence the
current upper bound on the actual diameter $L_7$ of $\bar{H}^7$ is about 14
TeV$^{- 1} \simeq 2.8 \times 10^{- 18}$ metres.

From between (\ref{second derivative of action wrt B}) and (\ref{Yang Mills
term}) above, the intrinsic diameter $\bar{L}_6$ of the closed hyperbolic
factor $\bar{H}^6$ of the HW boundary lies between about 5.7 and 6.0 if
$\bar{H}^6$ is reasonably isotropic, so if both $\bar{H}^7$ and $\bar{H}^6$
are reasonably isotropic, the current upper bound on the ratio $\bar{L}_7 /
\bar{L}_6$ of their intrinsic diameters lies between about 4.9 and 4.7. \ And
since the curvature radius $b_1$ of the HW boundary is $\simeq B$, this also
gives the current upper bound on the ratio $L_7 / L_6$ of their actual
diameters.

Closed hyperbolic 7-manifolds $\bar{\bar{H}}^7$ of intrinsic volume
$\bar{V}_7 \sim 10^{35}$ that have a closed hyperbolic minimal-area 6-cycle
$\bar{\bar{H}}^6$ of intrinsic volume $\sim 10^6$, such that in suitable
coordinates near $\bar{\bar{H}}^6$ the metric of sectional curvature $-
1$ on $\bar{\bar{H}}^7$ has the form of the last two terms in
(\ref{metric near boundary}) with $b = \mathrm{cosh} \left( \frac{y}{B}
\right)$, might be relatively rare among $\bar{\bar{H}}^7$ with
$\bar{V}_7 \sim 10^{35}$. \ For if cutting $\bar{\bar{H}}^7$ along
$\bar{\bar{H}}^6$ separates $\bar{\bar{H}}^7$ into two connected
components, let $\bar{\bar{H}}_{\left( 2 \right)}^7$ be formed by cutting
$\bar{\bar{H}}^7$ along $y = 0$ and joining two copies of the larger
volume component along this boundary, while if cutting $\bar{\bar{H}}^7$
along $\bar{\bar{H}}^6$ leaves $\bar{\bar{H}}^7$ connected, let
$\bar{\bar{H}}_{\left( 2 \right)}^7$ be formed from two copies of
$\bar{\bar{H}}^7$ cut along $y = 0$, by joining boundary $b$ of copy 1 to
boundary $a$ of copy 2, and boundary $b$ of copy 2 to boundary $a$ of copy 1.
\ Then the smallest non-zero intrinsic eigenvalue $\bar{\lambda}_1$ of the
negative of the Laplace-Beltrami operator $\Delta \equiv
\frac{1}{\sqrt{\bar{g}}} \partial_A \left( \sqrt{\bar{g}} \bar{g}^{A B}
\partial_B \cdot \right)$ on $\bar{\bar{H}}_{\left( 2 \right)}^7$ is
bounded above by $\sim 10^{- 29}$, since for any function $f \left( \bar{x}
\right)$ such that $\int_{\bar{\bar{H}}^7} \sqrt{\bar{g}} fd^7 \bar{x} =
0$:
\begin{equation}
  \label{upper bound on smallest nonzero eigenvalue} \bar{\lambda}_1 \leq
  \frac{\int_{\bar{\bar{H}}^7} \sqrt{\bar{g}} g^{A B} \left( \partial_A f
  \right) \left( \partial_B f \right) d^7 \bar{x}}{\int_{\bar{\bar{H}}^7}
  \sqrt{\bar{g}} f^2 d^7 \bar{x}},
\end{equation}
and we can choose $f$ to be $1$ on one of the two connected components of the
manifold obtained from $\bar{\bar{H}}^7$ by deleting the region with $|y|
< 1$, and $- 1$ on the other such component, with a smooth transition across
the region with $|y| < 1$ {\cite{Sarnak Selbergs Eigenvalue Conjecture}}. \
But from the discussion on pages 9 to 12 of {\cite{Almost Flat}}, it seems
possible that typical $\bar{\bar{H}}^n$, $n \geq 2$, of arbitrarily large
intrinsic volume $\bar{V}^n$, will have few or no nonzero intrinsic
eigenvalues $\bar{\lambda}$ of $- \Delta$ smaller than $\frac{\left( n - 1
\right)^2}{4}$ {\cite{KMST, Yau Isoperimetric, Agmon, Brooks Makover 1, Brooks
Makover 2, Brooks Makover 3, Brooks Makover 4, Cheeger, Cheeger Wikipedia,
Donnelly, Sarnak, Luo, Rudnick Sarnak, Orlando Park, Mazzeo Phillips, Donnelly
Xavier, Thurston, Colbois Courtois 1, Colbois Courtois 2}}.

Closed $\bar{\bar{H}}^7$ that have a closed hyperbolic minimal-area
6-cycle $\bar{\bar{H}}^6$, such that in suitable coordinates the metric
near $\bar{\bar{H}}^6$ is as above, exist with arbitrarily large values
of $\bar{V}_7 / \bar{V}_6$, for section 2.8.C of {\cite{Gromov Piatetski
Shapiro}} gives examples for all $n \geq 2$ of $\bar{\bar{H}}^n$ that
contain a 2-sided non-separating embedded closed hyperbolic hypersurface
$\bar{\bar{H}}^{n - 1}$. If we take $N$ copies of such an
$\bar{\bar{H}}^n$, cut each along that $\bar{\bar{H}}^{n - 1}$, and
join side $b$ of copy 1 to side $a$ of copy 2, side $b$ of copy 2 to side
$a$ of copy 3, $\ldots$, and side $b$ of copy $N$ to side $a$ of copy 1, we get
a closed hyperbolic $n$-manifold $\bar{\bar{H}}_{\left( N \right)}^n$
that is an $N$-fold cover of the original $\bar{\bar{H}}^n$, so the ratio
of the intrinsic volume $\bar{V}_{n \left( N \right)}$ of
$\bar{\bar{H}}_{\left( N \right)}^n$ to the intrinsic volume $\bar{V}_{n
- 1}$ of that $\bar{\bar{H}}^{n - 1}$ can be arbitrarily large. \ However
these $\bar{\bar{H}}_{\left( N \right)}^n$ are far from being reasonably
isotropic for large $N$, because their intrinsic diameters and intrinsic
volumes both grow in proportion to $N$, while from page 9 of {\cite{Almost
Flat}}, the intrinsic volume $\bar{V}_n$ of a reasonably isotropic
$\bar{\bar{H}}^n$ is approximately related to its intrinsic diameter
$\bar{L}_n$, for large $\bar{L}_n$, by $\bar{V}_n \simeq \frac{S_{n - 1}}{2^{n
- 1} \left( n - 1 \right)} e^{\left( n - 1 \right) \frac{\bar{L}_n}{2}}$,
where $S_{n - 1}$ is the area of the unit $\left( n - 1 \right)$-sphere.

\section{The bosonic Kaluza-Klein modes of the supergravity multiplet}
\label{The bosonic Kaluza Klein modes of the supergravity multiplet}

I shall continue to work to leading order in the Lukas-Ovrut-Waldram (LOW)
harmonic expansion of the energy-momentum tensor on $\bar{H}^7$ {\cite{Lukas
Ovrut Waldram}}, and to assume that the vacuum fluxes are approximately
uniformly distributed across $\bar{H}^7$, so that the LOW expansion only needs
to be applied over relatively small local regions of $\bar{H}^7$. \ In
addition to the assumption (\ref{flux bilinears}) on the vacuum flux
bilinears, I shall assume that to leading order in the LOW harmonic expansion,
expressions linear in the vacuum fluxes are zero. \ The Kaluza-Klein modes of
the metric $G_{I J}$ and the 3-form gauge field $C_{I J K}$ are then to a
first approximation decoupled from each other, and can thus be treated
separately. \ I shall use the convention stated between (\ref{t8 A B C D}) and
(\ref{t8 t8 X4}), that repeated lower coordinate indices are understood to be
contracted with an inverse metric $G^{I J}$.

\subsection{The Kaluza-Klein modes of the 3-form gauge field}
\label{The Kaluza Klein modes of the 3 form gauge field}

As stated just before (\ref{flux bilinears}), the vacuum 4-form fluxes are
assumed to be proportional to harmonic 4-forms on $\bar{H}^7$, and thus to
solve the classical CJS field equations (\ref{3 form field equations}) for
$H_{I J K L}$, and the quantum corrections to those field equations are
neglected, so for a first approximation to the Kaluza-Klein modes of $C_{I J
K}$, it is consistent to consider just the classical CJS action, whose bosonic
part is (\ref{CJS action}). \ With the above assumptions on terms linear or
bilinear in the vacuum fluxes, the vacuum fluxes do not affect the
Kaluza-Klein modes of $C_{I J K}$, and the $\epsilon_{11} CHH$ Chern-Simons
term in (\ref{CJS action}) also plays no role. \ Thus for a first, classical,
aproximation to the masses of the Kaluza-Klein modes of $C_{I J K}$, we can
neglect the vacuum fluxes completely, and consider just the $- \frac{1}{48}
H_{I J K L} H_{I J K L}$ term in (\ref{CJS action}). \ Then after adding
gauge-fixing terms as follows, and noting that for the metric ansatz
(\ref{metric ansatz for H7}), covariant derivatives $D_{\mu}$ are ordinary
derivatives $\partial_{\mu}$, and commute with each other and with $D_A$, we
have:
\begin{dmath*}[compact, spread=3pt]
 - \frac{1}{48} H_{I J K L} H_{I J K L} - \frac{1}{4} (a D_{\mu} C_{\mu \nu
   \sigma} + \frac{1}{a} D_A C_{A \nu \sigma}) (a D_{\tau} C_{\tau \nu \sigma}
   + \frac{1}{a} D_B C_{B \nu \sigma}) - \frac{1}{2} \hspace{0.25em} (b
   D_{\mu} C_{\mu A \sigma} + \frac{1}{b} D_B C_{B A \sigma}) (b D_{\nu}
   C_{\nu A \sigma} + \frac{1}{b} D_E C_{E A \sigma}) - \frac{1}{4}
   \hspace{0.25em} (c D_{\mu} C_{\mu A B} + \frac{1}{c} D_E C_{E A B}) (c
   D_{\nu} C_{\nu A B} + \frac{1}{c} D_F C_{F A B}) =
\end{dmath*}
\begin{dmath}[compact, spread=3pt]
  \label{gauge fixed 3 form action} = \frac{1}{12} \left\{ - \hspace{0.25em}
  \partial_{\mu} C_{\nu \sigma \tau} \partial_{\mu} C_{\nu \sigma \tau} + 3
  \left( 1 - a^2 \right) \hspace{0.25em} \partial_{\mu} C_{\mu \sigma \tau}
  \partial_{\nu} C_{\nu \sigma \tau} - \hspace{0.25em} D_A C_{\mu \nu \sigma}
  D_A C_{\mu \nu \sigma} \right\} + \frac{1}{4} \left\{ - \hspace{0.25em}
  \partial_{\mu} C_{\nu \sigma A} \partial_{\mu} C_{\nu \sigma A} + 2 \left( 1
  - \hspace{0.25em} b^2 \right) \partial_{\mu} C_{\mu \sigma A} \partial_{\nu}
  C_{\nu \sigma A} - \hspace{0.25em} D_A C_{\mu \nu B} D_A C_{\mu \nu B} + D_A
  C_{\mu \nu B} D_B C_{\mu \nu A} - \frac{1}{a^2} D_A C_{\mu \nu A} D_B C_{\mu
  \nu B} \right\} + \frac{1}{4} \left\{ - \hspace{0.25em} \partial_{\mu}
  C_{\nu A B} \partial_{\mu} C_{\nu A B} + \left( 1 - c^2 \right) 
  \hspace{0.25em} \partial_{\mu} C_{\mu A B} \partial_{\nu} C_{\nu A B} -
  \hspace{0.25em} D_A C_{\mu B E} D_A C_{\mu B E} + 2 D_A C_{\mu B E} D_B
  C_{\mu A E} - \frac{2}{b^2} D_A C_{\mu A E} D_B C_{\mu B E} \right\} +
  \frac{1}{12} \left\{ - \hspace{0.25em} \partial_{\mu} C_{A B E}
  \partial_{\mu} C_{A B E} - \hspace{0.25em} D_A C_{B E F} D_A C_{B E F} + 3
  D_A C_{B E F} D_B C_{A E F} - \frac{3}{c^2} D_A C_{A E F} D_B C_{B E F}
  \hspace{0.25em} \right\},
\end{dmath}
where $a$, $b$, and $c$ are gauge parameters. \ Derivatives in the right-hand
side of (\ref{gauge fixed 3 form action}) act only on the smallest object to
their immediate right. \ $C_{I J K}$ and $H_{I J K L}$ in (\ref{gauge fixed 3
form action}) refer to the Kaluza-Klein modes only. \ The modes of different
spin along the extended dimensions are decoupled in the right-hand side of
(\ref{gauge fixed 3 form action}). \ If we choose $a = b = c = 1$, which
corresponds to a gauge-fixing term $- \frac{1}{4} D_I C_{I K L} D_J C_{J K L}$
and is effectively Feynman gauge, then after making a Kaluza-Klein ansatz such
as $C_{\mu \nu A} = c_{\mu \nu} \left( \check{x} \right) \omega_A \left(
\bar{x} \right)$ in the corresponding field equations and separating the field
equations, the field equations on $\bar{H}^7$ in the metric $\bar{g}_{A B}$ of
sectional curvature $- 1$ have the form $- \left( \delta d + d \delta \right)
\omega = \bar{m}^2 \omega$, where $\delta d + d \delta$ is the Hodge - de Rham
Laplacian, so the intrinsic masses $\bar{m}$ of the modes with $p$ $A$-type
indices, $0 \leq p \leq 3$, are given by the spectrum of the negative of the
Hodge - de Rham Laplacian for $p$-forms on $\bar{H}^7$. \ From pages 42 to 43
of {\cite{Almost Flat}}, this means that their masses, as seen on the HW
boundary, are $m = \frac{A}{B} \bar{m}$, where $A$ and $B$ are the constants
in the metric ansatz (\ref{metric ansatz for H7}).

Choosing alternatively now the limiting gauge choice $a \rightarrow 0$, $b
\rightarrow 0$, $c \rightarrow 0$, we obtain Proca-type unitary gauges for the
massive antisymmetric tensor fields on the extended dimensions \cite{Proca 1,
Proca 2}, and
Landau-gauge-like restrictions such as $D_A C_{\mu \nu A} = 0$ on the
dependence of the modes on position on $\bar{H}^7$, which means that some of
the massive modes obtained in Feynman gauge are unphysical, and would be
cancelled by corresponding Faddeev-Popov ghosts in Feynman gauge.

From pages 9 to 12 and 16 to 17 of {\cite{Almost Flat}}, it seems likely that
classically, the lightest massive modes of a $p$-form gauge field on
$\bar{H}^7$, for $p < 3$, will have intrinsic mass $\bar{m} = 3 - p$
{\cite{KMST, Yau Isoperimetric, Agmon, Brooks Makover 1, Brooks Makover 2,
Brooks Makover 3, Brooks Makover 4, Cheeger, Cheeger Wikipedia, Donnelly,
Sarnak, Luo, Rudnick Sarnak, Orlando Park, Mazzeo Phillips, Donnelly Xavier,
Thurston, Colbois Courtois 1, Colbois Courtois 2}}. \ From pages 42 to 43 of
{\cite{Almost Flat}}, this means that their mass, as seen on the HW boundary,
is $m = \left( 3 - p \right) \frac{A}{B}$.

\subsubsection{The classically massless harmonic 3-form modes}
\label{The mass of the harmonic 3 form modes}

In addition to the classically massive modes of $C_{I J K}$, there are
classically massless modes $C_{A B C} = C \left( \check{x} \right) \omega_{A B
C} \left( \bar{x} \right)$, $C_{\mu A B} = C_{\mu} \left( \check{x} \right)
\omega_{A B} \left( \bar{x} \right)$, and $C_{\mu \nu A} = C_{\mu \nu} \left(
\check{x} \right) \omega_A \left( \bar{x} \right)$, corresponding respectively
to harmonic 3-forms $\omega_{A B C} \left( \bar{x} \right)$, 2-forms
$\omega_{A B} \left( \bar{x} \right)$, and 1-forms $\omega_A \left( \bar{x}
\right)$, on $\bar{H}^7$. \ The field strengths $H_{A B C D}$, $H_{\mu A B
C}$, and $H_{\mu \nu A B}$, that would occur respectively in their classical
mass terms, vanish identically, so they can only obtain masses from quantum
corrections that arise from interaction terms that contain $C_{I J K}$
explicitly, so the relevant terms in (\ref{Gamma SG}) are the CJS Chern-Simons
term $\epsilon_{11} CHH$ in (\ref{CJS action}), and the Green-Schwarz term
(\ref{eps11 t8 C R4}) in (\ref{Gamma 8 SG}). \ The harmonic 0-form mode
$C_{\mu \nu \sigma} \left( \check{x} \right)$ is classically a pure gauge
mode, with no physical degrees of freedom.

If these modes all acquire masses $\sim \kappa^{- 2 / 9}_{11}$ from quantum
corrections, then only the harmonic 3-form modes $C_{A B C}$ are expected to
be sufficiently numerous for their large number to compensate for the
gravitational suppression of their couplings enough for them to be seen at the
LHC, because from pages 17 to 19 of {\cite{Almost Flat}}, the Betti number
$B_3$ of $\bar{H}^7$ is estimated as $\sim \frac{\bar{V}_7}{\mathrm{\ln}
\bar{V}_7}$, while the Betti numbers $B_2$ and $B_1$ of $\bar{H}^7$ are
estimated as powers strictly less than 1 of $\bar{V}_7$ {\cite{Gromov Volume
Bounded Cohomology, Gromov Volume and Bounded Cohomology, Lueck, Clair Whyte,
Donnelly Xavier, Xue}}.

The leading contribution to the squared masses of the harmonic 3-form modes of
$C_{A B C}$, in the presence of the fluxes $H_{A B C D}$ proportional to
harmonic 4-forms on $\bar{H}^7$, arises from the CJS Chern-Simons term in
(\ref{CJS action}), on integrating out $H_{\mu \nu \sigma \tau}$
{\cite{Beasley Witten}}. \ The $\mu \nu \sigma \tau \rho$ component of the
Bianchi identity for $H_{I J K L}$ is satisfied automatically, so after making
a suitable choice of gauge for $C_{\mu \nu \sigma}$, we can change variables
from $C_{\mu \nu \sigma}$ to $H_{\mu \nu \sigma \tau}$, which is now an
unconstrained scalar multiple of $\epsilon_{\mu \nu \sigma \tau}$. \ In
particular, choosing the Lorentz gauge condition $\partial^{\mu} C_{\mu \nu
\sigma} = 0$, and suitable boundary conditions as $x^0 = \check{x}^0
\rightarrow \pm \infty$, we can write:
\begin{equation}
  \label{C in terms of H} C_{\mu \nu \sigma} \left( \check{x}, \bar{x} \right)
  = \int d^4 \check{x}' \frac{\partial}{\partial \check{x}_{\rho}} G_{4 F}
  \left( \check{x} - \check{x}' \right) H_{\mu \nu \sigma \rho} \left(
  \check{x}', \bar{x} \right),
\end{equation}
where $G_{4 F} \left( \check{x} - \check{x}' \right)$ is the Feynman
propagator for a massless scalar in $3 + 1$ dimensions, which satisfies \ $-
\frac{\partial^2}{\partial \check{x}^{\mu} \partial \check{x}_{\mu}} G_{4 F}
\left( \check{x} - \check{x}' \right) = \delta^4 \left( \check{x} - \check{x}'
\right)$.

Neglecting the leading quantum correction $\Gamma^{\left( 8, \mathrm{bos}
\right)}_{\mathrm{SG}}$, we can now integrate out $H_{\mu \nu \sigma \tau}$,
since it occurs quadratically in the CJS action (\ref{CJS action}). \ The
terms containing $H_{\mu \nu \sigma \tau}$ quadratically are the $- H_{\mu \nu
\sigma \tau} H^{\mu \nu \sigma \tau}$ and $- 4 H_{\mu \nu \sigma A} H^{\mu \nu
\sigma A}$ terms from $- H_{IJKL} H^{IJKL}$. \ The $- H_{\mu \nu \sigma \tau}
H^{\mu \nu \sigma \tau}$ term corresponds to a multiple of the identity matrix
in the $H_{\mu \nu \sigma \tau}$ Hilbert space, and we can expand the inverse
of the matrix defining the quadratic form corresponding to these two terms as
a power series in the matrix corresponding to the $- 4 \partial_A C_{\mu \nu
\sigma} \partial^A C^{\mu \nu \sigma}$ part of the second term.

However when we evaluate the expectation value of the resulting mass term in a
specific classical 3-form mode on $\bar{H}^7$, each derivative $\partial_A$,
(which as it occurs here is a covariant derivative for the metric (\ref{metric
ansatz for H7}), since the only non-vanishing Christoffel symbols are
$\Gamma_A \, \!^B \, \!_C$), will roughly give either a factor of the
classical mass of that mode, which is zero for the harmonic 3-form modes, or a
factor of $\frac{1}{B}$. \ The harmonic 3-forms are the most covariantly
smooth 3-form modes, so I shall assume that for them, any such factor of
$\frac{1}{B}$ is accompanied by a factor $\frac{1}{\bar{L}_7}$, where
$\bar{L}_7$, the intrinsic diameter of $\bar{H}^7$, is $\sim 27$ for
$\bar{V}_7 \sim 10^{34}$, if $\bar{H}^7$ is reasonably isotropic, in the sense
that it has a fundamental domain in $H^7$ that is approximately spherical. \
So for a first approximation to the mass of the harmonic 3-form modes, I shall
neglect the $- \partial_A C_{\mu \nu \sigma}$ term in $H_{\mu \nu \sigma A}$.

To extract the relevant part of the Chern-Simons term in (\ref{CJS action}),
we split each index $I_1 \ldots I_{11}$ independently into its $\mu$ range and
its $A$ range, and look for terms that can produce $H_{\mu \nu \sigma \tau}$,
after integrations by parts if necessary. \ To get a $C_{\mu \nu \sigma}$, one
of the three factors $C_{I_1 I_2 I_3} H_{I_4 \ldots I_7} H_{I_8 \ldots
I_{11}}$ has to have at least 3 $\mu$-type indices.

There are 2 terms like $\epsilon^{A B C D E F G \mu \nu \sigma \tau}_{\left(
11 \right)} C_{A B C} H_{D E F G} H_{\mu \nu \sigma \tau}$.

There are 8 terms like $\epsilon^{\mu \nu \sigma \tau A B C D E F G}_{\left(
11 \right)} C_{\mu \nu \sigma} H_{\tau A B C} H_{D E F G}$, which contains a
term

that on integration by parts, gives $\frac{1}{4} \epsilon_{\left( 11
\right)}^{A B C D E F G \mu \nu \sigma \tau} C_{A B C} H_{D E F G} H_{\mu \nu
\sigma \tau}$.

There are 32 terms like $\epsilon^{A B C \mu \nu \sigma D \tau E F G}_{\left(
11 \right)} C_{A B C} H_{\mu \nu \sigma D} H_{\tau E F G}$, which contains a
term that on integration by parts, gives\\
$- \frac{1}{4} \epsilon_{\left( 11 \right)}^{A B C \mu \nu \sigma D \tau E F
G} C_{A B C} \left( \partial_D H_{\mu \nu \sigma \tau} \right) C_{E F G} = 0$.

Thus the relevant terms in (\ref{CJS action}) containing $H_{\mu \nu \sigma
\tau}$ are:
\begin{equation}
  \label{relevant terms containing H} \frac{1}{96 \kappa_{11}^2} 
  \int_{\mathcal{B}} d^{11} xe \left( - H_{\mu \nu \sigma \tau} H^{\mu \nu
  \sigma \tau} - \frac{1}{108} \epsilon^{A B C D E F G \mu \nu \sigma
  \tau}_{\left( 11 \right)} C_{A B C} H_{D E F G} H_{\mu \nu \sigma \tau}
  \right) .
\end{equation}
After completing the square and integrating out $H_{\mu \nu \sigma \tau}$,
this becomes:
\begin{equation}
  \label{after integrating out H} - \frac{1}{96 \kappa_{11}^2} 
  \int_{\mathcal{B}} d^{11} xe \frac{4}{6^5} \epsilon^{A B C D E F G}_{\left(
  7 \right)} \epsilon_{\left( 7 \right) H I J K L M N} C_{A B C} H_{D E F G}
  C^{H I J} H^{K L M N}  .
\end{equation}
I shall now assume that to leading order in the LOW harmonic expansion, the
vacuum fluxes satisfy:
\begin{equation}
  \label{fluxes without contractions} H_{A B C D} H^{E F G H} = \frac{6 h^2}{5
  B^8} \delta_A \, \!\! \! \!^{\!\!\! \left[ E \right.} \delta_B \, \! \!^F
  \delta_C \, \! \!^G \delta_D \, \!\! \!^{\left. \!\!\! H \right]} =
  \frac{h^2}{120 B^8} \epsilon_{\left( 7 \right) A B C D I J K} \epsilon^{E F
  G H I J K}_{\left( 7 \right)} \!,
\end{equation}
where the coefficient is fixed by (\ref{flux bilinears}), and as with
(\ref{flux bilinears}), the LOW expansion only needs to be applied over
relatively small local regions of $\bar{H}^7$, due to the approximately
uniform distribution of the fluxes across $\bar{H}^7$. \ All indices in
(\ref{after integrating out H}) and (\ref{fluxes without contractions}) are
tangential to $\bar{H}^7$. \ After adding the kinetic term $- 4 H_{\mu A B C}
H^{\mu A B C}$ from $- H_{I J K L} H^{I J K L}$, (\ref{after integrating out
H}) becomes:
\begin{equation}
  \label{kinetic and mass term for C sub A B C} \frac{1}{96 \kappa_{11}^2} 
  \int_{\mathcal{B}} d^{11} xe \left( - 4 H_{\mu A B C} H^{\mu A B C} -
  \frac{4 h^2}{45 B^8} C^{A B C} C_{A B C}  \right) .
\end{equation}
Thus within the above approximations, all the harmonic 3-form modes of $C_{A B
C}$ obtain the same intrinsic mass $\bar{m} = \frac{h}{\sqrt{45} B^3}$. \ From
subsection \ref{subsection PMS} above, the best value of $\eta =
\frac{h}{B^3}$ chosen by the PMS is $\eta \simeq 1.425$, so $\bar{m} \simeq
0.2$. \ Thus the mass of these modes, as seen on the HW boundary, is $m
\simeq 0.2 \frac{A}{B}$.

\subsubsection{The coupling of the harmonic 3-form modes to the SM gauge bosons}
\label{The coupling of the harmonic 3 form modes to the SM gauge bosons}

The coupling of the harmonic 3-form modes of $C_{A B C}$ to the SM fields can
be obtained by integrating out $H_{\mu \nu \sigma \tau}$, in the same way as
was done above to calculate their mass. \ In Moss's improved form of
Ho\v{r}ava-Witten theory, the boundary condition for $H_{I J K L}$ has the
form {\cite{Moss 1, Moss 2, Moss 3, Moss 4}}:
\begin{equation}
  \label{bc for H} \left. H_{U V W X} \right|_{y = y_{1 +}} = \frac{1}{2 \pi}
  \left( \frac{\kappa_{11}}{4 \pi} \right)^{2 / 3} \left( - 3 F_{\left[ U V
  \right.}^{\mathcal{A}} F_{\left. W X \right]}^{\mathcal{A}} +
  \bar{\chi}^{\mathcal{A}} \Gamma_{\left[ U V W \right.} \left( D_{\left. X
  \right]} \chi \right)^{\mathcal{A}} \right) + \ldots,
\end{equation}
where $\chi^{\mathcal{A}}$ is the gaugino, and $\ldots$ denotes terms that
involve the gravitino or $R_{I J K L}$ or $H_{I J K L}$. \ This can be
integrated to:
\begin{equation}
  \label{bc for C} \left. C_{UVW} \right|_{y = y_{1 +}} = \frac{1}{4 \pi}
  \left( \frac{\kappa_{11}}{4 \pi} \right)^{2 / 3}  \left( - \frac{1}{30}
  \Omega_{UVW}^{\left( \mathrm{Y} \right)} + \frac{1}{4}
  \bar{\chi}^{\mathcal{A}} \Gamma_{U V W} \chi^{\mathcal{A}} \right) +
  \lambda_{UVW} + \ldots,
\end{equation}
where
\begin{equation}
  \label{Omega Y} \Omega_{UVW}^{\left( \mathrm{Y} \right)} = 6 \mathrm{tr}
  \left( A_{\left[ U \right.} \partial_V A_{\left. W \right]} + \frac{2}{3}
  iA_{\left[ U \right.} A_V A_{\left. W \right]} \right)
\end{equation}
is the Yang-Mills Chern-Simons 3-form, $\lambda_{UVW}$ is an arbitrary closed
3-form on the boundary, and $\ldots$ denotes terms that involve the gravitino
or the Lorentz Chern-Simons 3-form or $H_{I J K L}$.

With the notation of (\ref{metric near boundary}) above, let $\tilde{C}_{U V
W} \left( \check{x}, \hat{x} \right)$ denote the right-hand side of (\ref{bc
for C}). \ If we neglect $H_{\mu \nu \sigma \tau}$ and $H_{\mu \nu \sigma a}$
and the CJS Chern-Simons term, the CJS field equations (\ref{3 form field
equations}) for $H_{I J K L}$ include an equation $\frac{\partial}{\partial y}
\left( a^4 b^6 a^{- 6} H_{\mu \nu \sigma y} \right) = 0$, whose solution is
$H_{\mu \nu \sigma y} = \frac{a^2}{b^6} f_{\mu \nu \sigma} \left( \check{x},
\hat{x} \right)$. \ If we further neglect $C_{\mu \nu y}$, we then find that
the form of $C_{\mu \nu \sigma}$ induced by $\tilde{C}_{\mu \nu \sigma} \left(
\check{x}, \hat{x} \right)$ is $C_{\mu \nu \sigma} = f_{\mu \nu \sigma} \left(
\check{x}, \hat{x} \right) \int^{\infty}_y \frac{a^2 \left( y' \right)}{b^6
\left( y' \right)} dy'$. \ In the approximation of neglecting the
perturbations $p \left( y \right)$ and $q \left( y \right)$ defined between
(\ref{metric near boundary}) and (\ref{p dot eqn}) above, so that $a \left( y
\right) = A$, and $b \left( y \right) = B \mathrm{cosh} \frac{y}{B}$, we have:
\begin{equation}
  \label{integral of a2 over b6} \int^{\infty}_y \frac{a^2 \left( y'
  \right)}{b^6 \left( y' \right)} dy' \simeq \int_y^{\infty} \frac{A^2}{B^6
  \left( \mathrm{\cosh} \frac{y}{B} \right)^6} = \frac{A^2 \mathrm{e}^{-
  \frac{5 y}{B}}  \left( 10 \mathrm{e}^{\frac{4 y}{B}} + 5 \mathrm{e}^{\frac{2
  y}{B}} + 1 \right)}{30 B^5  \left( \mathrm{\cosh} \frac{y}{B} \right)^5},
\end{equation}
so in the further approximation of setting $y_1$, the value of $y$ at the
boundary, to 0, the flux $H^{\left( \mathrm{{{ind}}}
\right)}_{\mu \nu \sigma y}$ induced by $\tilde{C}_{\mu \nu \sigma} \left(
\check{x}, \hat{x} \right)$ is:
\begin{equation}
  \label{H mu nu sigma y induced by bc} H^{\left(
  \mathrm{{{ind}}} \right)}_{\mu \nu \sigma y} \left(
  \check{x}, \hat{x}, y \right) \simeq \frac{15 B^5 a^2}{8 A^2 b^6}
  \tilde{C}_{\mu \nu \sigma} \left( \check{x}, \hat{x} \right) .
\end{equation}
The principal coupling between the SM gauge bosons and the $C_{A B C}$ modes
arises from the cross term between $H^{\left(
\mathrm{{{ind}}} \right)}_{\mu \nu \sigma y}$, and the
flux $H^{\left( \mathrm{{{spont}}} \right)}_{\mu \nu
\sigma y}$ that originates from (\ref{C in terms of H}), in $- \frac{4}{96
\kappa_{11}^2}  \int_{\mathcal{B}} d^{11} xe \left( H_{\mu \nu \sigma y}
H^{\mu \nu \sigma y} \right)$, which is one of the $H_{\mu \nu \sigma A}
H^{\mu \nu \sigma A}$ terms neglected in deriving (\ref{relevant terms
containing H}). \ Using the algebraic field equation for $H_{\mu \nu \sigma
\tau}$ that follows from (\ref{relevant terms containing H}), we find:
\begin{dmath}[compact, spread=3pt]
  \label{H spont} H^{\left( \mathrm{{{spont}}}
  \right)}_{\mu \nu \sigma y} \left( \check{x}, \hat{x}, y \right) =
  \frac{1}{216}  \frac{\partial}{\partial y} \left( \epsilon^{a b c d e f
  y}_{\left( 11 \right)} \, \!_{\mu \nu \sigma \rho}  \int d^4 \check{x}'
  \frac{\partial}{\partial \check{x}_{\rho}} G_{4 F} \left( \check{x} -
  \check{x}' \right) \left( 3 C_{a b y} \left( \check{x}', \hat{x}, y \right)
  H_{c d e f} + 4 C_{a b c} \left( \check{x}', \hat{x}, y \right) H_{d e f y}
  \right) \right),
\end{dmath}
where $H_{c d e f}$ and $H_{d e f y}$ are the vacuum fluxes that to leading
order in the LOW harmonic expansion, applied over relatively small local
regions of $\bar{H}^7$, satisfy (\ref{flux bilinears}) and (\ref{fluxes
without contractions}). \ So considering just the Yang-Mills term in
$\tilde{C}_{\mu \nu \sigma} \left( \check{x}, \hat{x} \right)$, the principal
coupling between the SM gauge bosons and the $C_{A B C}$ modes is:
\begin{dmath}[compact, spread=3pt]
  \label{axion type coupling} - \frac{8}{96 \kappa_{11}^2}  \int_{\mathcal{B}}
  d^{11} xeH^{\left( \mathrm{{{ind}},
  \mathrm{{YM}}} \right)}_{\mu \nu \sigma y} H^{\left(
  \mathrm{{{spont}}} \right) \mu \nu \sigma y} \simeq
  \frac{5 B^5}{2^{11} 3^2 \pi A^2 \kappa_{11}^2} \left( \frac{\kappa_{11}}{4
  \pi} \right)^{2 / 3}  \int d^4 \check{x} \int_{\bar{H}^6} d^6 \hat{x}
  \sqrt{\hat{g}} \epsilon^{\mu \nu \sigma \rho}_{\left( 4 \right)} F_{\left[
  \mu \nu \right.}^{\mathcal{A}} F_{\left. \sigma \rho \right]}^{\mathcal{A}}
  \int_{y_1}^{\infty} dya^6  \frac{\partial}{\partial y} \int d^4 \check{x}'
  G_{4 F} \left( \check{x} - \check{x}' \right) \epsilon^{a b c d e f
  y}_{\left( 7 \right)} \left( 3 C_{a b y} \left( \check{x}', \hat{x}, y
  \right) H_{c d e f} + 4 C_{a b c} \left( \check{x}', \hat{x}, y \right) H_{d
  e f y} \right) .
\end{dmath}
The integral is strongly localized near the boundary, because $\epsilon^{a b c
d e f y}_{\left( 7 \right)}$ is $b^{- 6}$ times $\pm 1$ or 0, and $a
\rightarrow A$ and $b \rightarrow B \mathrm{cosh} \frac{y}{B}$ as $y
\rightarrow \infty$. \ If we again neglect the perturbation $p \left( y
\right)$, so that $a \left( y \right) = A$, and neglect the massive
Kaluza-Klein modes of the Yang-Mills gauge fields, then the coupling is
approximately:
\begin{dmath}[compact, spread=3pt]
  \label{simplified axion coupling} - \frac{5 A^4 B^5}{2^{11} 3^2 \pi
  \kappa_{11}^2} \left( \frac{\kappa_{11}}{4 \pi} \right)^{2 / 3} \sum_{\left(
  n \right)}  \int d^4 \check{x} \epsilon^{\mu \nu \sigma \rho}_{\left( 4
  \right)} F_{\left[ \mu \nu \right.}^{\mathcal{A}} F_{\left. \sigma \rho
  \right]}^{\mathcal{A}} \int d^4 \check{x}' G_{4 F} \left( \check{x} -
  \check{x}' \right) \times C_{\left( n \right)} \left( \check{x}' \right)
  \int_{\bar{H}^6} d^6 \hat{x} \sqrt{\hat{g}} \epsilon^{A B C D E F G}_{\left(
  7 \right)} \omega_{\left( n \right) A B C} \left( \hat{x}, y_1 \right) H_{D
  E F G} \left( \hat{x}, y_1 \right),
\end{dmath}
where $C_{A B C} \left( \check{x}, \hat{x}, y \right)$ has been expanded in
mass eigenmodes as:
\begin{equation}
\label{eigenmode expansion of C ABC}
C_{A B C} \left( \check{x}, \hat{x}, y \right)
= \sum_{\left( n \right)} C_{\left( n \right)} \left(
\check{x} \right) \omega_{\left( n \right) A B C} \left( \hat{x}, y \right).
\end{equation}
If we restrict this sum to the harmonic 3-form modes, with mass $m \simeq
0.2 \frac{A}{B}$, then when the coupling (\ref{simplified axion coupling})
is inserted into a momentum-space Feynman diagram for two gluons to turn into
a $C_{\left( n \right)}$, which then decays to 2 or 3 SM gauge bosons, the
massless propagator $G_{4 F}$ at each end of the $C_{\left( n \right)}$
propagator becomes a factor $\frac{1}{m^2}$ near the $C_{\left( n \right)}$
mass shell. \ Thus near the $C_{\left( n \right)}$ mass shell, the coupling
(\ref{simplified axion coupling}) has the standard form for the coupling of
the SM gauge bosons to axion fields $C_{\left( n \right)} \left( \check{x}
\right)$ {\cite{Beasley Witten}}. \ However the $C_{\left( n \right)}$ fields,
whose mass would be around a TeV, are very different from conventional axions,
which are extremely light {\cite{Weinberg Axion, Wilczek Axion}}.

If candidates for the $C_{\left( n \right)}$ modes are observed and their
decays to 3 gluon jets can be identified, the coupling (\ref{simplified axion
coupling}) could be tested by plotting the energies of the 3 gluon jets, in
the reconstructed rest frame of the candidate $C_{\left( n \right)}$ mode, on
a Dalitz plot {\cite{Dalitz, Fabri, Perkins}}. \ The coupling is proportional
to the 4-momentum of the $C_{\left( n \right)}$ mode because $\epsilon^{\mu
\nu \sigma \rho}_{\left( 4 \right)} F_{\left[ \mu \nu \right.}^{\mathcal{A}}
F^{\mathcal{A}}_{\left. \sigma \rho \right]}$ is a total derivative, so in
radiation gauge in the rest frame of the $C_{\left( n \right)}$ mode, the
polarizations of the gluons in the 3-gluon term in (\ref{Omega Y}) must be
linearly independent, and that is not possible if the 3 gluons are collinear.
\ Thus the amplitude vanishes for 3 collinear gluons, which means that the
distribution of events will be depleted near the edges of the Dalitz plot,
which are the lines $2 E_1 = m$, $2 E_2 = m$, and $2 E_3 = m$. \ The
background to this effect would include both the QCD background, and the
decays of the candidate $C_{\left( n \right)}$ modes to 2 gluons, where one of
the gluons radiates a third gluon.

\subsection{The Kaluza-Klein modes of the metric}
\label{KK modes of metric}

The background solution of the field equation for the metric depends
essentially on the presence of the leading quantum correction (\ref{Gamma 8
SG}) to the CJS action, so the presence of that term has to be taken into
account in studying the Kaluza-Klein modes of the metric. \ However
(\ref{Gamma 8 SG}) is 8th order in derivatives, so it has to be treated as a
perturbation of the momentum-dependent terms in the action of the Kaluza-Klein
modes. \ For a first approximation, I shall neglect the contribution of
(\ref{Gamma 8 SG}) to the momentum-dependent terms in the action of the
Kaluza-Klein modes, and calculate the contribution of (\ref{Gamma 8 SG}) to
the mass squared of the dilaton/radion. \ I shall then assume that the
contribution of (\ref{Gamma 8 SG}) to the mass squared of the other light
Kaluza-Klein modes of the metric is of similar magnitude to its contribution
to the mass squared of the dilaton/radion.

I shall write the perturbed metric as $\bar{\bar{G}}_{I J} \equiv G_{I J}
+ 2 h_{I J}$, where $G_{I J}$ is the metric defined by (\ref{metric ansatz for
H7}), and $h_{I J}$ is the perturbation tensor. \ Indices will still be raised
and lowered with $G_{I J}$, and covariant derivatives $D_I$ will still be
defined in terms of the unperturbed metric $G_{I J}$, and satisfy $D_I G_{J K}
\equiv 0$. \ Repeated lower coordinate indices are still understood to be
contracted with $G^{I J}$. \ The inverse of $\bar{\bar{G}}_{I J}$ will be
written as $\check{\bar{G}}^{I J} \equiv G^{I J} - 2 h^{I J} + 4 h^I \, \!_K
h^{K J} - \ldots$, to distinguish it from the tensor obtained from
$\bar{\bar{G}}_{I J}$ by raising both its indices with $G^{I J}$. \ Other
quantities defined in terms of $\bar{\bar{G}}_{I J}$ will be written with
a double bar above them. \ Repeated lower coordinate indices are still
understood to be contracted with an unperturbed inverse metric $G^{I J}$, in
accordance with the convention stated between (\ref{t8 A B C D}) and (\ref{t8
t8 X4}). \ Then we have the tensor:
\begin{equation}
  \label{Delta I J K} \Delta_I \, \!^J \, \!_K \equiv
  \bar{\bar{\Gamma}}_I \, \!^J \, \!_K - \Gamma_I \, \!^J \, \!_K =
  \check{\bar{G}}^{J L} \left( D_I h_{K L} + D_K h_{I L} - D_L h_{I K}
  \right),
\end{equation}
and the tensor:
\begin{equation}
  \label{R bar bar I J K L} \bar{\bar{R}}_{IJ} \hspace{0.25em}
  \hspace{-0.25em}^K \hspace{0.25em} \hspace{-0.25em}_L = R_{IJ}
  \hspace{0.25em} \hspace{-0.25em}^K \hspace{0.25em} \hspace{-0.25em}_L + D_I
  \Delta_J \hspace{0.25em} \hspace{-0.25em}^K \hspace{0.25em}
  \hspace{-0.25em}_L - D_J \Delta_I \hspace{0.25em} \hspace{-0.25em}^K
  \hspace{0.25em} \hspace{-0.25em}_L + \Delta_I  \hspace{0.25em}
  \hspace{-0.25em}^K \hspace{0.25em} \hspace{-0.25em}_M \Delta_J
  \hspace{0.25em} \hspace{-0.25em}^M \hspace{0.25em} \hspace{-0.25em}_L -
  \Delta_J  \hspace{0.25em} \hspace{-0.25em}^K \hspace{0.25em}
  \hspace{-0.25em}_M \Delta_I \hspace{0.25em} \hspace{-0.25em}^M
  \hspace{0.25em} \hspace{-0.25em}_L,
\end{equation}
and also:
\begin{equation}
  \label{e double bar} \bar{\bar{e}} = \sqrt{- \bar{\bar{G}}} = e
  \left( 1 + h_{I I} - h_{I J} h_{J I} + \frac{1}{2} h_{I I} h_{J J} \right),
\end{equation}
where the double-barred quantities in the above equations are defined in terms
of $\bar{\bar{G}}_{I J}$ with their indices in the positions shown. \ We
then find {\cite{Christensen Duff, Salam Strathdee Randjbar Daemi, Hoover
Burgess}}:
\begin{dmath}[compact, spread=3pt]
  \label{e bar bar R bar bar} \bar{\bar{e}} \bar{\bar{R}} =
  \bar{\bar{e}} \check{\bar{G}}^{I J} \bar{\bar{R}}_{I J} =
  \bar{\bar{e}} \check{\bar{G}}^{I J} \bar{\bar{R}}_{KI}
  \hspace{0.25em} \hspace{-0.25em}^K \hspace{0.25em} \hspace{-0.25em}_J \cong
  e \left( R + h_{I I} R - 2 h_{I J} R_{I J} + 2 h_{I K} h_{K J} R_{I J} + 2
  h_{I K} h_{J L} R_{I J K L} - 2 h_{K K} h_{I J} R_{I J} - h_{I J} h_{J I} R
  + \frac{1}{2} h_{I I} h_{J J} R - D_K h_{I J} D_K h_{I J} + 2 D_I h_{I K}
  D_J h_{J K} - 2 D_I h_{I J} D_J h_{K K} + D_K h_{I I} D_K h_{J J} \right),
\end{dmath}
where $\cong$ means up to the addition of total derivative terms, and the
identity:
\begin{equation}
\label{derivative commutator identity}
eD_I h_{J K} D_J h_{I K} \cong eD_I h_{I K} D_J h_{J K} - eh_{I K}
h_{K J} R_{I J} + eh_{I K} h_{J L} R_{I J K L}
\end{equation}
has been used.

Let $X_8$ be defined such that
$\Gamma_{\mathrm{{{SG}}}}^{\left( 8, \mathrm{bos}
\right)}$, in (\ref{Gamma 8 SG}), is $\frac{1}{2 \kappa_{11}^2}
\int_{\mathcal{B}} d^{11} xeX_8$. \ Let $\bar{\bar{X}}_8$ denote $X_8$ as
calculated from (\ref{Gamma 8 SG}), (\ref{R cup I J K L}), (\ref{t8 t8 X4}),
(\ref{eps11 t8 C R4}), (\ref{definition of Z}), and the definition of
$\tilde{Z}$ as explained after (\ref{definition of Z}), with $G_{I J}$
replaced by $\bar{\bar{G}}_{I J} = G_{I J} + 2 h_{I J}$, and let $\left[
\bar{\bar{X}}_8 \right]_{\mathrm{w.o.} \Delta}$ denote
$\bar{\bar{X}}_8$, but with $\Delta_I \, \!^J \, \!_K$ in (\ref{R bar bar
I J K L}) set to 0, so that $\left[ \bar{\bar{X}}_8
\right]_{\mathrm{w.o.} \Delta}$ does not contain any derivatives acting on
$h_{I J}$. \ Let $\left. \frac{\partial \left[ \bar{\bar{X}}_8
\right]_{\mathrm{w.o.} \Delta}}{\partial h_{I J}} \right|_G$ denote the
ordinary derivative of $\left[ \bar{\bar{X}}_8 \right]_{\mathrm{w.o.}
\Delta}$ with respect to $h_{I J}$ at $h_{I J} = 0$, where as usual in
differentiating a function that depends on a symmetric tensor, all components
of the tensor are treated as independent in the argument of the function, so
that $\left. \frac{\partial \left[ \bar{\bar{X}}_8 \right]_{\mathrm{w.o.}
\Delta}}{\partial h_{I J}} \right|_G h_{I J}$ is the linear term in the
expansion of $\left[ \bar{\bar{X}}_8 \right]_{\mathrm{w.o.} \Delta}$ in
powers of $h_{I J}$. \ Let $\left[ \bar{\bar{X}}_8 \right]_{1 \Delta}$
denote the part of the linear term in the expansion of $\bar{\bar{X}}_8$
in powers of $h_{I J}$ that arises from $\Delta_I \, \!^J \, \!_K$
{\itshape{only}}, and let $\left[ \bar{\bar{X}}_8 \right]_{2,
\mathrm{{{rd}}.}}$ denote the quadratic term in the
expansion of $\bar{\bar{X}}_8$ in powers of $h_{I J}$, but with the term
$\left. - 2 \frac{\partial \left[ \bar{\bar{X}}_8 \right]_{\mathrm{w.o.}
\Delta}}{\partial h_{I J}} \right|_G h_{I K} h_{K J}$, that originates from
the quadratic term in the expansion of $\check{\bar{G}}^{I J}$, removed. \ Let
$\Lambda^{\left( \mathrm{{{bos}}}
\right)}_{\mathrm{{{SG}}}}$ be the integrand of
$\Gamma^{\left( \mathrm{bos} \right)}_{\mathrm{SG}}$ in (\ref{Gamma SG}). \
Then the expansion of $2 \kappa^2_{11} \bar{\bar{\Lambda}}^{\left(
\mathrm{{{bos}}}
\right)}_{\mathrm{{{SG}}}}$ without the
metric-independent CJS Chern-Simons term, through quadratic order in $h_{I
J}$, up to total derivative terms, can be written:
\begin{dmath}[compact, spread=3pt]
  \label{action density through h squared} \bar{\bar{e}}
  \bar{\bar{R}} - \frac{1}{48} \bar{\bar{e}} \bar{\bar{G}}^{I
  M} \bar{\bar{G}}^{J N} \bar{\bar{G}}^{K O} \bar{\bar{G}}^{L
  P} H_{I J K L} H_{M N O P} + \bar{\bar{e}}  \bar{\bar{X}}_8 \cong
  e \left( 1 + h_{N N} - h_{M N} h_{N M} + \frac{1}{2} h_{M M} h_{N N} \right)
  \left\{ R - \frac{1}{48} H_{I J K L} H_{I J K L} + X_8 \right\} + e \left( 1
  + h_{N N} \right) \left( - 2 h_{I M} + 4 h_{I O} h_{O M} \right) \left\{
  R_{I M} - \frac{1}{12} H_{I J K L} H_{M J K L} - \frac{1}{2}  \left.
  \frac{\partial \left[ \bar{\bar{X}}_8 \right]_{\mathrm{w.o.}
  \Delta}}{\partial h_{I M}} \right|_G \right\} - eD_I h_{J K} D_I h_{J K} + 2
  eD_I h_{I K} D_J h_{J K} - 2 eD_I h_{I J} D_J h_{K K} + eD_I h_{J J} D_I
  h_{K K} - 2 eh_{I K} h_{K J} R_{I J} + 2 eh_{I K} h_{J L} R_{I J K L} -
  \frac{1}{2} eh_{I M} h_{J N} H_{I J K L} H_{M N K L} + eh_{N N} \left[
  \bar{\bar{X}}_8 \right]_{1 \Delta} + e \left[ \bar{\bar{X}}_8
  \right]_{2, \mathrm{{{rd}}.}}
\end{dmath}
For the unperturbed solution considered here, the field equation  (\ref{A
field equation})  resulting from varying $A$, or equivalently, the field
equation resulting from varying $G_{\mu \nu}$, states that the action is zero,
so the contents of the first pair of braces in  (\ref{action density through h
squared})  are 0. \ This requires fine-tuning the root mean square flux
strength $h$ in  (\ref{flux bilinears}) , and from {\cite{Bousso Polchinski}}
or page 34 of {\cite{Almost Flat}}, this can easily be achieved to the
required precision of about 1 part in $10^{90}$ of $h^2$, due to the large
flux numbers of the fluxes wrapping typical 4-cycles of $\bar{H}^7$, with
intrinsic 4-area $\sim 10^{30}$, even when the fine-tuning is required to hold
over the relatively small local regions over which the Lukas-Ovrut-Waldram
harmonic expansion {\cite{Lukas Ovrut Waldram}} is assumed to be applied. \
The observed cosmological vacuum energy density of about $\left( 2.3 \times
10^{- 3} \hspace{1ex} \mathrm{{{eV}}} \right)^4$
{\cite{cosmological constant 1, cosmological constant 2, cosmological
constant 3}} is negligible for terrestrial laboratory
experiments, and I shall here treat the contents of the first pair of braces
in  (\ref{action density through h squared})  as exactly 0. \ The contents of
the second pair of braces in  (\ref{action density through h squared})  are
then also exactly 0 in consequence of the field equation  (\ref{B field
equation})  resulting from varying $B$, or equivalently, the field equation
resulting from varying $G_{A B}$, since to first order in $h_{I J}$, the only
$h_{I J}$-dependent terms in the right-hand side of (\ref{R bar bar I J K L})
are the two $D \Delta$ terms, and when we integrate the $D$ away from the
$\Delta$ by parts in $\bar{\bar{\Gamma}}^{\left( 8, \mathrm{bos}
\right)}_{\mathrm{SG}}$, the $D$ can only act on an $R_{I J K L}$, whose
covariant derivatives are all 0, by the local symmetry of the metric ansatz
(\ref{metric ansatz for H7}).

When we split the index $I$ to $\mu$ and $A$, the third and fourth terms after
the second pair of braces in  (\ref{action density through h squared}) 
contain mixing terms between $h_{\mu \nu}$ and the dilaton/radion, which is
here proportional to $h_{A A}$. \ This mixing arises because the coefficient
of $\sqrt{- \check{g}} R \left( \check{g} \right)$ in (\ref{Einstein action}),
when (\ref{Einstein action}) is derived from (\ref{CJS action}) by integration
over $\bar{H}^7$, is proportional to the volume $\bar{\bar{V}}_7$ of
$\bar{\bar{H}}^7$. \ This mixing can always be removed in a manner
consistent with general covariance along the extended dimensions, by making a
dilaton-dependent conformal transformation of the metric $\check{g}_{\mu \nu}$
along the extended dimensions, of the form $\check{g}_{\mu \nu} = \left(
\frac{\bar{\bar{V}}_7}{V_7} \right)^{- \frac{2}{d - 2}}
\check{\check{g}}_{\mu \nu}$, where $d$ is the number of extended dimensions,
here 4. \ This is usually referred to as going to Einstein frame
{\cite{Wikipedia Einstein frame string frame}}.

To the relevant order for the mixing terms, $\frac{\bar{\bar{V}}_7}{V_7}
= 1 + h_{A A}$, from (\ref{e double bar}). \ Thus from $\check{g}_{\mu \nu} =
\bar{\bar{G}}_{\mu \nu} = G_{\mu \nu} + 2 h_{\mu \nu}$, and defining
$\check{\check{g}}_{\mu \nu} \equiv G_{\mu \nu} + 2 s_{\mu \nu}$, we have
$h_{\mu \nu} = s_{\mu \nu} - \frac{1}{d - 2} h_{A A} G_{\mu \nu}$. \ We also
define $t_{A B} \equiv h_{A B} - \frac{1}{n} h_{C C} G_{A B}$, where $n$ is
the number of compact dimensions, here 7, so that $t_{A A} = 0$. \ Then after
adding gauge-fixing terms as follows,  (\ref{action density through h
squared})  becomes:
\begin{dmath*}[compact, spread=3pt]
   - eD_I h_{JK} D_I h_{JK} + 2 \hspace{0.25em} eD_I h_{IK} D_J h_{JK} - 2
   \hspace{0.25em} eD_I h_{IJ} D_J h_{KK} + eD_I h_{JJ} D_I h_{KK} - 2
   \hspace{0.25em} eh_{IK} h_{KJ} R_{IJ} + 2 \hspace{0.25em} eh_{IK} h_{JL}
   R_{IJKL} - \frac{1}{2}  \hspace{0.25em} eh_{IM} h_{JN} H_{IJKL} H_{MNKL} +
   eh_{N N} \left[ \bar{\bar{X}}_8 \right]_{1 \Delta} + e \left[
   \bar{\bar{X}}_8 \right]_{2, \mathrm{{{rd}}.}}
   - 2 \hspace{0.25em} e \left( \tilde{a} D_{\mu} s_{\mu \nu} + \tilde{b}
   D_{\nu} s_{\mu \mu} + \frac{1}{\tilde{a}} D_A h_{\nu A} \right) \left(
   \tilde{a} D_{\sigma} s_{\sigma \nu} + \tilde{b} D_{\nu} s_{\sigma \sigma} +
   \frac{1}{\tilde{a}} D_B h_{\nu B} \right) - e \frac{\tilde{a} +
   \tilde{b}}{\tilde{a}} \left( - 2 \hspace{0.25em} D_{\mu} h_{\mu B} + D_B
   s_{\mu \mu} - \frac{\tilde{a}}{\tilde{a} + \tilde{b}} \left( D_A h_{AB} +
   \frac{1}{d - 2} D_B h_{AA} \right) \right) \left( - 2 \hspace{0.25em}
   D_{\nu} h_{\nu B} + D_B s_{\nu \nu} - \frac{\tilde{a}}{\tilde{a} +
   \tilde{b}} \left( D_C h_{CB} + \frac{1}{d - 2} D_B h_{CC} \right) \right)
   \cong
\end{dmath*}
\begin{dmath}[compact, spread=3pt]
  \label{gauge fixed hh action} \cong - e \partial_{\mu} s_{\nu \sigma}
  \partial_{\mu} s_{\nu \sigma} + 2 \left( 1 - \tilde{a}^2 \right)
  \hspace{0.25em} e \partial_{\mu} s_{\mu \sigma} \partial_{\nu} s_{\nu
  \sigma} - 2 \left( 1 + 2 \tilde{a}  \tilde{b} \hspace{0.25em} \right) e
  \partial_{\mu} s_{\mu \nu} \partial_{\nu} s_{\sigma \sigma} + \left( 1 - 2
  \hspace{0.25em} \tilde{b} ^2 \right) e \partial_{\mu} s_{\nu \nu}
  \partial_{\mu} s_{\sigma \sigma} - eD_A s_{\mu \nu} D_A s_{\mu \nu} -
  \frac{\tilde{b}}{\tilde{a}} eD_A s_{\mu \mu} D_A s_{\nu \nu} - 2
  \hspace{0.25em} e \partial_{\mu} h_{\nu A} \partial_{\mu} h_{\nu A} - 2
  \frac{\tilde{a} + 2 \tilde{b}}{\tilde{a}} e \partial_{\mu} h_{\mu A}
  \partial_{\nu} h_{\nu A} - 2 \hspace{0.25em} eD_A h_{\mu B} D_A h_{\mu B} +
  2 eD_A h_{\mu B} D_B h_{\mu A} - \frac{2}{\tilde{a}^2} eD_A h_{\mu A} D_B
  h_{\mu B} - e \partial_{\mu} t_{AB} \partial_{\mu} t_{AB} - eD_C t_{A B} D_C
  t_{A B} + 2 \hspace{0.25em} eR_{ABCD} t_{AC} t_{BD} - 2 \hspace{0.25em}
  eR_{AB} t_{AC} t_{BC} - \frac{1}{2}  \hspace{0.25em} eH_{ABCE} H_{ABDF}
  t_{CD} t_{EF} + \frac{\tilde{a} + 2 \tilde{b}}{\tilde{a} + \tilde{b}} e
  \left( D_A t_{AC} + \left( \frac{1}{d - 2} + \frac{1}{n} \hspace{0.25em}
  \right) D_C h_{A A}  \right) \left( D_B t_{BC} + \left( \frac{1}{d - 2} +
  \frac{1}{n} \hspace{0.25em} \right) D_C h_{BB}  \right) - \left( \frac{1}{d
  - 2} + \frac{1}{n} \right) e \partial_{\mu} h_{AA} \partial_{\mu} h_{BB} -
  \left( \frac{1}{d - 2} + \frac{1}{n} \right) eD_C h_{A A} D_C h_{BB} -
  \hspace{0.25em} \frac{1}{2 n^2} eH_{ABCD} H_{ABCD} h_{E E} h_{FF} + eh_{N N}
  \left[ \bar{\bar{X}}_8 \right]_{1 \Delta} + e \left[
  \bar{\bar{X}}_8 \right]_{2, \mathrm{{{rd}}.}}
\end{dmath}
Here $\tilde{a}$ and $\tilde{b}$ are gauge parameters, (\ref{flux bilinears})
has been used to set $H_{ABCD} H_{ABCE} t_{DE}$ to 0, and $eD_A h_{\mu A} D_B
h_{\mu B} - eR_{A B} h_{\mu A} h_{\mu B} \cong eD_A h_{\mu B} D_B h_{\mu A}$,
valid for the metric (\ref{metric ansatz for H7}), has been used. \ If we
choose $\tilde{a} = 1$, $\tilde{b} = - \frac{1}{2}$, we obtain de Donder gauge
for $s_{\mu \nu}$ and Feynman gauge for $h_{\mu A}$, and the traceless tensor
modes $t_{A B}$ on $\bar{H}^7$ are decoupled from $h_{A A}$, apart from
possible couplings coming from the last two terms.

The $t_{A B}$ and $h_{A A}$ modes are where tachyons are most likely to occur
{\cite{Duff Nilsson Pope, Salam Strathdee Randjbar Daemi}}. \ If we ignore the
last two terms in (\ref{gauge fixed hh action}), then after making a
Kaluza-Klein ansatz $t_{A B} = t \left( \check{x} \right) \omega_{A B} \left(
\bar{x} \right)$, the intrinsic masses $\bar{m}$ of the $t_{A B}$ modes are
determined by the spectrum on $\bar{H}^7$ in the metric $G_{A B} = B^2
\bar{g}_{A B}$ of the equation:
\begin{dmath}[compact, spread=3pt]
  \label{equation for tAB modes} - D_C D_C \omega_{A B} - 2 \hspace{0.25em}
  R_{ACBD} \omega_{CD} + R_{AC} \omega_{CB} + R_{B C} \omega_{C A} +
  \frac{1}{2} H_{AC EF} H_{BD EF} \omega_{CD} - \frac{1}{2 n} G_{A B} H_{C E F
  G} H_{D E F G} \omega_{C D} = \frac{\bar{m}^2}{B^2} \omega_{A B} .
\end{dmath}
The last term in the left-hand side of (\ref{equation for tAB modes}) results
from the tracelessness of $t_{A B}$ in (\ref{gauge fixed hh action}). \ From
pages 42 to 43 of {\cite{Almost Flat}}, the masses of these modes, as seen on
the HW boundary, are $m = \frac{A}{B} \bar{m}$, where $A$ and $B$ are the
constants in the metric ansatz (\ref{metric ansatz for H7}). \ The first 4
terms in the left-hand side of (\ref{equation for tAB modes}) are known as the
Lichnerowicz Laplacian acting on the traceless symmetric tensor $\omega_{A
B}$. \ On uncompactified $H^n$ of sectional curvature $- \frac{1}{B^2}$ with
$n > 2$, its spectrum extends from $\frac{\left( n - 1 \right) \left( n - 9
\right)}{4 B^2}$ to $+ \infty$ {\cite{Delay, Lee Lichnerowicz spectrum}}. \ If
similar eigenfunctions with approximately the same eigenvalues exist on
$\bar{H}^n$, then from (\ref{flux bilinears}), with the best estimate $\eta
\simeq 1.425$ from the second paragraph before (\ref{second derivative of
action wrt B}), where $\eta = \frac{h}{B^3}$ from (\ref{definition of eta}),
the spectrum of $\bar{m}^2$ in (\ref{equation for tAB modes}) on $\bar{H}^7$
would extend from about $- 3 - 1.02 = - 4.02$ to $+ \infty$. \ Thus the $t_{A
B}$ modes would include tachyons, unless the last two terms in (\ref{gauge
fixed hh action}) lift their squared masses sufficiently.

The dilaton/radion is the mode of $h_{A B}$ such that $h_{A B}$ is an
$\bar{x}$-independent multiple of $G_{A B}$, so that all covariant derivatives
$D_A h_{B C}$ are 0, and $t_{A B} = 0$. \ For this mode $\bar{\bar{G}}_{A
B} = G_{A B} + 2 h_{A B} = \left( B + \delta B \right)^2 \bar{g}_{A B}$, so
$h_{A A} = 7 \left( \frac{\delta B}{B} + \frac{\left( \delta B \right)^2}{2
B^2} \right)$. \ From just before (\ref{action density through h squared}),
$\Lambda^{\left( \mathrm{{{bos}}}
\right)}_{\mathrm{{{SG}}}}$ is the integrand of
$\Gamma^{\left( \mathrm{bos} \right)}_{\mathrm{SG}}$ in (\ref{Gamma SG}), and
after substituting for $\delta B$ in terms of $h_{A A}$, the expansion of
$\bar{\bar{\Lambda}}^{\left( \mathrm{{{bos}}}
\right)}_{\mathrm{{{SG}}}}$ in powers of $\delta B$
through quadratic order has the form:
\begin{equation}
  \label{Lambda bar bar SG}
  \bar{\bar{\Lambda}}_{\mathrm{{{SG}}}}^{\left(
  \mathrm{{{bos}}} \right)} = \Lambda^{\left(
  \mathrm{{{bos}}}
  \right)}_{\mathrm{{{SG}}}} + \frac{\partial
  \Lambda^{\left( \mathrm{{{bos}}}
  \right)}_{\mathrm{{{SG}}}}}{\partial B} B \left(
  \frac{1}{7} h_{A A} - \frac{1}{98} h_{A A} h_{B B} \right) + \frac{1}{98} 
  \frac{\partial^2 \Lambda_{\mathrm{{{SG}}}}^{\left(
  \mathrm{{{bos}}} \right)}}{\partial B^2} B^2 h_{A A}
  h_{B B}
\end{equation}
In the vacuum, $\Lambda^{\left( \mathrm{{{bos}}}
\right)}_{\mathrm{{{SG}}}}$ and $\frac{\partial
\Lambda^{\left( \mathrm{{{bos}}}
\right)}_{\mathrm{{{SG}}}}}{\partial B}$ vanish by
(\ref{A field equation}) and (\ref{B field equation}) respectively, and from
(\ref{second derivative of action wrt B}), with the best estimate $\eta \simeq
1.425$ from the second paragraph before (\ref{second derivative of action wrt
B}), $\frac{\partial^2
\Lambda_{\mathrm{{{SG}}}}^{\left(
\mathrm{{{bos}}} \right)}}{\partial B^2} \simeq - 1.18
\frac{A^4 \sqrt{\bar{g}}}{\kappa^{2 / 3}_{11} B^3}$. \ Thus after adding the
momentum-dependent part of the dilaton/radion's kinetic term from $\frac{1}{2
\kappa^2_{11}}$ times (\ref{gauge fixed hh action}) with $d = 4$ and $n = 7$,
neglecting any momentum-dependent contributions from the last two terms in
(\ref{gauge fixed hh action}), the terms quadratic in $h_{A A}$ in the
gauge-fixed $\bar{\bar{\Lambda}}^{\left(
\mathrm{{{bos}}}
\right)}_{\mathrm{{{SG}}, g.f.}}$, in a gauge with
$\tilde{a} + 2 \tilde{b} = 0$, are:
\begin{equation}
  \label{dilaton kinetic terms} \bar{\bar{\Lambda}}^{\left(
  \mathrm{{{bos}}}
  \right)}_{\mathrm{{{SG}}, g.f.}} \simeq - \frac{A^4
  B^7 \sqrt{\bar{g}}}{2 \kappa^2_{11}} \left( \frac{9}{14} \partial_{\mu}
  h_{AA} \partial_{\mu} h_{BB} + \frac{1}{49}  \frac{1.18 \kappa^{4 /
  3}_{11}}{B^8} h_{A A} h_{B B} \right) .
\end{equation}
Thus using the best estimate $B \simeq 0.28 \kappa^{2 / 9}_{11}$ from the
second paragraph before (\ref{second derivative of action wrt B}), the
intrinsic mass squared of the dilaton/radion is $\simeq 78$, so from pages 42
to 43 of {\cite{Almost Flat}}, the dilaton/radion's mass, as seen on the HW
boundary, is $m_{\mathrm{dil}} \simeq 9 \frac{A}{B}$.

The dilaton/radion's intrinsic mass squared receives a contribution $\simeq
1.4$ from the third from last term in (\ref{gauge fixed hh action}), and the
remaining $\simeq 76.3$ comes from the last term in (\ref{gauge fixed hh
action}), so its relatively large size suggests that the last two terms in
(\ref{gauge fixed hh action}) might be able to raise the squared intrinsic
masses of the $t_{A B}$ modes sufficiently to avoid the occurrence of
tachyons. \ The relatively large value of the mass term in (\ref{dilaton
kinetic terms}) is due to the relatively small value of $\frac{B}{\kappa^{2 /
9}_{11}} \simeq 0.28$, notwithstanding that this value of $\frac{B}{\kappa^{2
/ 9}_{11}}$ satisfies the Giudice-Rattazzi-Wells perturbativity criterion
{\cite{Giudice Rattazzi Wells}} by a substantial margin, as noted in the
paragraph before (\ref{second derivative of action wrt B}). \ It is
interesting to note that the above estimate of the dilaton/radion mass is
about 42 times larger than the mass $\simeq 0.2 \frac{A}{B}$ of the
classically massless harmonic 3-form modes found in subsection \ref{The mass
of the harmonic 3 form modes}.

If we ignore the last two terms in (\ref{gauge fixed hh action}), then after
making a Kaluza-Klein ansatz $h_{\mu A} = h_{\mu} \left( \check{x} \right)
\omega_A \left( \bar{x} \right)$, the intrinsic masses $\bar{m}$ of the
$h_{\mu A}$ modes are determined by the spectrum on $\bar{H}^7$ in the metric
$\bar{g}_{A B}$ of sectional curvature $- 1$ of the equation $- \left( \delta
d + d \delta \right) \omega = \bar{m}^2 \omega$, where $\delta d + d \delta$
is the Hodge - de Rham Laplacian for $1$-forms on $\bar{H}^7$. \ So from pages
9 to 12 and 16 to 17 of {\cite{Almost Flat}}, it seems likely that the
lightest massive modes of $h_{\mu A}$ will have intrinsic mass $\bar{m} = 2$,
up to corrections from the last two terms in (\ref{gauge fixed hh action})
{\cite{KMST, Yau Isoperimetric, Agmon, Brooks Makover 1, Brooks Makover 2,
Brooks Makover 3, Brooks Makover 4, Cheeger, Cheeger Wikipedia, Donnelly,
Sarnak, Luo, Rudnick Sarnak, Orlando Park, Mazzeo Phillips, Donnelly Xavier,
Thurston, Colbois Courtois 1, Colbois Courtois 2}}. \ And from pages 42 to 43
of {\cite{Almost Flat}}, their masses in this approximation, as seen on the HW
boundary, are $m = 2 \frac{A}{B}$, where $A$ and $B$ are the constants in the
metric ansatz (\ref{metric ansatz for H7}). \ However the relatively large
contribution of the last term in (\ref{gauge fixed hh action}) to the
dilaton/radion's mass squared suggests that the last two terms in (\ref{gauge
fixed hh action}) might give a larger contribution to $\bar{m}^2$ for these
modes than the value 4 obtained in this first approximation.

There are no massless vector modes $h_{\mu A}$ corresponding to continuous
symmetries of $\bar{H}^7$, because $\bar{H}^7$ is a smooth compact negatively
curved Einstein space, and therefore cannot have any continuous symmetries.
For a vector field $V^A$ that generates a continuous symmetry on a smooth
Riemannian manifold $\mathcal{M}$ satisfies the Killing vector equation $D_A
V_B + D_B V_A = 0$, and thus $0 = D^A  \left( D_A V_B + D_B V_A \right)$. But
from the definition (\ref{Riemann tensor}) of the Riemann tensor, on page
\pageref{Riemann tensor}, we have $D^A D_B V_A = D_B D^A V_A + R_{BD} V^D$,
and from the Killing vector equation, we have $D^A V_A = 0$. And if
$\mathcal{M}$ is a negatively curved $n$-dimensional Einstein space with $n >
2$, then $R_{BD} = - \alpha g_{BD}$, where $\alpha = - \frac{1}{n} R > 0$ is
independent of position by the contracted Bianchi identity, $D_A R - 2 D^B
R_{A B} = - \left( n - 2 \right) \partial_A \alpha = 0$. Thus we find $D^A D_A
V_B = \alpha g_{BD} V^D$, hence $V^B D^A D_A V_B = \alpha V^B g_{BD} V^D$.
Thus if $\mathcal{M}$ is compact, we find on integrating by parts that:
\begin{equation}
  \label{contradiction for generator of continuous symmetry on M}
  \int_{\mathcal{M}} d^d x \left( D^A V^B \right)  \left( D_A V_B \right) = -
  \alpha \int_{\mathcal{M}} d^d xV^B g_{BD} V^D
\end{equation}
The left-hand side of this equation is $\geq 0$, but for nonzero $V^A$, the
right-hand side is $< 0$, so there can be no such nonzero $V^A$.

The classically massless $h_{\mu A}$ modes corresonding to harmonic 1-forms on
$\bar{H}^7$ could obtain masses from terms in the last term in (\ref{gauge
fixed hh action}) built from $h_{\mu A} h_{\mu B}$ and the vacuum $R_{A B C
D}$ and $H_{A B C D}$, as well as from further quantum corrections, like the
harmonic $p$-form modes of $C_{I J K}$.

If we ignore the last two terms in (\ref{gauge fixed hh action}), then after
making a Kaluza-Klein ansatz $s_{\mu \nu} = s_{\mu \nu} \left( \check{x}
\right) \omega \left( \bar{x} \right)$, the intrinsic masses $\bar{m}$ of the
$s_{\mu \nu}$ modes are determined by the spectrum on $\bar{H}^7$ in the
metric $\bar{g}_{A B}$ of sectional curvature $- 1$ of the negative of the
Laplace-Beltrami operator: $- \frac{1}{\sqrt{\bar{g}}} \partial_A \left(
\sqrt{\bar{g}} g^{A B} \partial_B \omega \right) = \bar{m}^2 \omega$. \ So
from pages 9 to 12 and 16 to 17 of {\cite{Almost Flat}}, it seems likely that
in this approximation, the lightest massive modes of $s_{\mu \nu}$ will have
intrinsic mass $\bar{m} = 3$ {\cite{KMST, Yau Isoperimetric, Agmon, Brooks
Makover 1, Brooks Makover 2, Brooks Makover 3, Brooks Makover 4, Cheeger,
Cheeger Wikipedia, Donnelly, Sarnak, Luo, Rudnick Sarnak, Orlando Park, Mazzeo
Phillips, Donnelly Xavier, Thurston, Colbois Courtois 1, Colbois Courtois 2}},
and from pages 42 to 43 of {\cite{Almost Flat}}, their masses as seen on the
HW boundary will be $m = 3 \frac{A}{B}$, where $A$ and $B$ are the constants
in the metric ansatz (\ref{metric ansatz for H7}). \ But as for the $h_{\mu
A}$ modes, the relatively large contribution of the last term in (\ref{gauge
fixed hh action}) to the dilaton/radion's mass squared suggests that the last
two terms in (\ref{gauge fixed hh action}) might give a larger contribution to
$\bar{m}^2$ for these modes than the value 9 obtained in this first
approximation.

If we make the limiting gauge choice $\tilde{b} \rightarrow - \tilde{a} +
\tilde{a}^2$, $\tilde{a} \rightarrow 0$ in (\ref{gauge fixed hh action}), we
obtain the Fierz-Pauli unitary gauge for the massive $s_{\mu \nu}$ modes and
the Proca unitary gauge for the massive $h_{\mu A}$ modes along the extended
dimensions \cite{Fierz Pauli, Proca 1, Proca 2}, and Landau-gauge-like
restrictions $D_A h_{\mu A} = 0$ and $D_A
t_{AC} + \left( \frac{1}{d - 2} + \frac{1}{n} \hspace{0.25em} \right) D_C h_{A
A} = 0$ on the dependence of the modes on position on $\bar{H}^7$, which in
the same way as for the 3-form gauge field means that some of the massive
modes obtained in the de Donder/Feynman gauge are unphysical, and would be
cancelled by corresponding Faddeev-Popov ghosts in the de Donder/Feynman
gauge. \ The gauge invariance for the massless $s_{\mu \nu}$ modes is unfixed
in the limiting unitary gauge, and we are free to add additional gauge-fixing
terms just for these modes, which are independent of position on $\bar{H}^7$.

\section{Modes that decay along the beam line
outside the interaction region at the LHC}
\label{Modes that decay along the beam line}

The classically massless harmonic 3-form modes were found in subsection
\ref{The mass of the harmonic 3 form modes} above to aquire approximately
equal masses $\simeq 0.2 \frac{A}{B}$, and in subsection \ref{The coupling
of the harmonic 3 form modes to the SM gauge bosons} their leading couplings
(\ref{simplified axion coupling}) to the SM gauge bosons were found to be
axion-like near their mass shell. \ From pages 17 to 19 of {\cite{Almost
Flat}}, their number is expected to be $\sim \frac{\bar{V}_7}{\mathrm{\ln}
\bar{V}_7}$, so they could be sufficiently numerous for their large number to
compensate for the $\sim \frac{1}{\sqrt{\bar{V}_7}}$ suppression of their
couplings enough for them to be seen at the LHC.

From pages 12 to 13 of {\cite{Almost Flat}}, it seems possible, on the basis
of the results of {\cite{Inoue 1}} and {\cite{Leininger et al}}, that the
lightest generic
classically massive modes of $C_{\mu \nu \sigma}$, $C_{\mu \nu A}$,
$C_{\mu A B}$, $s_{\mu \nu}$, $h_{A A}$, and $h_{\mu A}$, with classical
masses $3 \frac{A}{B}$, $2 \frac{A}{B}$, $\frac{A}{B}$, $3 \frac{A}{B}$, $3
\frac{A}{B}$, and $2 \frac{A}{B}$ respectively, could have large degeneracies
$\sim \bar{V}_7$, that restore agreement between the spectral staircase and
the Weyl asymptotic formula for the number of modes up to mass $m$,
immediately above the generic spectral gap. \ Any non-generic lighter modes
would be too few to see at the LHC. \ However the much larger mass $\simeq 9
\frac{A}{B}$ calculated in subsection \ref{KK modes of metric} above for
the dilaton/radion mode of $h_{A A}$, which is classically massless, suggests
that when the contributions of the $\Gamma^{\left( 8, \mathrm{bos}
\right)}_{\mathrm{SG}}$ term (\ref{Gamma 8 SG}) in (\ref{Gamma SG}) are
included, the lightest classically massive modes of all the above types, and
also the lightest modes of $t_{A B}$ and the lightest modes of $C_{A B C}$
other than the harmonic 3-form modes, might all have masses $\sim 8
\frac{A}{B}$ or more.

The harmonic 1-form, 2-form, and 3-form modes of $C_{\mu \nu A}$, $C_{\mu A
B}$, and $C_{A B C}$ respectively have vanishing field strength $H_{I J K L}$,
and thus cannot get masses directly from the CJS action (\ref{CJS action}) or
$\Gamma^{\left( 8, \mathrm{bos} \right)}_{\mathrm{SG}}$. \ The leading
contributions to the masses of these modes come from loop diagrams that
contain two CJS Chern-Simons vertices, whose contribution to the mass of the
harmonic 3-form modes of $C_{A B C}$ was approximately calculated in
subsection \ref{The mass of the harmonic 3 form modes} above. \ It is the fact
that $C_{A B C}$ has no classical spectral gap {\cite{Donnelly, Donnelly
Xavier}} that results in the number of harmonic 3-form modes being $\sim
\frac{\bar{V}_7}{\mathrm{\ln} \bar{V}_7}$, while the numbers of harmonic
2-form and 1-form modes are $\sim \bar{V}^{\alpha}_7$, with $\alpha < 1$
{\cite{Lueck, Clair Whyte}}.

Thus it seems possible that the model considered here and in {\cite{Almost
Flat}} has several types of approximately degenerate bosonic modes of the
supergravity multiplet, such that the numbers of approximately degenerate
modes of each type are large enough to compensate for the $\sim
\frac{1}{\sqrt{\bar{V}_7}}$ suppression of their couplings to the SM states,
so that the modes could be seen at the LHC. \ The lightest such modes are
likely to be the classically massless harmonic 3-form modes of $C_{A B C}$,
with mass $\simeq 0.2 \frac{A}{B}$, and axion-like couplings to the SM gauge
bosons.

\subsection{Approximate distribution of decay lengths}
\label{Approximate distribution of decay lengths}

Considering now the modes of just one of these types, approximately degenerate
with mass $m$, the modes will not be exactly degenerate, because the root mean
square field strength $h$ of the vacuum 4-form fluxes is likely to vary
slightly from region to region on $\bar{H}^7$, and moreover for the harmonic
3-form modes, the degeneracy depends on the approximation discussed in the
second paragraph after (\ref{C in terms of H}). \ The variation of $h$ from
region to region on $\bar{H}^7$ will be random, so by Anderson localization
{\cite{Anderson localization}}, the modes will be approximately localized on
different regions of $\bar{H}^7$. \ For the following rough estimates I shall
treat the modes as if they were scalars, both along the 4 extended dimensions
and along $\bar{H}^7$. \ The spherically symmetric eigenmodes of the
Laplace-Beltrami operator on uncompactified $H^7$ behave like $e^{- 3
\bar{y}}$ times an oscillating factor at large intrinsic geodesic distance
$\bar{y}$ from their centre of spherical symmetry {\cite{Camporesi Higuchi 1,
Camporesi Higuchi 2}}, so I shall assume that the approximately localized
modes behave roughly as $e^{- \left( 3 +
\frac{1}{\bar{L}_{\mathrm{{{loc}}}}} \right)
\bar{y}}$, where $\bar{y}$ is the intrinsic geodesic distance from their
centre of localization, and the intrinsic localization length
$\bar{L}_{\mathrm{{{loc}}}}$ is likely to be $\sim
\frac{\bar{L}_7}{2}$ for the harmonic 3-form modes, which from the discussion
after (\ref{V7 in terms of GN}) above is $\sim 14$ if $\bar{H}^7$ is
reasonably isotropic, and $M_{11}$ is near its current lower limit of $2.3 \pm
0.7$ TeV for 7 flat extra dimensions.

The coupling constant $c$ of one of these modes to the SM fields is roughly
the amplitude $e^{- \left( 3 +
\frac{1}{\bar{L}_{\mathrm{{{loc}}}}} \right) \bar{y}}$
of the mode at the HW boundary times a constant that is the same for all the
modes of this type, as for example in (\ref{simplified axion coupling}) above,
where $\bar{y}$ is now the intrinsic geodesic distance from the HW boundary to
the localization centre of the mode. \ The $s$ channel production rate of such
a mode is proportional to $c^2$, and its width is also proportional to $c^2$,
so its lifetime is proportional to $\frac{1}{c^2}$. \ If such a mode is
produced at the interaction point (IP) in ATLAS or CMS, it will have a
longitudinal momentum along the beam direction equal to the net longitudinal
momentum of the two partons that produced it. \ The distribution of
longitudinal momentum of the mode, and of the corresponding relativistic
enhancement factor for its lifetime in the laboratory frame, are independent
of $c$, so for a rough estimate I shall treat the mode as having a fixed decay
length $l$ along the beam direction, that is $\frac{1}{c^2}$ times a constant
that is the same for all the modes of this type.

For reasonably isotropic $\bar{H}^7$, the intrinsic volume of $\bar{H}^7$
between intrinsic geodesic distances $\bar{y}$ and $\bar{y} + d \bar{y}$ from
a fixed point of $\bar{H}^7$ is from page 9 of {\cite{Almost Flat}} roughly a
constant times $e^{6 \bar{y}} d \bar{y}$, for $\bar{y}$ up to its maximum
value $\simeq \frac{\bar{L}_7}{2}$. \ Thus the number of modes whose coupling
constant to the SM fields is between $c$ and $c + dc$ is roughly $c^{- \frac{6
\bar{L}_{\mathrm{{{loc}}}}}{3
\bar{L}_{\mathrm{{{loc}}}} + 1}} \frac{dc}{c}$, times
a constant that is the same for all the modes of this type. \ So the number of
modes whose decay length along the beam direction is between $l$ and $l + dl$
is roughly $l^{\frac{3 \bar{L}_{\mathrm{{{loc}}}}}{3
\bar{L}_{\mathrm{{{loc}}}} + 1}} \frac{dl}{l}$, times
a constant that is the same for all the modes of this type.

The number of particles of decay length $l$ decaying between distances $z$ and
$z + dz$ along the beam line from the IP per unit time, summed over both
directions along the beam, is $\frac{dz}{l} e^{- z / l}$ times the production
rate of particles of decay length $l$. \ The production rate of one of the
approximately degenerate modes of mass $m$ at the LHC is proportional to
$c^2$, thus is $\frac{1}{l}$ times a constant that is the same for all the
modes of this type. \ Thus the number of particles of this type decaying
between distances $z$ and $z + dz$ along the beam line from the IP per unit
time is:
\begin{equation}
  \label{rate of particles decaying between z and z plus dz} k_1 dz
  \int_{l_{\mathrm{\min}}}^{l_{\mathrm{\max}}} \frac{1}{l} e^{- z / l}
  \frac{1}{l} l^{\frac{3 \bar{L}_{\mathrm{{{loc}}}}}{3
  \bar{L}_{\mathrm{{{loc}}}} + 1}} \frac{dl}{l} \simeq
  k_2 z^{- \frac{3 \bar{L}_{\mathrm{{{loc}}}} + 2}{3
  \bar{L}_{\mathrm{{{loc}}}} + 1}} \left( 1 - e^{- z /
  l_{\mathrm{\min}}} \right) dz,
\end{equation}
where $k_1$ and $k_2$ are constants that are the same for all modes of this
type, $l_{\mathrm{\min}} \sim 10^{- 19}$ metres, $l_{\mathrm{\max}}$ is
determined by the maximum value $\simeq \frac{\bar{L}_7}{2}$ of $\bar{y}$, and
in the right hand side of (\ref{rate of particles decaying between z and z
plus dz}) I have set $l_{\mathrm{\max}}$ to $\infty$ since the integral of
(\ref{rate of particles decaying between z and z plus dz}) over $z$ is
convergent at large $z$, and used an approximation for the incomplete $\Gamma$
function.

The integral of the right-hand side of (\ref{rate of particles decaying
between z and z plus dz}) over $z$ from $0$ to $\infty$ is convergent at both
limits, and by integration by parts is equal to:
\begin{equation}
  \label{integral of the decay rate along the beam line} k_2 \left( 3
  \bar{L}_{\mathrm{{{loc}}}} + 1 \right) \Gamma \left(
  \frac{3 \bar{L}_{\mathrm{{{loc}}}}}{3
  \bar{L}_{\mathrm{{{loc}}}} + 1} \right) l^{-
  \frac{1}{3 \bar{L}_{\mathrm{{{loc}}}} +
  1}}_{\mathrm{\min}} \simeq k_2 \left( 3
  \bar{L}_{\mathrm{{{loc}}}} + 1 \right) l^{-
  \frac{1}{3 \bar{L}_{\mathrm{{{loc}}}} +
  1}}_{\mathrm{\min}},
\end{equation}
where the right hand side of (\ref{integral of the decay rate along the beam
line}) is approximately valid for
$\bar{L}_{\mathrm{{{loc}}}} > 1$.

Thus even though the vast majority of the approximately degenerate modes of
mass $\simeq m$ couple only with gravitational strength to the SM fields, so
that their lifetimes are $\sim \frac{1}{G_N m^3} \sim
\mathrm{{{hours}}}$ {\cite{Giudice Rattazzi Wells}},
the approximate localization of the modes on $\bar{H}^7$, and the inverse
proportionality of the production rate of the modes to their decay length,
mean that most of the modes actually produced at the LHC decay in or near the
detectors, so that the Breit-Wigner formula {\cite{Perkins, Coleman Physics
253a}} can be used to calculate the total cross-section for the production and
decay of these modes in the $s$ channel, as implicitly assumed on pages 13 to
16 of {\cite{Almost Flat}}. \ The possibility that modes seen at the LHC could
be a relatively small number of linear combinations, localized near the HW
boundary, of the large number of approximately degenerate modes, and that the
modes localized near the HW boundary would have correspondingly large
couplings to the SM fields, was noted on page 43 of {\cite{Almost Flat}}.

\subsection{Approximate total cross-section}
\label{Approximate total cross-section}

For comparison with the LHC data, let $N$ be the number of approximately
degenerate modes of mass $\simeq m$, and $\mu$ be the width of the
distribution over which the modes are spread. \ The $N$ approximately
degenerate modes of mass $\simeq m$ will be labelled by an index $n$. \ I will
assume that $\mu \geq \Gamma_n$ for all the modes $\left| n \right\rangle$,
where $\Gamma_n$ is the total width of the mode $\left| n \right\rangle$.

A monoenergetic high energy beam of protons of energy $E$ is equivalent to a
beam of partons, such that the number of $u$ quarks per unit area per unit
time with energy between $xE$ and $\left( x + dx \right) E$ is $f_u \left( x
\right) dx$ times the number of protons per unit area per unit time, and
similarly for the other types of parton, where $f_p \left( x \right)$, $p = u,
d, g, \bar{u}, \bar{d}, \ldots$ are the parton distribution functions (PDFs).
\ The PDFs evolve logarithmically with $Q^2$, the square of the momentum
transferred in a scattering process, and for a rough estimate I shall use the
plot {\cite{MSTW}} with $Q^2 = 10^4$ GeV\ensuremath{^{\textrm{$2$}}}.

The plot {\cite{MSTW}} shows that to a good approximation, $f_g \left( x
\right) \geq 10 f_{\bar{u}} \left( x \right)$ for all $x$, $f_g \left( x
\right) \geq f_u \left( x \right)$ for all $x \leq 0.1$, $f_g \left( x \right)
\geq 0.1 f_u \left( x \right)$ for $x$ up to at least 0.4, $f_g \left( x
\right) \geq 5 f_{\bar{d}} \left( x \right)$ for all $x$, $f_g \left( x
\right) \geq f_d \left( x \right)$ for $x \leq 0.25$, and $f_g \left( x
\right) \geq 0.2 f_d \left( x \right)$ for $x$ up to at least 0.35. \ The
modes of mass $\simeq m$ are uncharged, and couple with approximately equal
strength to all the partons, so for a first approximation, valid for $m$ up to
at least about 2.5 TeV for the LHC with 7 TeV centre of mass energy, I shall
consider the $g g$ initial state only.

If $N$ was 1 and all the modes of mass $\simeq m$ produced in the $s$ channel
decayed within the detector, then the total cross-section for the process $g g
\rightarrow n \rightarrow f$, where $\left| n \right\rangle$ is the mode of
mass $\simeq m$, which for the rough estimates here I am treating as if it was
a scalar, and $f$ is an SM final state such as $u \bar{u}$, $g g$, $W^+ W^-$,
... , would be given in the $\left| n \right\rangle$ rest frame by the
Breit-Wigner formula {\cite{Perkins, Coleman Physics 253a}}:
\begin{equation}
  \label{Breit Wigner cross section} \sigma \left( g g \rightarrow n
  \rightarrow f \right) = \frac{1}{8} \cdot \frac{1}{2^2} \cdot \frac{1}{2}
  \cdot 4 \pi \cdot \frac{\Gamma_{n \rightarrow g g} \Gamma_{n \rightarrow
  f}}{E^2} \cdot \frac{1}{\left( \left( E - m_n \right)^2 + \Gamma_n^2 / 4
  \right)},
\end{equation}
where $E$ is the invariant mass of the $g g$ system, and $\Gamma_{n
\rightarrow g g}$ and $\Gamma_{n \rightarrow f}$ are the partial widths for
$\left| n \right\rangle$ decay to $g g$ and $f$. \ The factor $\frac{1}{8}$ is
the probability that the two initial gluons can form a colour singlet, the
factor $\frac{1}{2^2}$ is for the average over the helicity states of the
initial gluons, and the factor $\frac{1}{2}$ is the probability that the
helicities of the two initial gluons sum to 0.

For each type of SM final state $f$, $\Gamma_{n \rightarrow f} = \int \left|
\langle f \left| n \right\rangle \left|^2 d \rho_f \right. \right.$, where
$\int \ldots d \rho_f$ represents a phase space integration that is
independent of $n$ {\cite{Perkins, Coleman Physics 253a}}. \ The mode $\left|
n \right\rangle$ is not present in the final state, so for $N > 1$ we have to
sum the amplitude over all the modes $\left| n \right\rangle$. \ The amplitude
factor that leads to the final two factors in $\sigma \left( g g \rightarrow n
\rightarrow f \right)$, (\ref{Breit Wigner cross section}), after the phase
space integrations are done, is:
\begin{equation}
  \label{amplitude factor for last 2 BW factors} \frac{1}{E} \langle f|n
  \rangle \langle n|g g \rangle \frac{1}{E - m_n + i \Gamma_n / 2} .
\end{equation}
In the example (\ref{simplified axion coupling}), the only dependence on the
mode $\left| n \right\rangle$ of the matrix element $\langle f \left| n
\right\rangle$, for SM final states $f$ consisting of SM gauge bosons, whose
wave functions are independent of position on the closed hyperbolic factor
$\bar{H}^6$ of the HW boundary, is effectively via a single coupling constant
that measures the integral of $\left| n \right\rangle$ over the HW boundary,
weighted by the vacuum 4-form fluxes at the boundary. \ If the final state $f$
includes quarks or leptons, whose wave functions depend on position on
$\bar{H}^6$, different $\left| n \right\rangle$, whose wave functions are
larger or smaller in different regions of $\bar{H}^6$, could couple with
different effective coupling constants to different quarks and leptons. \ But
from the discussion after (\ref{V7 in terms of GN}), on page \pageref{V7 in
terms of GN}, the intrinsic diameter $\bar{L}_6$ of $\bar{H}^6$ lies between
about 5.7 and 6.0 if $\bar{H}^6$ is reasonably isotropic, and the current
upper bound on the intrinsic diameter $\bar{L}_7$ of $\bar{H}^7$ is about 28
if $\bar{H}^7$ is reasonably isotropic, so since the curvature radius $b_1$ of
$\bar{H}^6$ is $\simeq B$, the ratios $\bar{L}_7 / \bar{L}_6$ and $L_7 / L_6$
lie between about 4.9 and 4.7, if both $\bar{H}^6$ and $\bar{H}^7$ are
reasonably isotropic, and $\bar{L}_7$ is near its current upper bound.

I shall therefore assume that for a first approximation, the only dependence
of the matrix element $\langle f \left| n \right\rangle$ on the mode $\left| n
\right\rangle$ is via a single real-valued effective coupling constant $c \geq
0$ that measures how large the wave function of $\left| n \right\rangle$ is at
the HW boundary. \ We now choose a particular element of the eigenmode basis
that is localized close to $\bar{H}^6$, say $\left| 1 \right\rangle$. \ For
each mode $\left| n \right\rangle$, we define the coupling constant $c_n$ of
$\left| n \right\rangle$ to the SM states to be such that for one particular
SM final state $\langle f_0 |$, \ $\langle f_0 \left| n \right\rangle =
\langle f_0 \left| 1 \right\rangle c_n$. \ Then for all SM final states
$\langle f|$:
\begin{equation}
  \label{factorization of f n matrix element} \langle f \left| n \right\rangle
  \simeq \langle f \left| 1 \right\rangle c_n .
\end{equation}
Let $\hat{\rho}_2 \left( c, m' \right)$ be such that the number of elements
$\left| n \right\rangle$ of the eigenmode basis such that $c_n$ lies between
$c$ and $c + dc$, and the exact mass $m_n$ of $\left| n \right\rangle$ lies
between $m'$ and $m' + dm'$, is $\hat{\rho}_2 \left( c, m' \right) dcdm'$. \ I
shall assume that due to the random and uncorrelated nature of the slight
variations from region to region on $\bar{H}^7$ of the root mean square field
strength $h$ of the vacuum 4-form fluxes, $\hat{\rho}_2 \left( c, m' \right)$
approximately factorizes as:
\begin{equation}
  \label{factorization of rho hat sub 2} \hat{\rho}_2 \left( c, m' \right)
  \simeq \hat{\rho}_c \left( c \right) \hat{\rho}_m \left( m' \right),
\end{equation}
where
\begin{equation}
  \label{normalizations of rho hat c and rho hat m} \int_0^{\infty} dc
  \hat{\rho}_c \left( c \right) = 1, \hspace{2cm} \int_{m - \mu / 2}^{m + \mu
  / 2} \hat{\rho}_m \left( m' \right) dm' = N.
\end{equation}
Then:
\[ \sum_n \langle f \left| n \right\rangle \langle n \left| g g \right\rangle
   \simeq \langle f \left| 1 \right\rangle \langle 1 \left| g g \right\rangle
   \int_0^{\infty} dc \int_{m - \mu / 2}^{m + \mu / 2} dm' c^2 \hat{\rho}_2
   \left( c, m' \right) \simeq \]
\begin{equation}
  \label{mode sum in terms of rho hat sub c} \simeq N \langle f \left| 1
  \right\rangle \langle 1 \left| g g \right\rangle \int_0^{\infty} dcc^2
  \hat{\rho}_c \left( c \right) .
\end{equation}
The number $N$ of approximately degenerate modes $\left| n \right\rangle$ of
mass $\simeq m$ would be at most $\sim \bar{V}_7$, which for $m \sim \kappa^{-
2 / 9}_{11} \sim \mathrm{{{TeV}}}$ is $\sim 10^{32}$.
\ I shall assume that $\hat{\rho}_m \left( m' \right)$ is smooth, and is 0 for
$|m' - m| \geq \mu / 2$.

The only factor in the amplitude factor (\ref{amplitude factor for last 2 BW
factors}) that varies significantly with $m_n$ over the mass range $m - \mu /
2 \leq m_n \leq m + \mu / 2$ is the final factor:
\begin{equation}
  \label{m sub n dependent factor} \frac{1}{E - m_n + i \Gamma_n / 2},
\end{equation}
where by assumption $\Gamma_n \leq \mu$. \ The sum of the amplitude factor
(\ref{amplitude factor for last 2 BW factors}) over the modes $\left| n
\right\rangle$ can be replaced by integrals over $c = c_n$ and $m' = m_n$ as
in (\ref{mode sum in terms of rho hat sub c}). \ To a first approximation,
when multiplied by the smooth density of states $\hat{\rho}_2 \left( c, m'
\right) \simeq \hat{\rho}_c \left( c \right) \hat{\rho}_m \left( m' \right)$
and integrated over $m'$, the $m'$-dependent factor (\ref{m sub n dependent
factor}) is effectively $\simeq - i \pi \delta \left( E - m' \right)$, because
the $i \Gamma_n / 2$ means that the integration path has to go around the
singularity in the lower half of the complex $m'$ plane. \ Choosing the
integration path to be along the real axis except for a small semicircle
centred at $m' = E$, the contributions from the real axis cancel to a good
approximation for $\Gamma_n \ll \mu$, and roughly cancel for all $\Gamma_n
\leq \mu$, and the semicircle gives $- i \pi$ times the density of states
$\hat{\rho}_c \left( c \right) \hat{\rho}_m \left( m' \right)$ evaluated at
$m' = E$. \ Thus the sum of the amplitude factor (\ref{amplitude factor for
last 2 BW factors}) over the modes $\left| n \right\rangle$ is approximately:
\[ \sum_n \frac{1}{E} \langle f \left| n \right\rangle \langle n \left| g g
   \right\rangle \frac{1}{E - m_n + i \Gamma_n / 2} \simeq \]
\[ \simeq - i \pi \frac{1}{E} \langle f \left| 1 \right\rangle \langle 1
   \left| g g \right\rangle \int_0^{\infty} dcc^2 \hat{\rho}_c \left( c
   \right) \int_{m - \mu / 2}^{m + \mu / 2} dm' \hat{\rho}_m \left( m' \right)
   \delta \left( E - m' \right) \simeq \]
\[ \simeq - i \pi \frac{1}{E} \hat{\rho}_m \left( E \right) \langle f \left| 1
   \right\rangle \langle 1 \left| g g \right\rangle \int_0^{\infty} dcc^2
   \hat{\rho}_c \left( c \right) \simeq \]
\begin{equation}
  \label{BW weighted mode sum in terms of rho hat c} \simeq - i \pi
  \frac{1}{E} \hat{\rho}_m \left( E \right) \frac{1}{N} \sum_n \langle f
  \left| n \right\rangle \langle n \left| g g \right\rangle .
\end{equation}
where (\ref{mode sum in terms of rho hat sub c}) was used at the last step.

For $N$ modes, the final two factors in  (\ref{Breit Wigner cross section}) 
are replaced by the phase space integrals $\int \int \ldots d \rho_f d \rho_{g
g}$ of the squared magnitude of the mode sum (\ref{BW weighted mode sum in
terms of rho hat c}) of the amplitude factor (\ref{amplitude factor for last 2
BW factors}), which are approximately:
\[ \pi^2 \frac{1}{E^2} \hat{\rho}_m \left( E \right)^2 \frac{1}{N^2} \int \int
   \sum_n \langle f \left| n \right\rangle \langle n \left| g g \right\rangle
   \sum_{n'} \langle g g \left| n' \right\rangle \langle n' \left| f
   \right\rangle d \rho_f d \rho_{g g} \simeq \]
\[ \simeq \pi^2 \frac{1}{E^2} \hat{\rho}_m \left( E \right)^2 \frac{1}{N^2}
   \int \int \sum_n \sum_{n'} c^2_n c^2_{n'} \langle f \left| 1 \right\rangle
   \langle 1 \left| g g \right\rangle \langle g g \left| 1 \right\rangle
   \langle 1 \left| f \right\rangle d \rho_f d \rho_{g g} \simeq \]
\begin{equation}
  \label{phase space integrated squared mode sum} \simeq \pi^2 \frac{1}{E^2}
  \hat{\rho}_m \left( E \right)^2 \frac{1}{N^2} \int \sum_n \langle f \left| n
  \right\rangle \langle n \left| f \right\rangle d \rho_f \int \sum_{n'}
  \langle g g \left| n' \right\rangle \langle n' \left| g g \right\rangle d
  \rho_{g g},
\end{equation}
where the approximate factorization (\ref{factorization of f n matrix
element}) has been used at each step.

In addition to the eigenmode basis of the modes $\left| n \right\rangle$ of
mass $\simeq m$, we can consider a basis where all the modes are approximately
uniformly spread out over $\bar{H}^7$, and thus by (\ref{V7 in terms of GN})
couple with approximately equal, gravitational, strength to the SM fields. \
Let $\left\{ \left| \bar{n} \right\rangle \right\}$ be a basis of this type. \
It is related to the eigenmode basis by an $N \times N$ unitary
transformation.

The partial widths $\Gamma_{\bar{n} \rightarrow f} \equiv \int \left| \langle
f \left| \bar{n} \right\rangle \left|^2 d \rho_f \right. \right.$ of the modes
in the $\left\{ \left| \bar{n} \right\rangle \right\}$ basis are estimated in
order of magnitude by {\cite{Giudice Rattazzi Wells}}:
\begin{equation}
  \label{partial widths} \Gamma_{\bar{n} \rightarrow g g} \sim \Gamma_{\bar{n}
  \rightarrow g g g} \sim \ldots \sim \Gamma_{\bar{n} \rightarrow u \bar{u}}
  \sim \ldots \sim \Gamma_{\bar{n} \rightarrow \nu_{\tau}  \bar{\nu}_{\tau}}
  \sim m^3 G_N \sim 10^{- 32} {\mathrm{{TeV}}},
\end{equation}
where the final $\sim$ applies for $m \sim
\mathrm{{{TeV}}}$. \ Thus for all SM final states $f$:
\begin{equation}
  \label{mode sum in terms of GN} \int \sum_n \langle f \left| n \right\rangle
  \langle n \left| f \right\rangle d \rho_f = \int \sum_{\bar{n}} \langle f
  \left| \bar{n} \right\rangle \langle \bar{n} \left| f \right\rangle d \rho_f
  \sim Nm^3 G_N .
\end{equation}
Thus from (\ref{phase space integrated squared mode sum}), the last two
factors in (\ref{Breit Wigner cross section}) are for the $N$ modes replaced
by roughly:
\begin{equation}
  \label{last 2 factors for N modes in terms of GN} \sim \pi^2 \frac{1}{E^2}
  \hat{\rho}_m \left( E \right)^2 m^6 G^2_N .
\end{equation}
Thus the total cross-section is roughly:
\begin{equation}
  \label{cross section summed over phi modes} \sigma \left( g g \rightarrow
  \mathrm{{{any}}} \hspace{0.25em} n \rightarrow f
  \right) = \frac{1}{8} \cdot \frac{1}{2^2} \cdot \frac{1}{2} \cdot 4 \pi
  \cdot \pi^2 \frac{1}{E^2} \hat{\rho}_m \left( E \right)^2 m^6 G^2_N \sim m^4
  G^2_N \hat{\rho}_m \left( E \right)^2 .
\end{equation}
To produce a mode $\left| n \right\rangle$ with mass $\simeq m$ at rest with
$P = 3.5$ TeV per proton, each gluon needs $x \simeq \frac{m}{2 P}$. \ If
the 4-momenta of the protons are $\left( P, 0, 0, P \right)$ and $\left( P, 0,
0, - P \right)$ and the momentum fractions of the gluons are $x_1$ and $x_2$,
then their Mandelstam $s$ is $P^2 \left( \left( x_1 + x_2 \right)^2 - \left(
x_1 - x_2 \right)^2 \right) = 4 P^2 x_1 x_2$. \ From the plot {\cite{MSTW}} we
find that $f_g \left( x \right) \simeq 0.060 x^{- 2.17}$ for $0.05 \leq x \leq
0.2$, but substantially smaller than this for $x \geq 0.3$. \ For a first
approximation I shall use $f_g \left( x \right) \simeq 0.060 x^{- 2.17}$ for
$0 \leq x \leq 0.3$ and $f_g \left( x \right) \simeq 0$ for $0.3 < x \leq 1$.
\ The initial partons are massless, so the final form of the estimate
(\ref{cross section summed over phi modes}) of the total cross-section in the
centre of mass frame of the two gluons is also the approximate total
cross-section in the
laboratory frame {\cite{Coleman Physics 253a}}. Thus the total cross-section
for $\mathrm{{{proton}}} +
\mathrm{{{proton}}} \rightarrow
\mathrm{{{any}}} \hspace{0.25em} n + X \rightarrow f + X$ is
roughly:
\[ \int^{0.3}_0 dx_1  \int^{0.3}_0 dx_2  0.060^2 \left( x_1 x_2
   \right)^{- 2.17} m^4 G^2_N \hat{\rho}_m \left( 2 P \sqrt{x_1 x_2} \right)^2
\]
\[ \simeq 0.060^2 m^4 G^2_N \frac{N^2}{\mu^2} \int^{0.3}_{\frac{m^2}{1.2 P^2}}
   \frac{dx_1}{x_1}  \int^{\frac{2 m + \mu}{4 P}}_{\frac{2 m - \mu}{4 P}}
   \frac{2 xdx}{x^{4.34}} \]
\begin{equation}
  \label{cross section for bump} \simeq 0.1 m^3 G^2_N \frac{N^2}{\mu}  \left(
  \frac{P}{m} \right)^{2.34} \mathrm{\ln} \frac{0.6 P}{m},
\end{equation}
where I approximated $\hat{\rho}_m \left( E \right)$ as $\frac{N}{\mu}$ from
$m - \frac{\mu}{2}$ to $m + \frac{\mu}{2}$ and 0 outside this interval,
defined $x \equiv \sqrt{x_1 x_2}$, and assumed $\mu \ll m$ and $\frac{m}{2 P}
\leq 0.3$.

As a reference estimate of the number $N$ of approximately degenerate modes of
mass $\simeq m$, let $N_{\mathrm{Weyl},\bar{m}}$ be the
Weyl asymptotic formula for the number of eigenmodes of the negative of the
Laplace-Beltrami operator $\Delta = \frac{1}{\sqrt{\bar{g}}} \partial_A 
\left( \sqrt{\bar{g}}  \bar{g}^{AB} \partial_B \cdot \right)$ on $\bar{H}^7$,
in the metric $\bar{g}_{A B}$ of sectional curvature $- 1$, with eigenvalue up
to $\bar{m}^2$, where $\bar{m} = m \frac{B}{A}$ is the intrinsic mass
corresponding to $m$:
\begin{equation}
  \label{N Weyl} N_{\mathrm{Weyl},\bar{m}} =
  \frac{\bar{V}_7}{\left( 2 \pi \right)^7} S_6  \frac{\bar{m}^7}{7} =
  \frac{\bar{V}_7}{840 \pi^4}  \bar{m} ^7,
\end{equation}
where $S_6 = \frac{16}{15} \pi^3$ is the area of the unit 6-sphere. \ Thus
from (\ref{V7 in terms of GN}):
\begin{equation}
  \label{NW GN} N_{\mathrm{Weyl},\bar{m}} G_N =
  \frac{\kappa^2_{11} \bar{m} ^7}{6720 \pi^5 A^2 B^7} \simeq 0.046
  \frac{\bar{m}^9}{m^2},
\end{equation}
where the best value $B \simeq 0.28 \kappa^{2 / 9}_{11}$, from subsection
\ref{subsection PMS}, has been used.

In numerical studies of the spectrum of $- \Delta$ on compact hyperbolic
3-manifolds of small intrinsic volume, Inoue found that the Weyl asymptotic
formula is approximately valid down to the lowest non-zero eigenvalue
$\lambda_1$, so that if $\lambda_1$ occurs at a larger value of $\bar{m}$ than
would be expected from the Weyl formula, there is a degeneracy or approximate
degeneracy of eigenvalues near $\lambda_1$, that restores agreement with the
Weyl formula for $\bar{m}^2$ above $\lambda_1$ {\cite{Inoue 1}}. \ It is also
known that for $n \geq 2$, every $\bar{H}^n$ has pairs of finite covers of
arbitrarily large volume ratio, whose sets of eigenvalues of $- \Delta$,
ignoring multiplicities, are identical {\cite{Leininger et al}}, so since the
Weyl asymptotic formula is certainly valid for sufficiently large $\bar{m}$,
every $\bar{H}^n$ has finite covers whose eigenvalues have arbitrarily large
multiplicities, for sufficiently large $\bar{m}$.

It seems likely that $N_{\mathrm{Weyl},\bar{m}}$, (\ref{N
Weyl}), will give an under-estimate of $N$ for the classically massless
harmonic 3-form modes, whose intrinsic mass was calculated approximately as
$\bar{m} \simeq 0.2$ in subsection \ref{The mass of the harmonic 3 form
modes}, and an over-estimate for all the other types of mode. \ For the
harmonic 3-form modes, $N$ is the 3rd Betti number $B_3$ of $\bar{H}^7$, which
from pages 17 to 19 of {\cite{Almost Flat}}, is expected to be $\sim
\frac{\bar{V}_7}{\mathrm{\ln} \bar{V}_7}$. \ For a reference estimate of the
coefficient, the middle Betti number of an $\bar{H}^n$ with even $n$ and large
intrinsic volume $\bar{V}_n$ is from pages 18 to 19 of {\cite{Almost Flat}}
given roughly by $B_{n / 2} \simeq \frac{\left( n - 1 \right) !!}{\left( 2 \pi
\right)^{n / 2}} \bar{V}_n$, and in particular, for $n = 6$, $B_3 \simeq 0.060
\bar{V}_6$, and for $n = 8$, $B_4 \simeq 0.067 \bar{V}_8$ {\cite{Lueck, Clair
Whyte}}. \ I shall use $N_{3\textrm{-}\mathrm{form,ref}} \equiv 0.06
\frac{\bar{V}_7}{\mathrm{\ln} \bar{V}_7}$ as a
reference estimate of $B_3$ of $\bar{H}^7$, which for $\bar{V}^7 \sim 10^{34}$,
gives $N_{3\textrm{-}\mathrm{form,ref}} \simeq 8 \times 10^{- 4} \bar{V}_7$,
while from (\ref{N Weyl}),
$N_{\mathrm{Weyl},0.2}
\simeq 2 \times 10^{- 10} \bar{V}_7$.

The numbers of harmonic 2-form and 1-form modes are $\sim \bar{V}^{\alpha}_7$,
with $\alpha < 1$ {\cite{Lueck, Clair Whyte}}, so these modes are not expected
to be observable at the LHC.

For the remaining modes, the
much larger intrinsic mass $\bar{m}_{\mathrm{{{dil}}}}
\simeq 9$ calculated in subsection \ref{KK modes of metric} for the
dilaton/radion mode of $h_{A A}$, which is classically massless, suggests that
the $\Gamma^{\left( 8, \mathrm{bos} \right)}_{\mathrm{SG}}$ term (\ref{Gamma 8
SG}) in (\ref{Gamma SG}) might give an additive contribution $\sim 9^2$ to
the squares of their intrinsic masses, so that a better estimate of $N$ might
be obtained by replacing $\bar{m}$ in (\ref{N Weyl}) by $\sqrt{\bar{m}^2 -
\bar{m}^2_0}$, for some $\bar{m}_0 < \bar{m}$.

Substituting $N_{\mathrm{Weyl},\bar{m}}$ for $N$ in
(\ref{cross section for bump}) and using (\ref{NW GN}), we obtain:
\begin{equation}
  \label{cross section in terms of m bar} \sigma \left(
  \mathrm{{{prot}}} +
  \mathrm{{{prot}}} \rightarrow
  \mathrm{{{any}}} \: n + X \rightarrow f + X \right)
  \sim 10^{- 4}  \frac{\bar{m}^{18}}{\mu m} \left( \frac{P}{m} \right)^{2.34}
  \mathrm{ln} \frac{0.6 P}{m},
\end{equation}
as a rough reference estimate of the total cross-section for the modes of each
type except
the harmonic 3-forms, for $m < 0.6 P$, where
$P$ is the energy per proton, currently 3.5~TeV at the LHC.
For the harmonic 3-forms, the reference estimate of $N$ is
$N_{3\textrm{-}\mathrm{form,ref}}$ not $N_{\mathrm{Weyl},0.2}$, so the
coefficient $10^{-4}$ in the right-hand side of (\ref{cross section in terms
of m bar}) is replaced by $10^{9}$.

\subsection{LHC results and prospects}
\label{LHC results and prospects}

From (\ref{rate of particles decaying between z and z plus dz}) and
(\ref{integral of the decay rate along the beam line}), the fraction of the
events where one of the $N$ approximately degenerate modes $\left| n
\right\rangle$ of mass $\simeq m$ is produced in the $s$ channel, such that
the $\left| n \right\rangle$ decays further than
$z_{\mathrm{{{cut}}}} \gg l_{\mathrm{\min}}$ along the
beam line from the IP, is $\simeq \left(
\frac{z_{\mathrm{{{cut}}}}}{l_{\mathrm{\min}}}
\right)^{- \frac{1}{3 \bar{L}_{\mathrm{{{loc}}}} +
1}}$. \ Relevant searches at ATLAS and CMS have so far always accepted events
that pass all cuts not related to the position $z$ of the reconstructed
primary vertex along the beam line relative to the IP, and for which $|z| \leq
10$ centimetres {\cite{AC1088, ATLAS-CONF-2011-083, CMS PAS EXO-11-039,
11081582 ATLAS, CMS PAS EXO-11-004}}. \ Thus if the intrinsic localization
length $\bar{L}_{\mathrm{{{loc}}}}$ was 28, which from the
start of subsection \ref{Approximate distribution of decay lengths}, is twice
the largest expected value $\simeq \frac{\bar{L}_7}{2} \simeq 14$ for $M_{11}$
at its current lower limit for 7 flat extra dimensions, the fraction of
$\left| n \right\rangle$ production events that miss the $z$ cut would for
$l_{\mathrm{min}} \sim 10^{-19}$ metres be at
most about 0.6, and for smaller values of
$\bar{L}_{\mathrm{{{loc}}}}$, this fraction would be
smaller. \ Thus for comparison with the searches carried out so far at ATLAS
and CMS, the order-of-magnitude estimate (\ref{cross section for bump}) does
not require any correction for the $z$ cut.

An early candidate for such modes was a 2.8 sigma bump at 1.8 TeV seen in the
first 295 per nb of proton-proton collisions at 7 TeV centre-of-mass (c.o.m.)
energy in ATLAS-CONF-2010-088 {\cite{AC1088}}, which if real would have
corresponded to a 27 pb cross-section for the modes to be produced in the
$s$-channel and decay within the $15 + 15$ centimetres along the beam line
centred at
the nominal interaction point (IP) allowed by the $z$ cut on the primary
vertex.  However if the bump had been real there would have been a bump in
the dijet final state with a similar total cross-section, and from Table II
of \cite{11086311 ATLAS}, which used 1.0 per fb of proton-proton collisions at
7 TeV c.o.m. energy, the 95\% CL upper limit on the total cross-section of
such a bump at 1.8 TeV in the dijet final state is now about 0.1 pb.

In a recent search for narrow high-mass resonances decaying into $e^+ e^-$ or
$\mu^+ \mu^-$ final states in about 1.1 per fb of proton-proton collisions at
7 TeV c.o.m. energy, with each lepton having transverse momentum $p_T > 25$
GeV, ATLAS found no significant excess above the SM background in the search
region from about 110 GeV to 2 TeV {\cite{11081582 ATLAS}}. \ The signal
acceptances were around 65\% for electrons and 40\% for muons, and from Figure
1 of this article, the SM background in the $e^+ e^-$ final state from about
120 GeV to 2 TeV is roughly:
\begin{equation}
  \label{ATLAS ee background} \left. \frac{d \sigma}{d \left( \frac{m_{e^+
  e^-}}{\mathrm{{{TeV}}}} \right)}
  \right|_{\mathrm{{{ATLAS}}}} \simeq 6.0 \left( \frac{
  m_{e^+ e^-}}{\mathrm{{TeV}}} \right)^{- 5.14} \hspace{0.2ex} \mathrm{fb},
\end{equation}
and the SM background in the $\mu^+ \mu^-$ final state is roughly the same.

In a recent search for evidence of ADD large flat extra dimensions
{\cite{ADD1, AADD, ADD2}} in the $\mu^+ \mu^-$ final state in about 1.2 per fb
of proton-proton collisions at 7 TeV c.o.m. energy, with each muon having
transverse momentum $p_T > 35$ GeV, CMS found no significant excess above the
SM background in the search region from about 120 GeV to 3 TeV {\cite{CMS PAS
EXO-11-039}}. \ The simulated reconstruction efficiency for high mass
Drell-Yan dimuon events in the selected acceptance range was above 90\%, and
from Figure 1 of this article, the SM background from about 120 GeV to 2 TeV
is roughly:
\begin{equation}
  \label{CMS mu mu background} \left. \frac{d \sigma}{d \left( \frac{m_{\mu^+
  \mu^-}}{\mathrm{{{TeV}}}} \right)}
  \right|_{\mathrm{{{CMS}}}} \simeq 4.4 \left(
  \frac{m_{\mu^+ \mu^-}}{\mathrm{{{TeV}}}} \right)^{-
  5.60} \hspace{0.2ex} \mathrm{{fb}}.
\end{equation}

In a recent search for evidence of ADD or Randall-Sundrum extra dimensions
{\cite{RS1}} in the diphoton final state in 2.2 per fb of proton-proton
collisions at 7 TeV c.o.m. energy, with each photon having transverse energy
$E_T > 70$ GeV, CMS found no significant excess above the SM background in the
search region from about 150 GeV to 2 TeV {\cite{11120688 CMS}}. \ The
corresponding diphoton reconstruction and identification efficiency was about
76\%, and from Figure 1 of this article, the SM background from about 150 GeV
to 2 TeV is roughly:
\begin{equation}
  \label{CMS diphoton background} \left. \frac{d \sigma}{d \left(
  \frac{m_{\gamma \gamma}}{\mathrm{{{TeV}}}} \right)}
  \right|_{\mathrm{{{CMS}}}} \simeq 2.1 \left(
  \frac{m_{\gamma \gamma}}{\mathrm{{{TeV}}}}
  \right)^{- 5.31} \hspace{0.2ex} \mathrm{{fb}}.
\end{equation}

In a recent search for evidence of ADD or Randall-Sundrum extra dimensions in
the diphoton final state in 2.12 per fb of proton-proton collisions at 7 TeV
c.o.m. energy, with each photon having transverse energy $E_T > 25$ GeV, ATLAS
found no significant excess above the SM background in the search region from
about 150 GeV to 2 TeV {\cite{11122194 ATLAS}}. \ The selection efficiency for
events within the detector acceptance was about 70\%, and from Figure 1 of
this article, the SM background from about 150 GeV to 2 TeV is roughly:
\begin{equation}
  \label{ATLAS diphoton background} \left. \frac{d \sigma}{d \left(
  \frac{m_{\gamma \gamma}}{\mathrm{{{TeV}}}} \right)}
  \right|_{\mathrm{{{ATLAS}}}} = 4.3 \left(
  \frac{m_{\gamma \gamma}}{\mathrm{{{TeV}}}}
  \right)^{- 5.25} \; \mathrm{{{fb}}} .
\end{equation}

Let $t$ denote one of the types of mode for which there might be a
sufficiently large number of approximately degenerate modes of intrinsic mass
$\simeq \bar{m}_t$ for them to produce a bump in the above cross-sections if
they were produced in the $s$-channel at the LHC. \ Thus $t$ denotes either
the harmonic 3-forms $C_{A B C}$ of intrinsic mass $\bar{m}_{3 \mathrm{f}}
\simeq 0.2$, or the lightest generic classically massive modes of one of the
types $C_{\mu \nu \sigma}$, $C_{\mu \nu A}$, $C_{\mu AB}$, $s_{\mu \nu}$,
$h_{AA}$, and $h_{\mu A}$, with classical intrinsic masses 3, 2, 1, 3, 3, and
2 respectively, or the lightest generic modes of $t_{A B}$, which from the
discussion after (\ref{equation for tAB modes}) would be tachyonic unless the
last two terms in (\ref{gauge fixed hh action}) lift their squared masses
sufficiently. \ The much larger intrinsic mass $\simeq 9$ calculated after
(\ref{dilaton kinetic terms}) for the dilaton/radion mode of $h_{AA}$, which
is classically massless, suggests that when the contributions of the
$\Gamma^{\left( 8, \mathrm{bos} \right)}_{\mathrm{SG}}$ term (\ref{Gamma 8
SG}) in (\ref{Gamma SG}) are included, the intrinsic masses of all modes other
than the harmonic 3-forms, harmonic 2-forms, and harmonic 1-forms might be
$\sim 8$ or more.

The harmonic 2-forms and harmonic 1-forms, and also any other non-generic
modes, sometimes called supercurvature modes {\cite{Inoue 1}}, such as the light
modes in the far-from-isotropic closed hyperbolic 7-manifolds considered in
the paragraph after (\ref{upper bound on smallest nonzero eigenvalue}), whose
classical squared intrinsic masses are less than the minimum value of the
classical squared intrinsic mass of the corresponding type of mode on
uncompactified $H^7$, are expected to be too few in number to be seen at the
LHC.

If the actual number of approximately degenerate modes of type $t$ and
intrinsic mass $\simeq \bar{m}_t$ is $N_t = x_t
N_{\mathrm{{{Weyl}}}, \bar{m}_t}$, where $x_t$ is
expected from the discussion before (\ref{cross section in terms of m bar}) to
be $< 1$ except for the harmonic 3-form modes, the estimated total
cross-section for $\mathrm{{{proton}}} +
\mathrm{{{proton}}} \rightarrow {\textrm{{any}
{mode} {of} {type} }} t + X \rightarrow f + X$ is by (\ref{cross
section for bump}) obtained from the reference estimate (\ref{cross section in
terms of m bar}) by multiplying by $x^2_t$. \ Thus for $f = e^+ + e^-$, the
requirement that the total cross-section for this process, spread over a peak
of width $\mu_t$ centred at $m_{e^+ e^-} = m_t$, should be less than
(\ref{ATLAS ee background}), gives on using $P = 3.5$ TeV and 1fb$= 2.569
\times 10^{- 6}$~TeV$^{- 2}$:
\begin{equation}
  \label{first limit from search results} x^2_t  \bar{m}_t^{18} \mathrm{ln}
  \frac{2.1 \: \mathrm{{{TeV}}}}{m_t} < 0.01
  \frac{\mu^2_t }{m^2_t} \left(
  \frac{m_t}{\mathrm{{{TeV}}}} \right)^{0.20}
\end{equation}
in order of magnitude, which would have applied from about 120 GeV to 2 TeV if
the statistics had been sufficient. \ However the total number of background
events expected for $m_{e^+ e^-} > 1$ TeV is only about 1, so the limit from
{\cite{11081582 ATLAS}} is weaker than (\ref{first limit from search results})
for $m_t > 1$ TeV.

For the harmonic 3-forms, the reference estimate of the number $N$ of modes is
$N_{3\textrm{-}\mathrm{form,ref}} = 0.06 \frac{\bar{V}_7}{\mathrm{\ln}
\bar{V}_7}$, which for $\bar{V}_7 \sim 10^{34}$ is $\simeq 4 \times 10^6
N_{\mathrm{{{Weyl}}}, 0.2}$, so if the actual number
of approximately degenerate harmonic 3-form modes of intrinsic mass
$\bar{m}_{3 \mathrm{f}} \simeq 0.2$ is $N_{\mathrm{3 f}} = \tilde{x}_{3
\mathrm{f}} N_{3\textrm{-}\mathrm{form,ref}}$, the limit (\ref{first limit from
search results}) becomes:
\begin{equation}
  \label{second limit from search results} \tilde{x}_{3 \mathrm{f}}^2
  \mathrm{ln} \frac{2.1 \; \mathrm{{{TeV}}}}{m_{3
  \mathrm{f}}} < 0.01 \frac{\mu_{3 \mathrm{f}}^2}{m_{3 \mathrm{f}}^2} \left(
  \frac{m_{3 \mathrm{f}}}{\mathrm{{{TeV}}}}
  \right)^{0.20}
\end{equation}
in order of magnitude.

The logarithmic factor in the left-hand sides of (\ref{first limit from search
results}) and (\ref{second limit from search results}) decreases from about 3
at $m_t \simeq 120$ GeV to 0 at $m_t = 2.1$ TeV, so allowing for the low
statistics for $m_t > 1$ TeV, the limits from the search {\cite{11081582
ATLAS}} are that if $m_{\mathrm{3 f}}$ or an $m_t$ lies in the range from
about 120 GeV to about 1.5 TeV, then the corresponding adjustment factor
$\tilde{x}_{\mathrm{3 f}} = N_{\mathrm{3 f}} / N_{3
\textrm{-}\mathrm{form,ref}}$
or $x_t = N_t / N_{\mathrm{{{Weyl}}}, \bar{m}_t}$ is
bounded in order of magnitude by:
\begin{equation}
  \label{third limit from search results} \tilde{x}_{\mathrm{3 f}} < 0.1
  \frac{\mu_{\mathrm{3 f}}}{m_{\mathrm{3 f}}}, \hspace{2.5cm} x_t \bar{m}^9_t <
  0.1 \frac{\mu_t}{m_t} .
\end{equation}
The limits (\ref{third limit from search results}) give absolute bounds in
order of magnitude on the adjustment factors $\tilde{x}_{\mathrm{3 f}}$ and
$x_t$ if the corresponding mass $m_{\mathrm{3 f}}$ or $m_t$ lies in the range
from about 120 GeV to about 1.5 TeV, since $\frac{\mu_t}{m_t} \leq 1$ in order
of magnitude, and for $t$ other than the harmonic 3-forms, the intrinsic mass
$\bar{m}_t$ seems likely to be larger than 1, and possibly as large as $\sim
8$ or more.

The backgrounds (\ref{CMS mu mu background}), (\ref{CMS diphoton background}),
and (\ref{ATLAS diphoton background}) are equal in order of magnitude to the
background (\ref{ATLAS ee background}) at corresponding final state masses
$m_f$, and cover roughly the same range of $m_f$ from about 120 GeV to about 2
TeV, and the corresponding searches have the same lack of statistics for $m_f$
above 1 TeV as the search {\cite{11081582 ATLAS}}. \ Thus the limits from the
searches {\cite{CMS PAS EXO-11-039}}, {\cite{11120688 CMS}}, and
{\cite{11122194 ATLAS}} are also that if $m_{\mathrm{3 f}}$ or an $m_t$ lies
in the range from about 120 GeV to about 1.5 TeV, then the corresponding
adjustment factor $\tilde{x}_{\mathrm{3 f}}$ or $x_t$ is bounded in order of
magnitude by (\ref{third limit from search results}).

If the order of magnitude bound $\tilde{x}_{\mathrm{3 f}} < 0.1$ was not
satisfied, and $\tilde{x}_{\mathrm{3 f}}$ was also sufficiently large for
$m_{\mathrm{3 f}} < 120$ GeV to be excluded by earlier searches, for example
at the Tevatron and LEP, then since $m_{3 \mathrm{f}} \simeq 0.2 \frac{A}{B}$
from subsection \ref{The mass of the harmonic 3 form modes}, where the
constant $A$ in the metric ansatz (\ref{metric ansatz for H7}) lies between
about 0.7 and 0.9, from the discussion following (\ref{second bc in terms of
x}), and the best value of the constant $B$ in the metric ansatz (\ref{metric
ansatz for H7}) is $B \simeq 0.28 \kappa^{2 / 9}_{11} \simeq 1.2 M^{-
1}_{11}$, from subsection \ref{subsection PMS}, $m_{3 \mathrm{f}} > 1.5$ TeV
would imply $M_{11} > 10$ TeV, corresponding to $\kappa^{- 2 / 9}_{11} > 2.3$
TeV, which is a stronger limit than the current experimental lower bound on
$M_{11}$ for 7 flat extra dimensions, which is roughly $M_{11} \geq 2.3 \pm
0.7$ TeV, corresponding to $\kappa^{- 2 / 9}_{11} \geq 0.55 \pm 0.2$ TeV
{\cite{Franceschini et al, CMS-EXO-10-026, ATLAS-CONF-2011-096, CMS PAS
EXO-11-038, CMS PAS EXO-11-039, CMS PAS EXO-11-058, 11120688 CMS diphoton, CMS
PAS EXO-11-071, ATLAS-CONF-2011-065, ATLAS-CONF-2011-068}}.

If $m_{3 \mathrm{f}}$ is smaller than about 300 GeV, and $\tilde{x}_{3
\mathrm{f}}$ satisfies (\ref{third limit from search results}) if $m_{3
\mathrm{f}} > 120$ GeV, and is sufficiently small to have allowed the harmonic
3-form modes to have escaped discovery at the Tevatron, and at LEP if $m_{3
\mathrm{f}} < 209$ GeV, then $m_t$ could be under 1.5 TeV for some of the
other types of mode, if the second bound in (\ref{third limit from search
results}) is satisfied for that $t$. \ From the discussion before (\ref{cross
section in terms of m bar}), it seems likely that for $t$ other than the
harmonic 3-forms, a better estimate of the number $N_t$ of modes than
$N_{\mathrm{{{Weyl}}}, \bar{m}_t}$ might be $N_{\mathrm{Weyl},
\sqrt{\bar{m}_t^2 - \bar{m}_0^2}}$ for some $\bar{m}_0 < \bar{m}_t$, where
$N_{\mathrm{{{Weyl}}}, \bar{m}}$ was defined in (\ref{N Weyl}). \ Then the
second bound in (\ref{third limit from search results}) becomes:
\begin{equation}
  \label{fourth limit from search results} \left( \bar{m}^2_t - \bar{m}^2_0
  \right)^{\frac{7}{2}} \bar{m}^2_t < 0.1 \frac{\mu_t}{m_t} .
\end{equation}
This form of the bound cannot be used if $t$ denotes the lightest generic
modes of $t_{AB}$, which would be tachyonic unless the last two terms in
(\ref{gauge fixed hh action}) lift their squared masses sufficiently, but if
$t$ denotes the lightest generic classically massive modes of one of the
remaining types $C_{\mu \nu \sigma}$, $C_{\mu \nu A}$, $C_{\mu AB}$, $s_{\mu
\nu}$, $h_{AA}$, and $h_{\mu A}$, whose classical intrinsic masses
$\sqrt{\bar{m}^2_t - \bar{m}^2_0}$ lie in the range 1 to 3, then it implies
that $m_t > 1.5$ TeV. \ If we then assume that for at least one of these types
of mode, $m_t$ is not much larger than the mass
$m_{\mathrm{{{dil}}}}$ of the dilaton/radion mode of
$h_{AA}$, which from the paragraph after (\ref{dilaton kinetic terms}) is
$\simeq 9 \frac{A}{B}$, we find $M_{11} > 0.2$ TeV, which corresponds to
$\kappa^{- 2 / 9}_{11} > 50$ GeV. \ This then implies $m_{3 \mathrm{f}} \simeq
0.2 \frac{A}{B} > 30$ GeV. \ These limits do not depend on the value of
$\tilde{x}_{3 \mathrm{f}}$.

The tachyonic $\bar{m}^2$ at the bottom of the $\bar{m}^2$ spectrum of the
generic modes of $t_{AB}$, when the last two terms in (\ref{gauge fixed hh
action}) are neglected, is $\simeq - 4$ from the discussion after
(\ref{equation for tAB modes}), so if the last two terms in (\ref{gauge fixed
hh action}) contribute a term $\sim
\bar{m}^2_{\mathrm{{{dil}}}} \simeq 80$ to $\bar{m}^2$
for $t_{A B}$, the lightest generic modes of $t_{A B}$ will not be much
lighter than the lightest generic classically massive modes of the other types
other than $C_{A B C}$, so will also be heavier than around 1.5 TeV.

Thus\hspace{-0.47pt} it\hspace{-0.47pt} seems\hspace{-0.47pt}
likely\hspace{-0.47pt} that\hspace{-0.47pt} if\hspace{-0.47pt}
modes\hspace{-0.47pt} decaying\hspace{-0.47pt} along\hspace{-0.47pt}
the\hspace{-0.47pt} beam\hspace{-0.47pt} line\hspace{-0.47pt}
outside\hspace{-0.47pt} the\hspace{-0.47pt}
\mbox{interaction} region are to be observable at the LHC with 7 TeV or 8 TeV
c.o.m.\hspace{-0.3ex}
energy, these modes must be the $C_{A B C}$ harmonic 3-form modes whose mass
was approximately calculated in subsection \ref{The mass of the harmonic 3
form modes} as $m_{3 \mathrm{f}} \simeq 0.2 \frac{A}{B}$, and whose number is
$N_{3 \mathrm{f}} = \tilde{x}_{3 \mathrm{f}}
N_{3\textrm{-}\mathrm{form,ref}}$,
where $N_{3\textrm{-}\mathrm{form,ref}} = 0.06 \frac{\bar{V}_7}{\mathrm{\ln}
\bar{V}_7}$ from the discussion before (\ref{cross section in terms of m
bar}), and $\tilde{x}_{3 \mathrm{f}}$ satisfies the first bound in
(\ref{third limit from search results}) in order of magnitude. \ These modes
are pseudo-scalars along the extended dimensions, and were shown in subsection
\ref{The coupling of the harmonic 3 form modes to the SM gauge bosons} to have
axion-like couplings to the SM gauge bosons.

For these modes, the estimates in subsection \ref{Approximate distribution of
decay lengths} can be put on a slightly firmer foundation. \ Let $\bar{H}^n$
be a closed hyperbolic $n$-manifold, $n \geq 2$, and $\bar{H}^p$ be a closed
$p$-manifold that is embedded as a minimal-area $p$-cycle in $\bar{H}^n$,
where $1 \leq p \leq n - 1$, and $\bar{H}^p$ is closed hyperbolic for $p \geq
2$. \ Near $\bar{H}^p$ we can choose the coordinates on $\bar{H}^n$ to be
$\bar{x}^A
= \left( \hat{x}^a, \theta^i, \bar{y} \right)$, where $\hat{x}^a$ are
coordinates on $\bar{H}^p$, $\theta^i$ are coordinates on the unit $\left( n -
p - 1 \right)$-sphere, and $\bar{y}$ is the intrinsic geodesic distance from
$\bar{H}^p$. \ The metric is:
\begin{equation}
  \label{metric near minimal area p cycle} ds_n^2 = B^2 \bar{g}_{A B} d
  \bar{x}^A d \bar{x}^B = B^2 \left( \mathrm{\cosh}^2 \bar{y}\ \hat{g}_{a b} d
  \hat{x}^a d \hat{x}^b + \mathrm{\sinh}^2 \bar{y}\ \breve{g}_{i j} d \theta^i
  d \theta^j + d \bar{y}^2 \right),
\end{equation}
where $\hat{g}_{a b}$ is a metric on $\bar{H}^p$ of sectional curvature $- 1$,
and $\breve{g}_{i j}$ is a metric on the unit $\left( n - p - 1
\right)$-sphere.

Let $\omega_{A_1 \ldots A_p}$ be a harmonic $p$-form on $\bar{H}^n$ that
coincides with the $p$-volume form on $\bar{H}^p$ at $\bar{y} = 0$, and does
not closely coincide with a nonzero multiple of the $p$-volume form on any
other minimal-area $p$-cycle in $\bar{H}^n$. \ The integral $\int d
\hat{x}^{a_1} \ldots d \hat{x}^{a_p} \omega_{a_1 \ldots a_p} \left( \hat{x},
\theta, \bar{y} \right)$ at fixed $\theta^i$ and $\bar{y}$ is independent of
the $\theta^i$ and $\bar{y}$ by the generalized Stokes's theorem
{\cite{Wikipedia Stokess theorem}}, so for $\bar{y}$ up to the smallest value
at which a point of $\bar{H}^n$ has two different representations in these
coordinates, $\omega_{a_1 \ldots a_p} \left( \hat{x}, \theta, \bar{y} \right)$
will be approximately independent of the $\theta^i$ and $\bar{y}$. \ If
$\bar{H}^n$ has intrinsic diameter substantially larger than 1 and is
reasonably isotropic, in the sense that it has an approximately spherical
Dirichlet domain in $n$-dimensional hyperbolic space $H^n$, then $\omega_{a_1
\ldots a_p} \left( \hat{x}, \theta, \bar{y} \right)$ could be approximately
independent of the $\theta^i$ and $\bar{y}$ up to values of $\bar{y}$ that are
substantially larger than 1. \ In that case the integral $\int_{\bar{H}^n} d^n
\bar{x} \sqrt{\bar{g}} \bar{g}^{A_1 B_1} \ldots \bar{g}^{A_p B_p} \omega_{A_1
\ldots A_p} \omega_{B_1 \ldots B_p}$ will be approximately equal to the
contribution to it from the region with $\bar{y}$ less than about 2 or 3 if $2
p > n - 1$, since the factor $e^{- 2 p \bar{y}}$ from the inverse metrics then
outweighs the factor $e^{\left( n - 1 \right) \bar{y}}$ in $\sqrt{\bar{g}}$
for $\bar{y}$ larger than about 1, so for $2 p > n - 1$ the harmonic $p$-form
$\omega_{A_1 \ldots A_p}$ is effectively localized in a region of intrinsic
half-thickness $\bar{y} \sim 1$ centred at $\bar{H}^p$.

The case $p = 3$, $n = 7$ is on the borderline where this form of geometric
localization just fails to occur. \ If we convert the coordinate indices of
$\omega_{A_1 A_2 A_3}$ to local Lorentz indices by contraction with a vielbein
$\bar{e}^A \, \!_{\hat{B}}$, where hatted indices are local Lorentz indices
and $\delta^{\hat{C} \hat{D}} \bar{e}^A \, \!_{\hat{C}} \bar{e}^B \,
\!_{\hat{D}} = \bar{g}^{A B}$, then the coordinate scalar $\omega_{\hat{A}_1
\hat{A}_2 \hat{A}_3}$ has the same $\bar{y}$-dependence $e^{- 3 \bar{y}}$ for
moderate $\bar{y} > 1$ as the amplitude of the spherically symmetric
eigenmodes of the Laplace-Beltrami operator on uncompactified $H^7$
{\cite{Camporesi Higuchi 1, Camporesi Higuchi 2}}. \ However the effective
rate of decrease of $\omega_{\hat{A}_1 \hat{A}_2 \hat{A}_3}$ with increasing
$\bar{y}$ is expected to be more rapid than $e^{- 3 \bar{y}}$ due to Anderson
localization, which is an interference effect in which waves fail to propagate
in a disordered medium, due to interference between multiple scattering paths
{\cite{Anderson localization, Hundertmark, Lagendijk et al, 11055368 Kondov et
al}}.

The Ioffe-Regel criterion for Anderson localization of single-particle
wavefunctions in a disordered potential is that wavefunctions are localized
when the mean free path between scatterings is smaller than the wavelength
{\cite{Anderson localization, Ioffe Regel}}. \ The harmonic $3$-forms are
classically massless, so if $\bar{H}^7$ is reasonably isotropic, their
intrinsic wavelength on $\bar{H}^7$
is roughly the intrinsic diameter $\bar{L}_7$ of $\bar{H}^7$. \ The
classical dynamics of a free particle in a compact hyperbolic space is
strongly chaotic, and the Gutzwiller trace formula, which gives the
semiclassical correspondence for classically chaotic systems and relates a set
of periodic orbits along closed geodesics to a set of energy eigenstates,
becomes for compact hyperbolic spaces an exact relation known as the Selberg
trace formula {\cite{Inoue 1, Gutzwiller trace formula, Selberg trace
formula}}. \ Thus it seems likely that both classically and quantum
mechanically, the effective mean free path between scatterings on $\bar{H}^7$
will be at most $\bar{L}_7$, so that harmonic 3-forms on $\bar{H}^7$ will
behave roughly as $e^{- \left( 3 +
\frac{1}{\bar{L}_{\mathrm{{{loc}}}}} \right) \bar{y}}$
for $\bar{y} > 1$, where $\bar{L}_{\mathrm{{{loc}}}} >
0$ is the intrinsic localization length.

The intrinsic diameter $\bar{L}_3$ of a minimal-area $3$-cycle $\bar{H}^3$ in
$\bar{H}^7$ cannot be more than the intrinsic diameter $\bar{L}_7$ of
$\bar{H}^7$, so if $\bar{H}^3$ is reasonably isotropic, it cannot have
intrinsic $3$-volume $\bar{V}_3$ greater than $\sim e^{\left( 3 - 1 \right)
\bar{L}_7 / 2}$, from page 9 of {\cite{Almost Flat}}. \ For
$\bar{L}_7 \simeq 28$, from the discussion following (\ref{V7 in terms of GN}),
on page \pageref{V7 in terms of GN},
this gives $\bar{V}_3$ not above $\sim 10^{12}$, so that the
intrinsic 7-volume of the region of $\bar{H}^7$ within intrinsic distance
$\bar{y} < 1$ from $\bar{H}^3$ is not above
$\sim 10^{12}$, which is very small in
comparison to the intrinsic 7-volume $\bar{V}_7 \sim 10^{35}$ of $\bar{H}^7$,
if $\kappa_{11}^{-2/9}$ is comparable to its current experimental lower limit.
\ Thus for a rough first approximation we can treat $\bar{H}^3$ as a point,
and $\bar{y}$ as the intrinsic geodesic distance from that point, and to this
approximation the behaviour $e^{- \left( 3 +
\frac{1}{\bar{L}_{\mathrm{{{loc}}}}} \right) \bar{y}}$
of the harmonic 3-forms is the behaviour assumed in subsection
\ref{Approximate distribution of decay lengths}.

The radio-frequency (RF) cavities that accelerate the protons in the LHC
beams operate at 400 MHz, so the separation between adjacent RF ``buckets'' is
2.5 ns, which corresponds to a separation of 75 cm in the laboratory frame
{\cite{LHC Conceptual Design Report}}. \ The r.m.s.\hspace{-2.7px} length of
the bunch of
protons in a single RF bucket is 7.5 cm in the laboratory frame {\cite{LHC
Conceptual Design Report, CMS Technical Design Report}}, and during the 2011
proton-proton runs, one RF bucket in 20 was actually filled with a bunch, so
the actual separation between adjacent bunches was 15 metres in the laboratory
frame. \ This is also the planned separation between adjacent bunches for the
2012 proton-proton runs, for which the energy of a proton in one of the beams
is to be 4 TeV {\cite{Chamonix 2012}}.

From page 44 of {\cite{LHC Conceptual Design Report}}, the
r.m.s.\hspace{-3.0px} beam radius at
the interaction point (IP) of one of the two principal experiments was
initially planned to be 16$\mu$m, with the r.m.s.\hspace{-2.7px} divergence of
a beam at the IP
set at 32$\mu$rad, and the crossing angle set at 200$\mu$rad. \ Thus the
collisions would take place in the middle 7.5 cm of a beam crossing region of
length about 32 cm, that is about 32$\mu$m in diameter at its centre, and tapers to
a point at each end. \ ATLAS and CMS appear to use approximately these
parameters {\cite{ATLAS Technical Design Report Vol 1, ATLAS-CONF-2011-049,
ATLAS-CONF-2011-116, CMS Technical Design Report}}, except that from page 273
of {\cite{CMS Technical Design Report}}, the crossing angle in CMS is 285
$\mu$rad, and from pages 2 to 3 of {\cite{ATLAS-CONF-2011-137}},
the crossing angle in ATLAS might
also be 285~$\mu$rad. \ Thus in both ATLAS and CMS, the collisions take place
in an approximately cylindrical region of diameter $\simeq 32 \mu$m and length
$\simeq 7.5$ cm centred at the IP, and the experiments must detect jets and
charged leptons emitted from any point in this region, so as not to waste part
of the available luminosity.

In practice during 2011 ATLAS appears to have imposed a cut requiring the
distance $\left| z \right|$ along the beam line from the primary interaction
vertex to the IP to be less than 20 cm for inclusive final states or final
states containing muons, in order to reduce the background from cosmic ray
muons {\cite{11090934 ATLAS Searches for high mass dilepton resonances,
ATLAS-CONF-2011-047 Inclusive Jet and Dijet Cross Sections,
ATLAS-CONF-2011-082, ATLAS-CONF-2011-083}},
and CMS has sometimes imposed a cut requiring $\left| z \right|
< 12$ cm to reduce the background from cosmic ray muons {\cite{CMS PAS
EXO-11-004}}, while for dijet final states, ATLAS does not appear to impose
any cut on $\left| z \right|$ {\cite{11033864 ATLAS dijets,
ATLAS-CONF-2011-081, ATLAS-CONF-2011-095, 11086311 ATLAS}}, although in
practice a limit of roughly $\left| z \right| < 6$ cm might arise from finding
the event vertex or vertices using tracks that originate in the beam collision
spot {\cite{ATLAS-CONF-2011-043 multi-jet cross sections}}, since for 7 TeV
c.o.m.\hspace{-2.3px} energy, the $z$-distribution of primary interaction vertices is a
Gaussian with $\sigma \simeq 2.2$ cm {\cite{ATLAS-CONF-2010-027 Interaction
Point Beam Parameters}}. \ For a rough estimate at 7 TeV or 8~TeV c.o.m.
energy, I shall treat the interaction region as extending for 6 cm in each
direction along the beam line from the IP.

The ATLAS Inner Detector is 7 metres in length along the beam line
{\cite{ATLAS Technical Design Report Vol 1}}, and the CMS Inner Tracking
System is 5.4 metres in length along the beam line {\cite{CMS Technical Design
Report}}. \ The central barrel part of the ATLAS Inner Detector is
1.6 metres in length, with the remainder of the length of the Inner Detector
consisting of two identical end caps, and the CMS Tracker Inner Barrel is 1.3
metres in length, surrounded by the Tracker Outer Barrel which is 2.2 metres
in length. \ From page 24 of {\cite{ATLAS-CONF-2011-137}},
the ATLAS detector is capable of measuring the $z$ values of
tracks roughly perpendicular to the beam line up to at least $\left| z \right|
= 1$ metre, and thus beyond the central barrel part of the ATLAS Inner
Detector, and from page 3 of {\cite{CMS PAS EXO-11-004}}, CMS is capable of
reconstructing tracks from decays that occur up to 50 cm from the beam line,
although with significantly less than 100\% efficiency. \ I shall assume that
both ATLAS and CMS can approximately measure the $z$ values of tracks roughly
perpendicular to the beam line, over the whole length of their Inner Detector
or Inner Tracking System, although with substantially less than 100\%
efficiency for finding tracks at the larger $\left| z \right|$ values.

For a reference estimate I shall consider the ATLAS Inner Detector, and thus
consider modes that decay along the beam line at a distance between 6 cm and
3.5 metres from the IP.
From (\ref{rate of particles decaying between z and z plus dz}) and
(\ref{integral of the decay rate along the beam line}), on page \pageref{rate
of particles decaying between z and z plus dz}, the fraction of the harmonic
3-form modes, of intrinsic mass $\bar{m}_{\mathrm{3f}}\simeq 0.2$, that
decay further than a
distance $z \gg l_{\mathrm{\min}} \sim 10^{- 19}$ metres along the beam line
from the IP, is approximately $\left( \frac{l_{\mathrm{\min}}}{z}
\right)^{\frac{1}{3 \bar{L}_{\mathrm{{{loc}}}} + 1}}
\sim \left( \frac{10^{- 19} \; \mathrm{{{metres}}}}{z}
\right)^{\frac{1}{3 \bar{L}_{\mathrm{{{loc}}}} + 1}}$,
for $\bar{L}_{\mathrm{{{loc}}}} > 1$, where
$\bar{L}_{\mathrm{{{loc}}}}$ is the intrinsic
localization length of the harmonic 3-forms on $\bar{H}^7$. \ For reasonably
isotropic $\bar{H}^7$, whose intrinsic diameter $\bar{L}_7$ would from the
discussion following (\ref{V7 in terms of GN}), on page \pageref{V7 in terms
of GN}, be about 28, if $\kappa^{- 2 / 9}_{11}$ and $M_{11}$ are close to
their current experimental lower limits, for 7 flat extra dimensions, of about
0.55 TeV and 2.3 TeV respectively,
$\bar{L}_{\mathrm{{{loc}}}}$ would be expected, from
the above discussion of Anderson localization, to be somewhere in the range
from about 4 to about 28. \ Let
\begin{equation}
  \label{f00635} f_{0.06, 3.5} \left(
  \bar{L}_{\mathrm{{{loc}}}} \right) \equiv \left(
  \frac{10^{- 19} \; \mathrm{{{metres}}}}{0.06 \;
  \mathrm{{{metres}}}} \right)^{\frac{1}{3
  \bar{L}_{\mathrm{{{loc}}}} + 1}} - \left(
  \frac{10^{- 19} \; \mathrm{{{metres}}}}{3.5 \;
  \mathrm{{{metres}}}} \right)^{\frac{1}{3
  \bar{L}_{\mathrm{{{loc}}}} + 1}}
\end{equation}
be the fraction of the harmonic 3-form modes that decay between 6 cm
and 3.5 metres along the beam line from the IP. \ We then find the values:
\vspace{0.3cm}

\hspace{-0.23cm}\begin{tabular}{|c|c|c|c|c|c|c|c|c|c|}
\hline
  $\bar{L}_{\mathrm{{{loc}}}}$ & 1 & 2 & 4 & 8 & 12 &
  16 & 20 & 24 & 28\\
\hline
  $f_{0.06, 3.5}$ & 0.000023 & 0.0013 & 0.012 & 0.029 & 0.034 & 0.035 & 0.033
  & 0.031 & 0.029\\
\hline
\end{tabular}
\vspace{0.3cm}

Thus for $\bar{L}_{\mathrm{{{loc}}}}$ throughout most
of the expected range, $f_{0.06, 3.5} \left(
\bar{L}_{\mathrm{{{loc}}}} \right) \simeq 0.03$, and
this is valid within a factor of 3 throughout the whole expected range. \ Thus
from (\ref{cross section in terms of m bar}) and the following discussion, on
page \pageref{cross section in terms of m bar}, with $\bar{m} = 0.2$, and the
discussion around (\ref{first limit from search results}) and (\ref{second
limit from search results}), on page \pageref{first limit from search
results}, we obtain:
\begin{equation}
  \label{cross section for decay outside interaction region} \sigma \left(
  \mathrm{prot} + \mathrm{prot} \rightarrow \mathrm{any} \hspace{0.5em} n + X
  \rightarrow f + X \right)_{0.06, 3.5} \sim 10^{- 5} 
  \frac{\tilde{x}^2_{\mathrm{3 f}}}{\mu_{\mathrm{3 f}} m_{\mathrm{3 f}}}
  \left( \frac{P}{m_{\mathrm{3 f}}} \right)^{2.34} \mathrm{ln} \frac{0.6
  P}{m_{\mathrm{3 f}}},
\end{equation}
as an order of magnitude estimate of the cross section for a harmonic 3-form
mode of intrinsic mass $\bar{m}_{\mathrm{3 f}} \simeq 0.2$ and mass
$m_{\mathrm{3 f}} = \bar{m}_{\mathrm{3 f}} \frac{A}{B} < 0.6 P$ to be produced
in the $s$-channel and decay between 0.6 cm and 3.5 metres along the
beamline from the IP, where the number of approximately degenerate harmonic
3-form modes of intrinsic mass $\bar{m}_{\mathrm{3 f}} \simeq 0.2$ is
$N_{\mathrm{3 f}} = \tilde{x}_{\mathrm{3 f}} N_{3\textrm{-}
\mathrm{{{form}}, {{ref}}}} =
\tilde{x}_{\mathrm{3 f}} \, 0.06 \frac{\bar{V}_7}{\mathrm{\ln}
\bar{V}_7}$, $P$ is the energy per proton, which was 3.5 TeV at the LHC in
2011, and is to be 4.0 TeV at the LHC in 2012 {\cite{Chamonix 2012}}, the warp
factor $A$ lies between about 0.7 and 0.9, from the discussion between
(\ref{second bc in terms of x}), on page \pageref{second bc in terms of x},
and (\ref{Einstein action}), on page \pageref{Einstein action}, the curvature
radius $B$ of $\bar{H}^7$ is $B \simeq 0.28 \kappa^{2 / 9}_{11}$, from
subsection \ref{subsection PMS}, starting on page \pageref{subsection PMS},
and if $m_{\mathrm{3 f}}$ lies in the range from about 120 GeV to about 1.5
TeV, then $\tilde{x}_{\mathrm{3 f}}$ is bounded in order of magnitude by
$\tilde{x}_{\mathrm{3 f}} < 0.1 \frac{\mu_{\mathrm{3 f}}}{m_{\mathrm{3 f}}}$,
from (\ref{third limit from search results}), on page \pageref{third limit
from search results}. \ $\mu_{\mathrm{3 f}}$ is the width of the distribution
of the masses of the harmonic 3-form modes, which was assumed in subsection
\ref{Approximate total cross-section}, starting on page \pageref{Approximate
total cross-section}, to be $\geq \Gamma_n$ for all the harmonic 3-form modes
$\left| n \right\rangle$, where $\Gamma_n$ is the total width of the mode
$\left| n \right\rangle$, in order to derive the total cross-section estimate
(\ref{cross section in terms of m bar}), on page \pageref{cross section in
terms of m bar}.

Using 1 $\mathrm{{{TeV}}^{- 2} = 0.3893}$ nb and the
limit (\ref{third limit from search results}), (\ref{cross section for decay
outside interaction region}) becomes:
\begin{equation}
  \label{bound on cross section for decay outside IR} \sigma \left(
  \mathrm{prot} + \mathrm{prot} \rightarrow \mathrm{any} \hspace{0.5em} n + X
  \rightarrow f + X \right)_{0.06, 3.5} < \frac{\mu_{\mathrm{3
  f}}}{m_{\mathrm{3 f}}}  \left( \frac{m_{\mathrm{3
  f}}}{\mathrm{{{TeV}}}} \right)^{- 4.34} \;
  \mathrm{{{fb}}},
\end{equation}
as an order of magnitude upper limit on the cross-section for a harmonic
3-form mode of mass $\simeq m_{\mathrm{3 f}}$ to be produced in the
$s$-channel and decay between 6 cm and 3.5 metres along the beam line
from the IP, for $P = 3.5$ or 4 TeV per proton, and $m_{\mathrm{3 f}}$ between
about 120 GeV and 1.5 TeV. \ Thus if $\tilde{x}_{\mathrm{3 f}}$ is at the
upper limit allowed by (\ref{third limit from search results}), and
$\mu_{\mathrm{3 f}} \sim m_{\mathrm{3 f}}$, then at the LHC design luminosity
of 10 per nb per second {\cite{LHC Conceptual Design Report}}, there would be
about 0.1 such events per second if $m_{\mathrm{3 f}}$ is 120 GeV, and about
$10^{- 6}$ such events per second if $m_{\mathrm{3 f}}$ is 1.5 TeV, and if the
LHC delivers the expected 15 to \mbox{19 per fb} to ATLAS and CMS during 2012
{\cite{Chamonix 2012}}, there would be about $10^5$ such events in ATLAS and
CMS during 2012 if $m_{\mathrm{3 f}}$ is 120 GeV, and about 1 such event in
ATLAS and CMS during 2012 if $m_{\mathrm{3 f}}$ is 1.5 TeV.

\begin{figure}[t]
\includegraphics[width=0.995\textwidth]{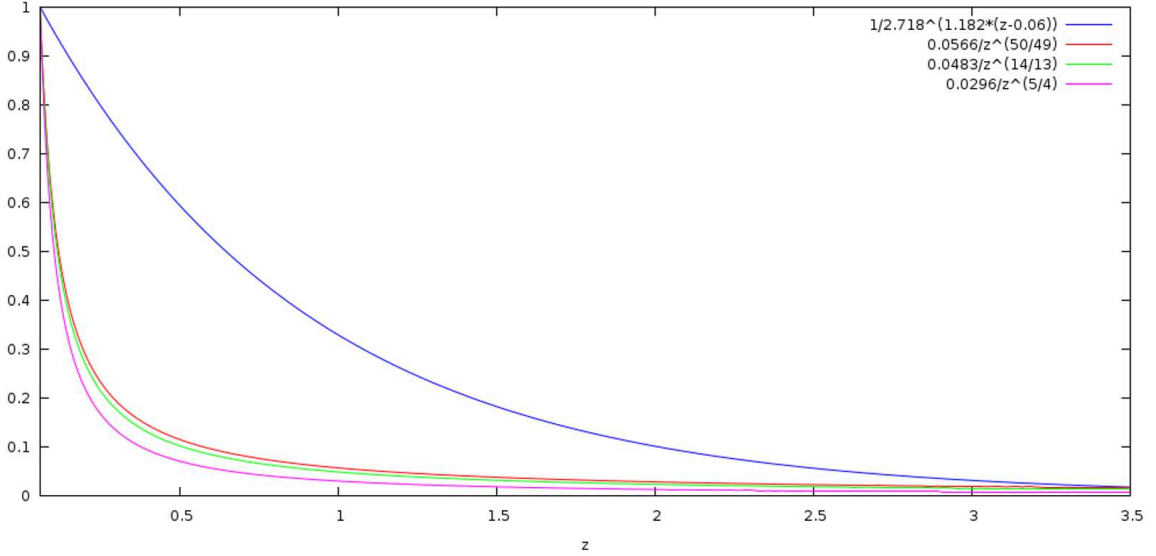}
\caption{Comparison of exponential and power dependence on $z$, for $z \geq 0.06$}
\label{Compare exponential with power}
\end{figure}

Figure \ref{Compare exponential with power}
shows the $z$-dependence $z^{- \frac{3
\bar{L}_{\mathrm{{{loc}}}} + 2}{3
\bar{L}_{\mathrm{{{loc}}}} + 1}}$ of the number of
harmonic 3-form modes decaying between distances $z$ and $z + \mathrm{d} z$
along the beam line from the IP, for $6 \hspace{0.2em}
\mathrm{{{cm}}} \leq z \leq 3.5 \hspace{0.2em} \mathrm{{{metres}}}$ and fixed
$\mathrm{d} z$, normalized to 1 at $z = 6 \hspace{0.2em} \mathrm{{{cm}}}$, for
$\bar{L}_{\mathrm{{{loc}}}} = 16$, 4, and 1, together
with an exponential curve that matches the limiting case of large
$\bar{L}_{\mathrm{{{loc}}}}$ at $z = 6 \hspace{0.25em}
\mathrm{{{cm}}}$ and $z = 3.5 \hspace{0.25em}
\mathrm{{{metres}}}$. \ This figure shows that the
power-law $z$-dependence could be distinguished from a single exponential with
a relatively small number of events, but it could be difficult to distinguish
different values of $\bar{L}_{\mathrm{{{loc}}}}$ in
the relevant range of about $4 \leq
\bar{L}_{\mathrm{{{loc}}}} \leq 28$, even with the
$\sim 10^5$ events expected during 2012 if $m_{\mathrm{3 f}}$ is 120 GeV.

The principal backgrounds to this process are beam-induced backgrounds and
cos-mic-ray showers {\cite{ATLAS-CONF-2011-137}}. \
Beam-induced backgrounds are due to proton losses upstream of the IP. \ These
result in cascades of secondary particles that fly through the detectors
almost parallel to the beam line. \ The cosmic-ray showers are produced by
cosmic rays, mostly protons and heavier nuclei, colliding with atoms in
the Earth's atmosphere, and
muons produced in these showers can penetrate down to the ATLAS and CMS
detectors, which are situated in caverns about 100 metres underground
{\cite{Performance of the CMS Hadron Calorimeter with Cosmic Ray Muons}}. \
The cosmic ray muons that reach ATLAS come mostly from above, and arrive
mainly via two large access shafts that were used for the detector
installation {\cite{10116665 ATLAS Cosmic Ray Muons}}.

The harmonic 3-form modes are pseudo-scalars along the $3 + 1$ extended
dimensions, so their decay is isotropic in their rest frame. \ Their decay
products will be boosted in the direction away from the IP in the laboratory
frame, so the background from both beam-induced backgrounds and cosmic-ray
muons could be reduced by selecting events where at least 2 charged leptons or
2 jets originate from a primary vertex that is at least 6 cm from the
IP along the beam line, but within a few mm of the beam line in the
transverse directions, with no missing transverse momentum, and a significant
net longitudinal momentum in the direction away from the IP.

The initial, hardware-based stages of the ATLAS and CMS trigger systems use
information only from the from the muon systems and calorimeters, so they
accept events of this type. \ Approximate track reconstruction is not carried
out until the later, software-based stages of the trigger systems, which can
use the high-resolution position data from the inner detectors, in addition to
the data from the muon systems and calorimeters {\cite{ATLAS Technical Design
Report Vol 1, CMS Technical Design Report, ATLAS-CONF-2010-027
Interaction Point Beam Parameters, ATLAS-CONF-2010-069 Performance Of Primary
Vertex Reconstruction}}. From the discussion before equation (\ref{f00635})
above, ATLAS and CMS are able to reconstruct approximately the tracks from
primary
interaction vertices up to around 50 cm to 1 metre from the IP along the beam
line, and their high-level triggers can accept and store these events for
offline analysis. \ If sufficient rejection of the beam-induced backgrounds and
the cosmic ray background could be achieved without reducing the signal too
much, and $\tilde{x}_{\mathrm{3 f}}$ is at the upper limit allowed by
(\ref{third limit from search results}), and $\mu_{\mathrm{3 f}} \sim
m_{\mathrm{3 f}}$, then the order of magnitude estimate (\ref{bound on cross
section for decay outside IR}) suggests that a 5-sigma discovery of the
harmonic 3-form modes decaying more than 6 cm along the beam line from the IP
could be achieved in 2012, if their central mass $m_{\mathrm{3 f}}$ is not
more than about 900 GeV,
which corresponds roughly to $\kappa^{- 2 / 9}_{11} < 1.6$ TeV and $M_{11} <
7$~TeV.
\vspace{1cm}

\begin{center}
  {\bfseries{Acknowledgements}}
\end{center}

\noindent I would like to thank the organizers of the 2007 CERN BSM Institute,
in particular Nima Arkani-Hamed, Savas Dimopolous, and Christophe Grojean, for
arranging for me to give a talk and spend a very enjoyable and helpful week at
CERN with financial support, Asimina Arvanitaki, Savas Dimopoulos, Philip
Schuster, Jesse Thaler, Natalia Toro, and Jay Wacker for very interesting
discussions, Greg Moore for a helpful email about flux quantization, Kasper
Peeters for correspondence about Cadabra both on and off the mailing list,
bloggers Philip Gibbs, ``Jester'', Lubo\v{s} Motl, Matt Strassler,
Tommaso Dorigo, and Peter Woit for providing timely updates and discussion on
current developments in high energy physics, and Jeff McGowan for a discussion
on Peter Woit's blog that led me to find the examples in the last paragraph of
section \ref{Introduction}. \ The calculations made heavy use of Maxima
{\cite{Maxima}} and Cadabra \cite{Cadabra 1, Cadabra 3, Cadabra 6},
the diagram was prepared using Maxima and GNUPlot \cite{GNUPlot},
the bibliography was sequenced with help from Ordercite \cite{Ordercite},
the work was done in notebooks written with GNU TeXmacs {\cite{TeXmacs}}
running in \mbox{KDE 4.4} \cite{KDE} in Debian GNU/Linux \cite{Debian}, and the
article was written with GNU TeXmacs and completed with Kile \cite{Kile}.

\end{document}